\documentclass[twocolumn]{aastex63}
\usepackage{amsmath}
\usepackage[utf8]{inputenc}
\usepackage{multirow}
\usepackage{dcolumn}
\usepackage{color}

\newcommand{\Msunyr}{\hbox{$M_\odot \,\hbox{yr}^{-1}$}}

\received{February 16, 2021}
\revised{June 25, 2021}
\accepted{June 28, 2021}
\submitjournal{ApJ}

\shorttitle{New hydrodynamic solutions for line-driven winds using Lambert $W$-function}
\shortauthors{Gormaz-Matamala et al.}

\begin{document}

\title{New hydrodynamic solutions for line-driven winds of hot massive stars using Lambert $W$-function}

\correspondingauthor{Alex C. Gormaz-Matamala}
\email{alex.gormaz@uv.cl}

\author[0000-0002-0786-7307]{A. C. Gormaz-Matamala}
\affil{Instituto de F\'isica y Astronom\'ia, Universidad de Valpara\'iso, Av. Gran Breta\~na 1111, Casilla 5030, Valpara\'iso, Chile}
\affiliation{Centro de Astrofísica, Universidad de Valparaíso. Av. Gran Breta\~na 1111, Casilla 5030, Valpara\'io, Chile.}
\affiliation{Departamento de Ciencias, Facultad de Artes Liberales, Universidad Adolfo Ib\'a\~nez, Av. Padre Hurtado 750, Vi\~na del Mar, Chile}

\author[0000-0002-2191-8692]{M. Cur\'e}
\affil{Instituto de F\'isica y Astronom\'ia, Universidad de Valpara\'iso, Av. Gran Breta\~na 1111, Casilla 5030, Valpara\'iso, Chile}
\affiliation{Centro de Astrofísica, Universidad de Valparaíso. Av. Gran Breta\~na 1111, Casilla 5030, Valpara\'io, Chile.}

\author{D. J. Hillier}
\affil{Department of Physics and Astronomy \& Pittsburgh Particle Physics, Astrophysics, and Cosmology Center (PITT PACC),\\University of Pittsburgh, Pittsburgh, PA 15260, USA}

\author{F. Najarro}
\affil{Centro de Astrobiología (CSIC/INTA), ctra. de Ajalvir km. 4, 28850 Torrejón de Ardoz, Madrid, Spain}

\author{B. Kubátová}
\affil{Astronomický ústav, Akademie věd České republiky, 251 65 Ondřejov, Czech Republic}

\author{J. Kubát}
\affil{Astronomický ústav, Akademie věd České republiky, 251 65 Ondřejov, Czech Republic}



\begin{abstract}

Hot massive stars present strong stellar winds that are driven by absorption, scattering and re\-emission of photons by the ions of the atmosphere (\textit{line-driven winds}).
A better comprehension of this phenomenon, and a more accurate calculation of hydrodynamics and radiative acceleration is required to reduce the number of free parameters in spectral fitting, to determine accurate wind parameters such as mass-loss rates and velocity profiles.

We use the non-LTE model-atmosphere code CMFGEN to numerically solve the radiative transfer equation in the stellar atmosphere and to calculate the radiative acceleration $g_\text{rad}(r)$.
Under the assumption that the radiative acceleration depends only on the radial coordinate, we solve analytically the equation of motion by means of the Lambert $W$-function.
An iterative procedure between the solution of the radiative transfer and the equation of motion is executed in order to obtain a final self-consistent velocity field that is no longer based on any $\beta$-law.

We apply the Lambert-procedure to three O supergiant stars ($\zeta$-Puppis, HD~165763 and $\alpha$-Cam) and discuss the Lambert-solutions for the velocity profiles.
It is found that, even without recalculation of the mass-loss rate, the Lambert-procedure allows the calculation of consistent velocity profiles that reduce the number of free parameters when a spectral fitting using CMFGEN is performed.
Synthetic spectra calculated from our Lambert-solutions show significant differences compared to the initial $\beta$-law CMFGEN models.
The results indicate the importance of consistent velocity profile calculation in the CMFGEN code and its usage in a fitting procedure and interpretation of observed spectra.

\end{abstract}

\keywords{Radiative transfer -- Hydrodynamics -- Stars: winds -- Stars: atmospheres -- Stars: early-type -- Stars: individual}


\section{Introduction} \label{sec:intro}
	Massive stars play an important role in astrophysics because they deposit momentum, energy and nuclear processed material into the interstellar medium on a relatively short timescale by means of their powerful stellar winds \citep[see reviews from][]{kudritzki00,puls08}.
	The amount of matter outflowing from the star (set by the mass-loss rate, $\dot M$) is also directly linked with the evolution of these stars: it has been found, for example, that differences on a factor of two in $\dot M$ considerably affect the final fate of a massive star \citep{meynet94,georgy12}.
	Therefore, a better understanding about massive stars and their evolution  requires an accurate determination of their fundamental wind parameters.
	
	The wind parameters are commonly determined by comparing observed and synthetic spectra, since some spectral features (e.g., H$_\alpha$ and UV resonance transitions) are sensitive to the adopted wind parameters.
	Great advances in the accuracy of synthetic spectral calculations have been achieved with the help of NLTE radiative transfer codes such as WMBASIC \citep[e.g.,][]{pauldrach01}, FASTWIND \citep{santolaya97,repolust04,puls05}, PoWR \citep{grafener02,hamann03} and CMFGEN \citep{hillier90a,hillier90b,hillier98,hillier01}.
	
	Many of the codes use the so-called $\beta$-law to prescribe the velocity field; with this approach the $\beta$ parameter, the terminal velocity ($v_\infty$) and the mass-loss rate ($\dot M$) are treated as fitting parameters.
	However such a procedure does not guarantee consistency between the wind hydrodynamics and the stellar radiation field; every wind parameter is treated as a separate entity instead of considering them as linked parts of a hydrodynamic system.
	The inclusion of a self-consistent hydrodynamic solution (i.e., a solution where the hydrodynamic properties of the wind come from the radiation force and vice versa) in the spectral fit procedure is important because it reduces the number of free parameters to be determined, potentially leading to greater confidence in results derived from spectral fitting.

	Self-consistent hydrodynamic solutions require the calculation of the acceleration that drives the wind.
	Thanks to the pioneering work of \citet{lucy70} we know that the winds of hot stars are accelerated by the radiation field emanating from the photosphere of the star.
	Momentum is delivered to the wind primarily by scattering (and absorption/re-emission) of photons in bound-bound (i.e., line) transitions.
	A quantitative description of line-driven winds was first performed by \citet[][hereafter called CAK theory]{cak} and later improved by \citet{abbott82}, \citet{ppk} and \citet[][called modified or m-CAK theory]{friend86}.
	The approximations and assumptions underlying the m-CAK theory, and their implications for the wind dynamics, have been extensively discussed \citep{schaerer94,puls00,kk10}.
%
%

	Over the last decade there has been a renewed effort to improve our understanding of line-driven winds.
	\cite{hillier03} stressed the importance of the adopted Doppler velocity (with contribution from both thermal motions and turbulence) on setting the line acceleration around the sonic point, and \cite{lucy07} studied the wind dynamics around the sonic point, and argued that the sonic point was the true critical point for the wind, and confirmed the importance of the adopted Doppler velocity for setting the mass-loss rate and obtained mass-loss rates roughly a factor of 2 lower than those obtained by \citet{vink00,vink01}.
	
	More recently there was the study of \citet{sander17}, who calculated consistent line-acceleration and hydrodynamics for the O4 I(n)f star $\zeta$-Puppis using the PoWR code.
	The advantage of using the PoWR code is that the line-force does not rely on the use of the Sobolev approximation,  and thus it computes correctly the line force in the neighbourhood of the sonic point.
	\citet{kk17} have provided new confident theoretical values for wind parameters of O dwarfs, giants and supergiants by solving the radiative transfer equation and hydrodynamics simultaneously.
	Along the same lines, we also mention the recent studies of \citet{sundqvist19} and \citet{bjorklund21}, who used their full NLTE CMF radiative transfer solutions from the code FASTWIND \citep{santolaya97,sundqvist18} for calculation of the radiative acceleration responsible for the wind driving.
		
	Based on the m-CAK theory, \citet[][hereafter Paper I]{alex19} developed a prescription to derive self-consistent solutions given different sets of stellar parameters for hot massive stars, following the quasi-NLTE treatment for atomic populations of \citet{mazzali93}, the radiation field calculated from \textsc{Tlusty} \citep{hubeny95} and the hydrodynamic solutions performed by \textsc{HydWind} \citep{michel04,michel07}. 
	Paper I demonstrated that the prescription provided useful theoretical wind parameters for different sets of temperature, surface gravities, abundances and even the rotational velocity.
	
	In this work, we aim to obtain a self-consistent solution without using the m-CAK model, and employing a full NLTE treatment for atomic populations to calculate the line-acceleration $g_\text{line}(r)$.
	To obtain a reliable expression for the line-acceleration we use the radiative transfer code CMFGEN.
	We then use the Lambert $W$-function \citep{corless93,corless96,muller08,araya14}, which is a mathematical function that arises as a solution of the wind's stationary equation of motion (see Appendix \ref{lambertdefinition} for a formal definition of Lambert $W$-function) to obtain a new self-consistent velocity profile no longer based in the $\beta$-law formalism.
	
	The ultimate aim of our studies is to obtain a hydrodynamic solution consistent with the stellar parameters, and which gives synthetic spectra comparable with observation.
	However, due to our underlying assumptions this will not always be realisable.
	In particular the hydrodynamics will depend on the adopted abundances and atomic models.
	Further, the wind structure depends on the (unknown) clumping and porosity structure of the wind, and the photospheric microturbulence (especially in the neighbourhood of the sonic point).
	Photospheric convection cells, if they exist, will also affect the wind dynamics.
	Large scale motions (macroturbulence) exceeding the sound speed are also known to exist in the photosphere of O stars \cite[e.g.,][]{howarth97,simon17}, and their origins are not understood.
	However the use of theoretical velocity laws, even if only partially consistent with the hydrodynamics, will provide useful insights to aid spectral modelling and the study of wind dynamics.

	The focus of this work is to derive a self-consistent velocity profile for the wind assuming a given mass-loss rate and volume-filling factor.
	The advantage of this approach is flexibility; the mass-loss rate can still be obtained from spectral fitting, unconstrained by uncertainties in the wind dynamics near the sonic point.
	In a forthcoming paper, we will explore alternate methods to constrain mass-loss rates in order to obtain a fully consistent description of the wind.
	
	The structure of this work is as  follows: the general theoretical framework of wind hydrodynamics is introduced in Section~\ref{windhydrodynamics}, while detailed information about CMFGEN, and its use in our calculations, is outlined in Section~\ref{cmfgen}.
	In Section~\ref{lambertprocedure}, the methodology of the Lambert-procedure to calculate the self-consistent velocity profiles is described.
	An evaluation of the converged results is presented in Section~\ref{results} while a brief comparison of synthetic spectra with observational data is presented in Section~\ref{obsdata}.
	A comparison of our results with earlier studies, and the consequences of instabilities for the determination of mass-loss rates, are presented in Section~\ref{discussion}.
	Finally, our summary and conclusions are given in Section~\ref{conclusions}.

\section{Wind hydrodynamics}\label{windhydrodynamics}	
	The standard line-driven theory relies on the assumption that the winds of massive stars are spherically symmetric, stationary, smooth, non-rotating, and that the effects of viscosity, heat conduction and magnetic fields can be neglected.
	
	Under the above assumptions the velocity $v(r)$ and density $\rho(r)$ profiles are coupled by the equation of mass conservation, which connects them with the mass-loss rate
	
	\begin{equation}\label{continuity}
		\dot M=4\pi\rho(r)r^2 v(r)\;,
	\end{equation}
	and by the radial equation of momentum
	
	\begin{equation}\label{momentumgrad}
		v\frac{dv}{dr}=-\frac{1}{\rho}\frac{dp}{dr}-\frac{GM_*}{r^2}+g_\text{rad}\;,
	\end{equation}
	where $p$ is the gas pressure, $ v$ is the wind velocity and $GM_*/r^2$ is the gravitational acceleration $g_\text{grav}$.
	The term $g_\text{rad}$ is the radiative acceleration, and it corresponds to the \textit{total} acceleration due to all radiative processes, i.e., it not only considers the effects of absorption and reemission of photons by line transitions, but also electron scattering and continuum opacity.
	Acceleration due to electron scattering is expressed as
	
	\begin{equation}\label{gelec}
		g_\text{elec}(r)=\frac{GM_*\Gamma_e}{r^2}\;\;,
	\end{equation}
	with $\Gamma_e$ being the Eddington parameter:
	
	\begin{equation}\label{eddington}
		\Gamma_e=\frac{\sigma_\text{e}L_*}{4\pi c\,GM_*}\;,
	\end{equation}
	where $\sigma_\text{e}$ is the Thomson scattering cross section, $L_*$ is the stellar luminosity, $M_*$ is the stellar mass, $c$ is the light speed, and $G$ is the gravitational constant.
	
	Because Eq.~\ref{gelec} has the same radial dependence to the acceleration due to gravitation, $g_\text{grav}=GM_*/r^2$ (indeed, electron scattering acceleration acts to counter gravitational acceleration), it is possible to subtract $g_\text{elec}$ from the radiative acceleration and include it by means of an acceleration due to \textit{effective} gravity defined as
	
	\begin{eqnarray}\label{effectivegravity}
		g_\text{eff}&=&g_\text{grav}-g_\text{elec}\;,\nonumber\\
		&=&\frac{GM_\text{eff}}{r^2}\;,\nonumber\\
		&=&-\frac{GM_*(1-\Gamma_e)}{r^2}\;.
	\end{eqnarray} 
	
	The acceleration due to line-effects only (i.e., line-acceleration) corresponds to the total radiative acceleration minus the acceleration coming from the electron scattering\footnote{In Eq.~\ref{glinedef}, the contribution by continuum acceleration is also included. However, these processes have been found to be negligible for the winds of O, B and A stars \citep{gagnier19}.}:
	
	\begin{equation}\label{glinedef}
		g_\text{line}=g_\text{rad}-g_\text{elec}(r)\;.
	\end{equation}

	The  solution of the equation of momentum is not straightforward because the line-acceleration is a non-linear function of the radius, density, velocity, and the velocity gradient.
	Thus, in general, numerical techniques are needed to find solutions \citep[e.g.][]{michel04}.
	However, if we can express the line-acceleration as a function of radius only, the equation of momentum can be analytically solved.
	
\subsection{Solution of the momentum equation}\label{solutionmomentum}
	Taking into account the differentiation between radiation and line-acceleration outlined in the previous paragraphs, we rewrite the equation of momentum as
	
	\begin{equation}\label{motion1}
		 v\frac{d v}{dr}=-\frac{1}{\rho}\frac{dp}{dr}-\frac{GM_\text{eff}}{r^2}+g_\text{line}\;\;,
	\end{equation}
	where $p$ is the gas pressure, $ v$ is the wind velocity and $GM_\text{eff}/r^2$ is the effective gravitational acceleration introduced in Eq.~\ref{effectivegravity}.
		
	Following \citet{muller08} and \citet{araya14}, we express Eq. \ref{motion1} in dimensionless form by introducing the dimensionless variables
	
	\begin{equation}\label{changeofvariables}
		\hat r=\frac{r}{R_*}\;,\;\;\;\hat v=\frac{v}{a}\;,
	\end{equation}
	where $R_*$ is the stellar radius and $a$ is the isothermal sound speed given by
	
	\begin{equation}\label{sound_plus_vturb}
		a^2=\frac{k_BT_\text{eff}}{\mu\,m_H}+\frac{1}{2} v_\text{turb}^2\;.
	\end{equation}
	
	Unlike \citet{sander17} we need to assume a constant sound speed.
	However the assumption of an isothermal wind near the sonic point is generally an excellent approximation for O stars, and above the sonic point the sound speed quickly becomes irrelevant for the dynamics.
	The turbulent velocity $v_\text{turb}$ is included to solve the equation of motion (hereafter e.o.m.), and is commonly assumed to be in the order of $\sim10-15$ km s$^{-1}$ \citep{villamariz00,puls05}.
	In principle the turbulent velocity can be derived by spectral fitting.
	
	Defining the dimensionless line acceleration
	
	\begin{equation}
		\hat g_\text{line}=\frac{R_*}{a^2}g_\text{line}\;,
	\end{equation}
	the equation of motion reads
	
	\begin{equation}\label{motion2}
		\hat v\frac{d\hat v}{d\hat r}=-\frac{1}{\rho}\frac{dp}{d\hat r}-\frac{\hat v_\text{crit}^2}{\hat r^2}+\hat g_\text{line}\;,
	\end{equation}	
	with $v_\text{crit}$ being the rotational break-up velocity in the case of a corresponding rotating star \citep{araya14}
	
	\begin{equation}\label{changeofvariables2}
		\hat v_\text{crit}=\frac{ v_\text{esc}}{a\sqrt{2}}=\frac{1}{a}\sqrt{\frac{GM_\text{eff}}{R_*}}\;.
	\end{equation}
	
	With the use of the equation of state for an ideal gas ($p=a^2\rho$), Eq.~\ref{motion2} becomes
	
	\begin{equation}\label{motion3}
		\left(\hat v-\frac{1}{\hat v}\right)\frac{d\hat v}{d\hat r}=-\frac{\hat v_\text{crit}^2}{\hat r^2}+\frac{2}{\hat r}+\hat g_\text{line}\;.
	\end{equation}
	and is formally independent on density (and therefore independent on mass-loss rate).
	
	This expression has the advantage that if the line-acceleration is expressed as a function of the radius only $g_\text{line}(\hat r)$, the velocity and radial terms are separated to either side of the equation.
	Therefore, by means of the  Lambert $W$-Function (see mathematical definition and its main properties in Appendix~\ref{lambertdefinition}), Eq. \ref{motion3} can be solved integrating both sides along the atmosphere, and expressing $\hat v(\hat r)$ in terms of the defined $W(z)$ (see Eq.~\ref{lambertdef}).
	
	In Eq.~\ref{motion3} the explicit dependence on the density $\rho(r)$ has vanished and as a consequence the Lambert $W$-function cannot provide a new value for the mass-loss rate.
	Thus, $\dot M$ can be considered as a \textit{free parameter} of the e.o.m. 
	Our solution to the e.o.m. will be consistent with the assumed line-force but this will (in general) be inconsistent with the new line-force computed by a subsequent CMFGEN iteration.
	As discussed more in detail in Section~\ref{massloss}, a fully consistent solution mass-loss rate will require an iterative adjustment for the mass-loss rate.
	However, given the uncertainties and assumptions, it can be desirable to leave the mass-loss rate fixed, and simply iterate until our solution for the velocity law stabilises (Section~\ref{lambertprocedure}, Appendix~\ref{iter_append}).

\subsubsection{Subsonic vs. supersonic region}
	Assuming a monotonic behaviour of the velocity field throughout the atmosphere ($d\hat v/d\hat r>0$), it is seen that the left hand side of Eq. \ref{motion3} becomes zero when $v=a$ because then $\hat v=1$.
	Thus, at the \textit{sonic (or critical) point} $\hat r_c$, the right hand side must also be zero, i.e.,
	
	\begin{equation}\label{sonicpoint}
		-\frac{\hat v_\text{crit}^2}{\hat r_c^2}+\frac{2}{\hat r_c}+\hat g_\text{line}(\hat r_c)=0\;.
	\end{equation}
	
	Because we are assuming that the line force is only a function of $r$, the formal critical point of the equation of motion is the sonic point, which was also assumed by \citet{lucy70}.
	However, using the Sobolev theory, the line force depends on $dv/dr$ and this moves the critical point above the sonic point. 
	The requirement that the wind moves smoothly through the critical point sets the velocity gradient at the critical point, and (potentially) yields a unique solution for the mass-loss rate.
	In more recent work, \citet{lucy07} assumed that radiative acceleration in the neighbourhood of the sonic point was a function of $v$, and this also leads to the critical point being at the sonic point. 
		
	Due to the monotonic behaviour of $v(r)$ (and consequently $\hat v$), the sonic point becomes a boundary between two regions.
	The first of them is for $\hat v<1$, which makes both sides of Eq.~\ref{motion3} negative. It is called the \textit{subsonic region}.
	The second, for $\hat v>1$, is then called the \textit{supersonic region}.
	The difference between them is not only due to mathematical analysis but also to physical reasons: in a one dimensional fluid moving at velocities below sound speed, perturbations are propagated both inwards and outwards, whereas perturbations occurring in a supersonic fluid are only propagated downstream.
	
	This means that we actually are searching solutions for the equation of momentum \textit{in two different branches} that merge at the critical point: the analytical solution for the velocity profile implies the integration from the sonic point both outwards and inwards.
	
	In terms of the Lambert-function the general solution is
	
	\begin{eqnarray}\label{v2expv2}
		 \hat v^2e^{-\hat v^2}&=&\left(\frac{\hat r_c}{\hat r}\right)^4\exp\left[-1-2\,\hat v_\text{crit}^2\left(\frac{1}{\hat r}-\frac{1}{\hat r_c}\right)\right]\nonumber\\
		 &\times&\exp\left[-2\int_{\hat r_c}^{\hat r}\hat g_\text{line}\,d\hat r\right]\;,
	\end{eqnarray}
	See Appendix~\ref{analyticaleom} for a detailed step-by-step of the integration approach.
	
\section{Basic concepts of CMFGEN}\label{cmfgen}

\subsection{Calculation of line-acceleration $g_\text{line}(r)$}

	The radiative acceleration, as a function of wind location, is obtained from a converged CMFGEN model. It is evaluated using the formula:
	
	\begin{equation}\label{glinecalc}
		g_\text{rad}(r)=\frac{\chi_\nu L_\nu}{4\pi c\rho r^2}\;\; =\frac{\chi_\text{f}L_*}{4\pi c\rho r^2}, 
	\end{equation}
	where $\chi_\text{f}$ is the flux mean opacity
				
	CMFGEN provides the radiative acceleration as a function of radius, and this can be used directly with our Lambert-procedure to obtain a new velocity law.
	However it is widely known that the radiative  acceleration is also a function of the velocity and velocity gradient, and hence, in general, the radiative acceleration in CMFGEN will change when we use the new  velocity law.
	Thus an iterative procedure is adopted.
	
\subsection{Mass-loss rate and clumping factor}\label{massloss}

\begin{figure}[t]
		\centering
		\includegraphics[width=\linewidth]{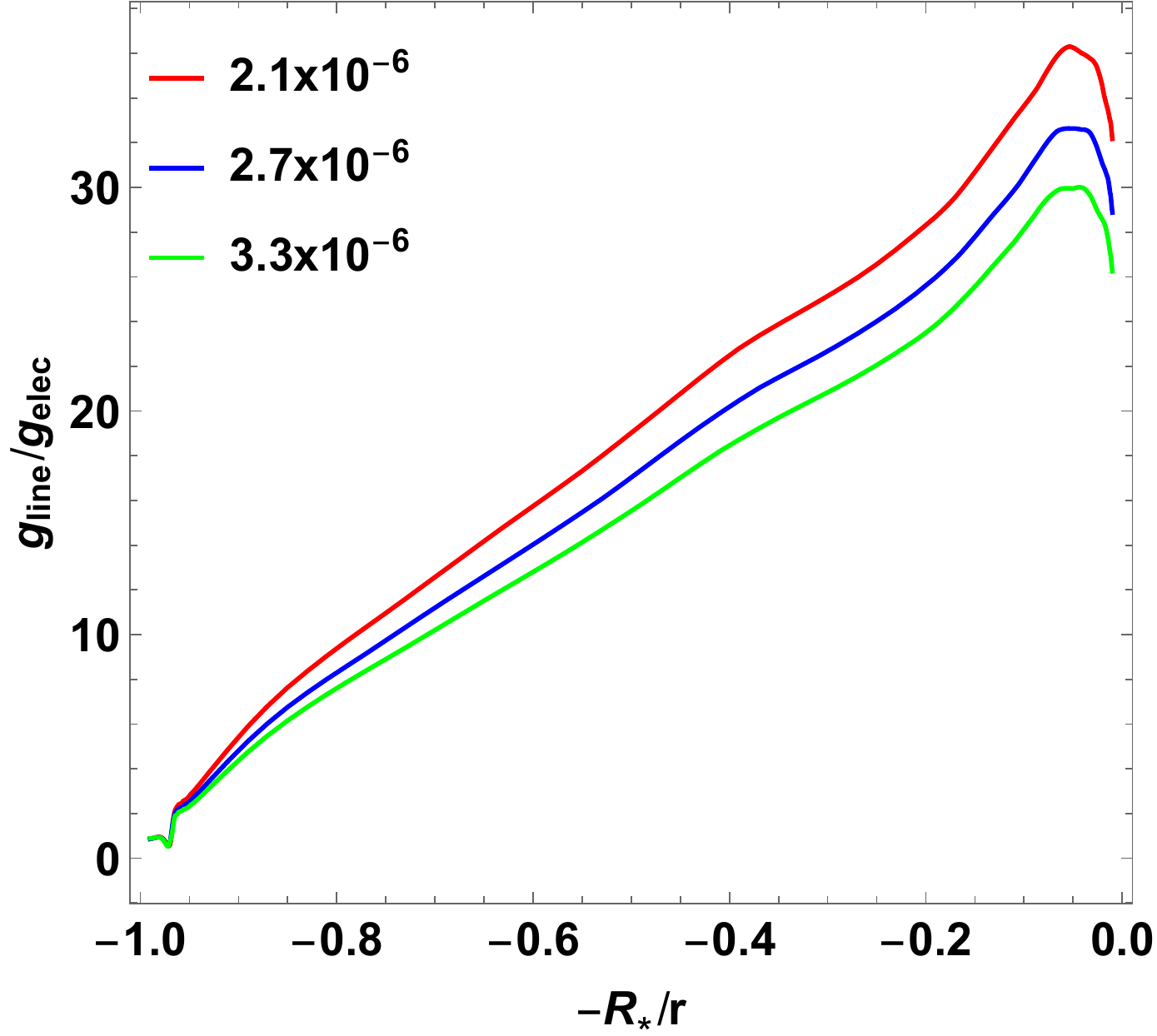}
		\caption{Radial dependent line accelerations $g_\text{line}$, normalised by $g_\text{elec}$, from initial CMFGEN models of $\zeta$-Puppis with different values for $\dot M$. All the other stellar and wind parameters are the same and they are given in Table \ref{initialmodel}.}
		\label{mdotsinitial}
	\end{figure}

	The dimensionless equation of momentum (Eq.~\ref{motion3}) is not explicitly dependent on density because $\rho(r)$ terms are mathematically cancelled.
	As a consequence, the Lambert $W$-function cannot calculate a new mass-loss rate and hence $\dot M$ here is considered as a free parameter.
	However, the dependence on density, and therefore on mass-loss rate, is implicitly included in Eq.~\ref{motion3} via the $g_\text{line}$ term.
	From Eq.~\ref{glinecalc}, the line-acceleration appears to be inversely proportional to the mass-loss rate $\dot M$, and hence also inversely proportional to the density $\rho(r)$.
	However there is also a further hidden dependence arising from line driving.
	For O-type stellar winds, Sobolev theory shows that $g_\text{line}$ scales as $\sim\dot M^{-\alpha}$ where $\alpha$ is the CAK parameter, and is around $2/3$ for O-star winds \citep{puls08}.
	Because the line force is density dependent, different mass-loss rates will lead to different final Lambert-hydrodynamics. 	Higher mass-loss rates produce slower line-accelerations (see Fig.~\ref{mdotsinitial}), which yield velocity profiles with lower terminal velocities.
	
	Line-acceleration is also affected by small-scale inhomogeneities (i.e., clumping) present throughout the wind.
	Clumping alters the line force, and modifies spectral features produced in the wind.
	Clumping is implemented in CMFGEN in terms of the \textit{volume filling factor} $f$.
	CMFGEN assumes a void interclump medium, and that the clumps are much smaller than the mean free path of a photon. 
	The density inside the clumps, $\rho_\text{cl}$, satisfies $\rho_\text{cl}=\rho_0/f$, where $\rho_0$ is the homogeneous (smooth) wind density at the same radius \citep{bouret05}.
	The volume filling factor is defined in terms of the velocity field as
	
	\begin{equation}\label{fillingfactorv}
		f( v(r))=f_\infty+(1-f_\infty)e^{- v(r)/ v_\text{cl}}
	\end{equation}
	although other options are also available.
	 For some of the simulations presented in this paper we used a slightly modified form which forced $f$ to unity at the sonic point.
	$f_\infty$ can be interpreted as the volume-filling factor at larger distance, while $v_\text{cl}$ indicates the \textit{onset velocity of clumping} and can be interpreted as the point in the velocity field where clumping effects start.
	
	When clumping is included in stellar winds inconsistencies will potentially arise in the solution of the steady-state hydrodynamic equations which assume a smooth flow.
	With the inclusion of clumping in CMFGEN the mass conservation equation becomes 
	
	\begin{equation}\label{clumpedmassconservation}
		\rho_\text{cl}=\frac{\dot M}{4\pi r^2v(r)f}\;,
	\end{equation}
	and the equation of motion is
	
	\begin{equation}\label{clumpedeom}
		\left(v-\frac{a^2}{v}\right)\frac{dv}{dr}=g_\text{rad}-\frac{GM_*}{r^2}+2\frac{a^2}{r}+\frac{a^2}{f}\frac{df}{dr}\;.
	\end{equation}
	This last expression is a more general articulation from Eq.~\ref{momentumgrad}, using the clumped density for the equation of state of the ideal gas, $p=a^2f\rho_\text{cl}$ and relaxing the isothermal condition previously imposed\footnote{This more complex expression for the e.o.m. cannot be solved by Lambert-procedure because the application of Lambert $W$-function requires that variables $v$ and $r$ must be separated in order to perform the integration of both sides of the equation, such as in the case of Eq.~\ref{motion3}.}.
	Similar to the case of Eq.~\ref{motion3}, we neglect the radial derivative over the sound speed term $a^2(r)$ (Eq.~\ref{sound_plus_vturb}) because it has been empirically demonstrated that its value is almost constant over the wind \citep[see figure~1 from][]{sander17}.
	
	If we consider an homogeneous wind, the last term of the right hand side (hereafter RHS) of Eq.~\ref{clumpedeom} vanishes and we can recover the original Eq.~\ref{motion3}.
	The term will also become unimportant above the sonic point where gas pressure forces become negligible. 
	Unfortunately studies of O-star spectra suggest that clumping effects extend down to the photosphere \citep{hillier03,bouret03,puls06}. 
	However a more important issue is that the hydrodynamic equation as written is inconsistent, and takes no account of the interclump medium, and clump formation.
			
\begin{figure}[t]
		\centering
		\includegraphics[width=\linewidth]{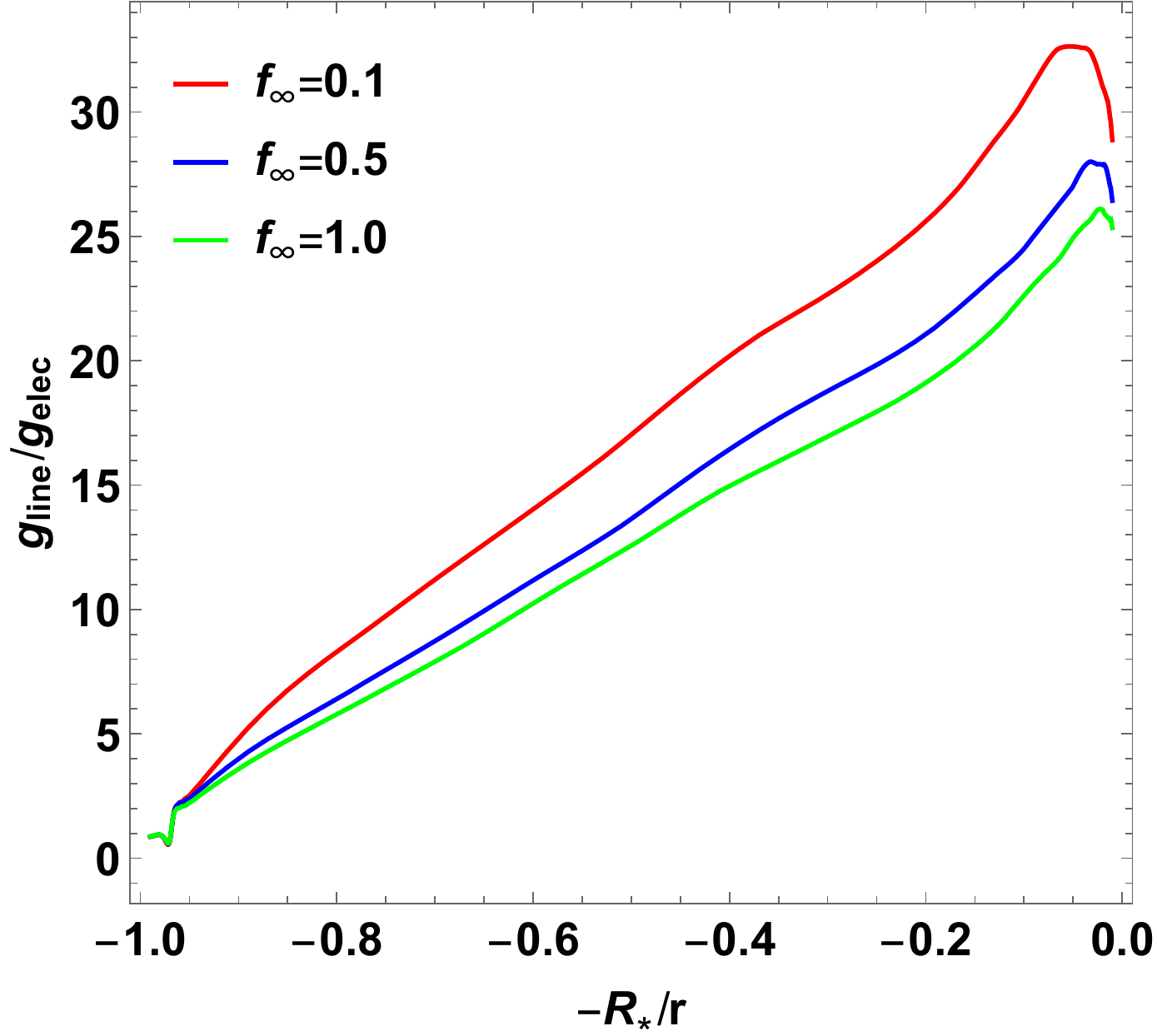}
		\caption{{Radial dependent line-accelerations $g_\text{line}$, normalised by $g_\text{elec}$, from initial CMFGEN models of $\zeta$-Puppis with different values for $f_\infty$. The other stellar and wind parameters are given in Table~\ref{initialmodel} for all three models.}}
		\label{clumpinitial}
	\end{figure}

	Above the sonic point, a stronger clumping factor, expressed by a smaller filling factor $f_\infty$, gives a stronger line-acceleration (see Fig.~\ref{clumpinitial}), which, for a given mass-loss rate, will lead to higher terminal velocities.
	Clumping boosts line acceleration since clumping enhances recombination which enhances line acceleration since the atomic levels producing the lines responsible for driving the flow will have a larger population.
	This effect can also be seen in m-CAK theory through the $\delta$ parameter which indicates that the line force has an additional dependence proportional to $(N_\text{e}/W(r))^\delta$, where $N_\text{e}$ is the electron density and $W(r)$ the dilution factor\footnote{Do not confuse this $W(r)$ term with the $W$ from the Lambert $W$-function.}\citep{abbott82}.
	
	As readily apparent, and which has been noted many times previously \citep[e.g.,][]{lucy70}, Eq.~\ref{clumpedeom} implies that the radiative force is equal to the gravitational force at the sonic point, i.e.,
	
	\begin{equation}\label{gammaeq}
		g_\text{grav}-g_\text{rad}=0\;\;\;\text{when}\;\;\;v=a\;,
	\end{equation}
	\citep[see also][]{lucy07}\footnote{In this equation the influence of the term $2a^2/r$ has been neglected. This is an excellent approximation because the gravitational acceleration can be rewritten in terms of the escape velocity as $g_\text{grav}=GM_*/r^2=v_\text{esc}^2/2$, and for O stars  $v_\text{esc}\gg a$.}.
	This singularity condition is also introduced in the self-consistent studies of \citet{sander17} and \citet{sundqvist19} by means of the new factor $\Gamma=g_\text{rad}/g_\text{grav}$, which must be 1 at the sonic point\footnote{Do not confuse this new $\Gamma$ factor with the Eddington parameter $\Gamma_\text{e}$.}.
	It becomes evident then that the initial value given in our CMFGEN models for mass-loss rate and for clumping will alter the equilibrium of Eq.~\ref{clumpedeom}, especially around the sonic point singularity, and then the derived hydrodynamic solution.

	Following this, the accuracy of the mass-loss rate could be checked using the singularity criterium (Eq.~\ref{gammaeq}), and the hydrodynamic error determined from the CMFGEN modelling.
	Typically for CMFGEN models this is defined by
	
	\begin{equation}\label{errorcmfgen}
		\text{Err}(r)=\frac{200\times\left(v\frac{dv}{dr}+\frac{1}{\rho}\frac{dP}{dr}+\frac{P}{\rho}\frac{dT}{dr}-g_\text{tot}\right)}{\biggr|v\frac{dv}{dr}\biggr|+\biggr|\frac{1}{\rho}\frac{dP}{dr}\biggr|+\biggr|\frac{P}{\rho}\frac{dT}{dr}\biggr|+|g_\text{tot}|}\;.
	\end{equation}
	Other error prescriptions are also possible.
	In the above, for example, the use of $|g_\text{tot}|$ rather than $|g| +|g_r|$ exacerbates the error at the sonic point (since $|g_\text{tot}|$=0 at the sonic point).
	Another alternative is to express the error relative to the radiative force.

	The two criteria discussed above can be used to constrain the mass-loss rate to be used in the Lambert-procedure.
	In principle we can adjust $\dot M$ to look for the value which minimises the error.
	However, the hydrodynamic solution is influenced by several other (uncertain) free parameters besides the mass-loss rate (e.g., $v_{\rm turb}$, X-rays and the parameters associated with clumping), and the neglect of other factors that will influence the hydrodynamics (e.g., rotation).
	In practice we have adjusted some of these parameters to improve the hydrodynamics, but given the uncertainties we did not attempt to get perfect agreement with the hydrodynamics at all velocities.
	Importantly, we wanted to retain some consistency of the parameters, particularly the mass-loss rate, with values obtained from empirical model fitting.
	
	\begin{figure*}[t!]
		\centering
		\includegraphics[width=0.85\linewidth]{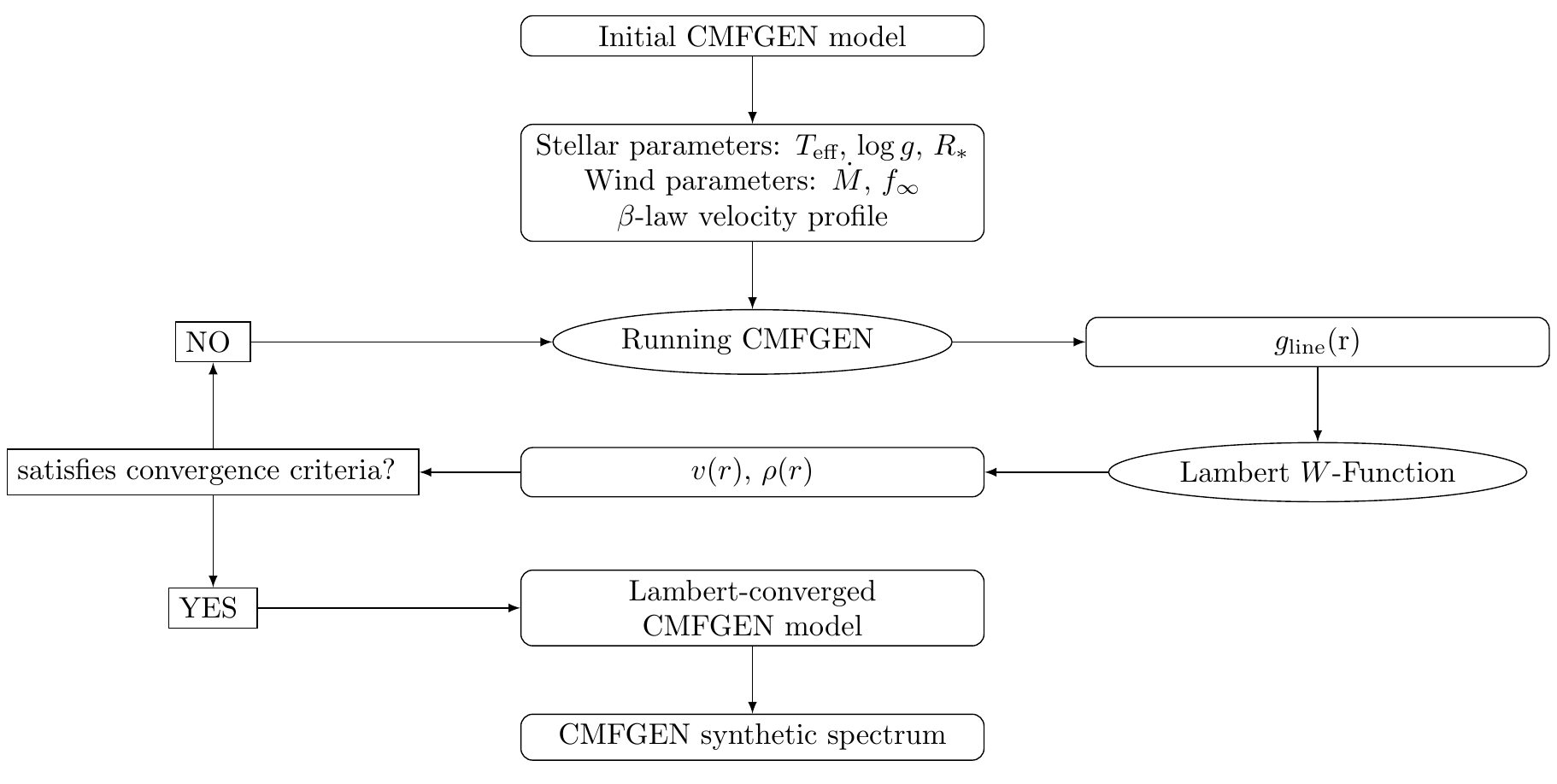}
		\caption{{Flowchart of the Lambert-procedure. The stellar parameters are the same during the entire procedure, as well as mass-loss rate and clumping factor.}}
		\label{flowchart}
	\end{figure*}

\section{Lambert procedure}\label{lambertprocedure}
	In the following we will refer to the iterative process that uses the line-acceleration computed  by CMFGEN and the velocity profile obtained from the Lambert $W$-Function, as the Lambert-procedure.
	In this procedure, the computed $g_\text{line}$ is used as input to recalculate a new velocity profile, whereas the calculated $v(r)$ is reintroduced in CMFGEN to recompute a new model with a new line acceleration.
	The iterative loop is executed until  convergence criteria are satisfied.
	A flowchart for the Lambert-procedure is presented in Fig.~\ref{flowchart}.
	The Lambert-procedure only searches for an agreement between the velocity field and the line-acceleration –the mass-loss rate is outside the loop and it is considered as a free parameter.
	If desirable, the mass-loss rate can be modified to better fulfil the criteria mentioned in Section~\ref{massloss}, but those changes are outside the circuit involving the Lambert $W$-function.

	The Lambert-procedure in this work is introduced with the aim to evaluate: i) whether there is an hydrodynamic solution able to properly couple the line acceleration and velocity field, ii) whether this solution for $v(r)$ is unique (i.e., does not depend on the initial onset for terminal velocity and $\beta$-law), and iii) whether this solution does reproduce a spectrum in agreement with observations.
	
	The great advantage of using the Lambert $W$-function is that we can solve the equation of motion analytically to obtain a new velocity profile.
	\citet{muller08} and \citet{araya14} have also used the Lambert $W$-function for a solution of the equation of motion, but they used a parametrised form of $g_\text{line}$, whereas our line acceleration comes directly from the detailed solution of NLTE line formation given by CMFGEN.

\subsection{Calculation of velocity field}
	Calculation of the new velocity profile is done by solving Eq.~\ref{v2expv2}, using for $\hat g_\text{line}(\hat r)$ the output of a CMFGEN model.
	The formal expression for $v(r)$ comes from Eq.~\ref{lambertvprofile} (see Appendix~\ref{analyticaleom}).
	
	From the two real branches of the Lambert $W$-function (see Appendix~\ref{lambertdefinition}), the one with asymptotic behaviour is $W_{-1}$.
	We use this branch to obtain the velocity profile in the supersonic region, i.e., from the sonic point outwards.
	As a consequence, each new step in the Lambert-procedure will have a new terminal velocity and a new exponential behaviour different from the initial $\beta$-law.

	To obtain the solution in the subsonic region, the $W_0$ branch of the Lambert $W$-function can be used.
	However, we decided not to do so, as large errors were produced for the acceleration and velocity profiles in the lowest part of the wind (in the hydrostatic region) when using the CMFGEN solution.
	Rather we used a modification of $v(r)$ below the sonic point by rescaling the velocity profile with respect to the initial model to ensure a smooth connection to the supersonic solution.
	
\subsubsection{Junction of subsonic and supersonic regions}

	\begin{figure}[t!]
		\centering
		\includegraphics[width=0.95\linewidth]{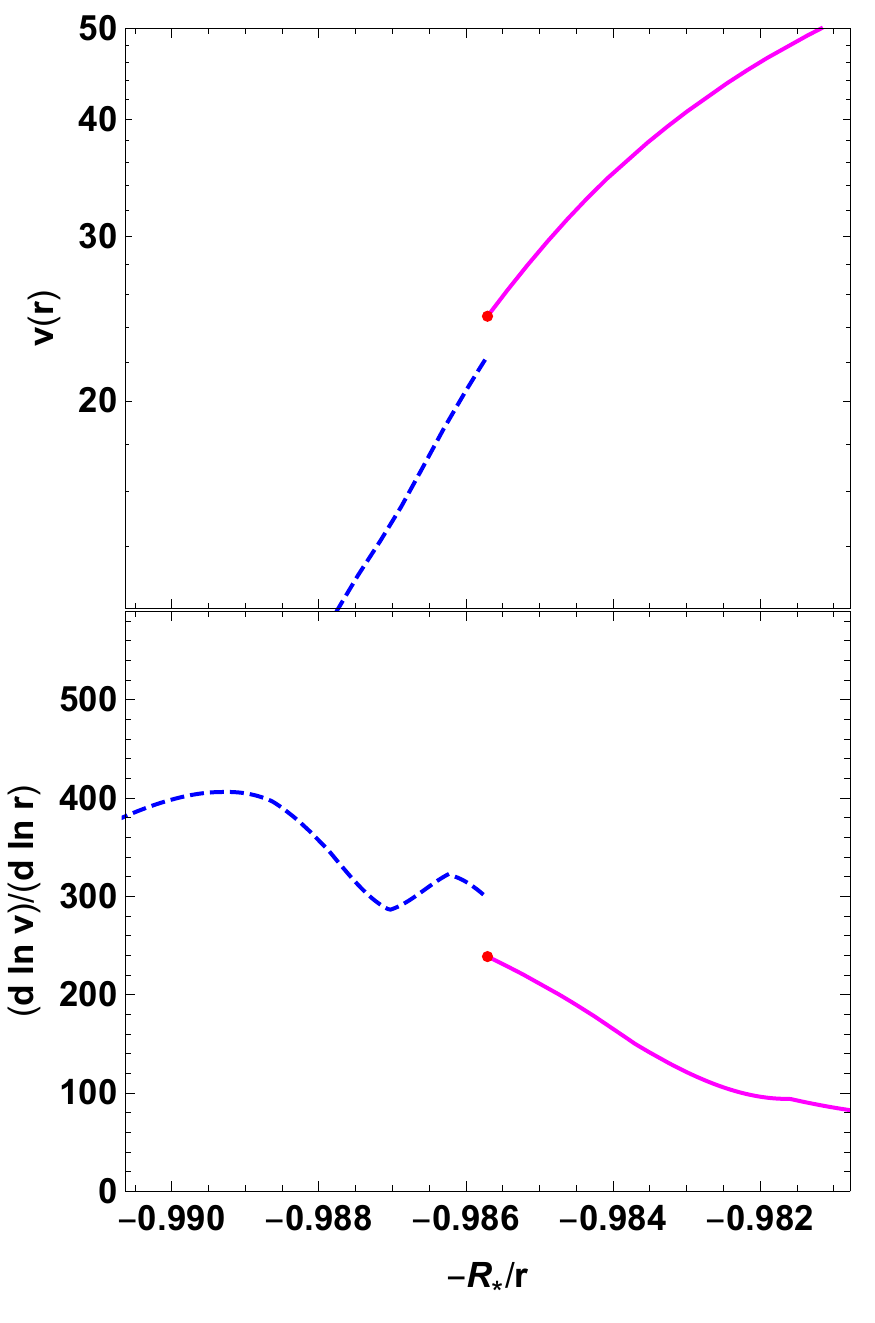}
		\caption{{Zoom over the sonic point (red point), where subsonic and supersonic regions are coupled during the Lambert-procedure. Magenta solid line represents the new velocity profile and velocity gradient for the supersonic region obtained with Lambert $W$-function; the blue dashed line represents the old velocity profile and velocity gradient for the subsonic region.}}
		\label{coupling}
	\end{figure}
	After determining the critical point $r_c$ from Eq.~\ref{sonicpoint} and calculating the new Lambert velocity profile, we spline both regions (new-supersonic and old-subsonic) at the critical point where $v(r_c)=a$.
	\textcolor{black}{This coupling must ensure a smooth transition between subsonic and supersonic zones.}
	The conditions:
	
	\begin{equation}\label{fitv}
		\lim_{r\rightarrow r_\text{c}-} v(r)=\lim_{r\rightarrow r_\text{c}+} v(r)\;\;,
	\end{equation}
	and
	
	\begin{equation}\label{fitdv}
		\lim_{r\rightarrow r_\text{c}}\frac{d v(r)}{dr}\Biggr\rvert_{r<r_\text{c}}=\lim_{r\rightarrow r_\text{c}}\frac{d v(r)}{dr}\Biggr\rvert_{r>r_\text{c}}\;\;,
	\end{equation}
	must be fulfilled.
	
	To be in agreement with the CMFGEN units system (where the derivative of the velocity profile is expressed by the term $\sigma\equiv d\ln v/d\ln r+1$), the second limit is easier to evaluate if we use  logarithm derivatives which are related by:
	
	\begin{equation}
		\frac{d\ln v}{d\ln r}=\frac{r}{v(r)}\frac{dv}{dr}\;.
	\end{equation}
	
	Then:
	
	\begin{equation}\label{fitdlnv}
		\lim_{r\rightarrow r_\text{c}}\frac{d \ln v(r)}{d\ln r}\Biggr\rvert_{r<r_\text{c}}=\lim_{r\rightarrow r_\text{c}}\frac{d\ln v(r)}{d\ln r}\Biggr\rvert_{r>r_\text{c}}\;\;.
	\end{equation}
	
	As a consequence of these conditions, the subsonic region must be readapted too, to have a smooth transition and avoid instabilities.
	Because our $r_c$ is determined from the singularity of the equation of motion, this does not even coincide with the old velocity profile at $v=a$ \citep[differences on these two sonic points were also previously referred by][see their figure~3]{sander17}.
	Fig.~\ref{coupling} shows the disconnection of $v(r)$ and $d\ln v/d\ln r$ around the sonic point for a velocity profile after solving the equation of motion via the Lambert $W$-function.
	Therefore, we proceed to rescale the radius in the next model to  match $v(r_c)=a$ .
	The value of the rescaling depends on where in the old $v(r)$ we get $v=a$ from both sides as in Eq. 23.
	Because it can happen that the differences of the velocity gradients between subsonic and supersonic regions are too high, the spline can be also done below or over the critical point (up to 3 ND depth points depending of the case).
	In any case, the value of the rescaling is typically around 1.001 or 1.003 (i.e., radius is never modified more than a 0.3\%).

\subsection{Convergence of models}\label{lambertconvergence}
	Given the line-acceleration as an input, the Lambert $W$-function provides a new velocity profile, whereas CMFGEN provides a new line acceleration given a prescribed velocity profile.
	Therefore, a self-consistent solution requires the combination of both these calculations in an iterative loop until reaching some convergence criteria.

	The convergence of the Lambert-iterations is checked by evaluating both the velocity field $v(r)$ and the line acceleration $g_\text{line}(r)$.
	To accept a Lambert model as well-converged, we require that the conditions:
	
	\begin{equation}\label{thresholdv}
		\Biggr\lvert\log\frac{ v(r)_n}{ v(r)_{n-1}}\Biggr\rvert\le0.01\;\;,
	\end{equation}
	and
	
	\begin{equation}\label{thresholdg}
		\Biggr\lvert\log\frac{ g_\text{line}(r)_n}{g_\text{line}(r)_{n-1}}\Biggr\rvert\le0.01\;\;,
	\end{equation}
	are fulfilled for all $r$.
	Here, $n$ denotes the $n$-th Lambert-iteration executed\footnote{Notice that, if $f(r)_n=f(r)_{n-1}$ the value of the absolute expression would be zero, because the ratio between both functions would be 1.}.

	The definition of convergence we apply in this Section only relates with the execution of the Lambert-procedure to solve the dimensionless e.o.m.
	We assume then, conditions Eq.~\ref{thresholdv} and Eq.~\ref{thresholdg} imply that Eq.~\ref{motion3} is satisfied thought the full wind.
	Unfortunately, we cannot extend this implication to the full e.o.m introduced in Eq.~\ref{clumpedeom}, because Lambert $W$-function is applied over the simplified version.
	As a consequence, the convergence of a Lambert-procedure does not assure the fulfilment of the e.o.m. inside CMFGEN.
	Given the explicit independence on density, we can mathematically find a wide family of well converged solutions for velocity profiles from different sets of mass-loss rate and clumping factors, independent of whether only one specific set of $\dot M$ and $f_\infty$ would fully satisfy Eq.~\ref{clumpedeom}.
	
	The thresholds from Eq.~\ref{thresholdv} and \ref{thresholdg} are applied over all the range of the radial coordinate $r$, from the photosphere outwards and with special consideration around the sonic point.
	Because this is the zone where the solution coupling is done, it is also the zone with more instability at the moment of the convergence.

	Besides purporting to be a solution for the velocity profile, these solutions under the Lambert $W$-function also must be proven to be unique.
	Wind hydrodynamics expressed under the $\beta$-law depends on two parameters: terminal velocity $ v_\infty$ and the $\beta$ exponent.
	If the Lambert-procedure is able to generate a well-converged hydrodynamic solution, it should be the same independently of any initial velocity profiles (i.e., initial $v_\infty$ and $\beta$ parameters) chosen.
	
	Typically, the Lambert-procedure converges after 5 or 6 Lambert-iterations, each one of them corresponding to a CMFGEN model executed with about 12 inner iterations.
	In some cases the Lambert-procedure may not converge, and in such cases the mass-loss rate is adjusted to facilitate convergence.

	\begin{table*}[t!]
		\caption{Stellar and wind parameters of initial CMFGEN models and the final Lambert models (bold characters). $R_*$ is set for optical depth $\tau=2/3$. Spectral types are taken from \citet{sota11,sota14}, whereas solar abundances are taken from \citet{asplund09}.}
		\label{initialmodel}
		\centering
		\begin{tabular}{cccc}
			\hline\hline
			& $\zeta$-Puppis & HD 163758 & $\alpha$-Cam\\
			& O4 I(n)f & O6.5 Iafp & O9 Ia\\
			\hline
			$T_\text{eff}\;[\text{K}]$ & 41\,000 & 34\,500 & 28\,200\\
			$\log g$ & 3.6 & 3.41 & 2.975\\
			$R_*/R_\odot$ & 17.9 & 19.1 & 30.3\\
			$\dot M_\text{ini}\;[10^{-6}M_\odot\text{ yr}^{-1}]$ & $2.7$ & $1.5$ & $0.64$\\
			$\mathbf{\dot M_\text{Lamb}\;[10^{-6}M_\odot\text{ yr}^{-1}]}$ & $\mathbf{2.7}$ & $\mathbf{1.2}$ & $\mathbf{0.7}$\\
			$f_{\infty,\text{ini}}$ & 0.1 & 0.05 & 0.02\\
			$\mathbf{f_{\infty,\text{Lamb}}}$ & $\mathbf{0.1}$ & $\mathbf{0.05}$ & $\mathbf{0.1}$\\
			$v_{\infty,\text{ini}}\;[\text{km s}^{-1}]$ & 2\,300 & 2\,100 & 1\,550\\
			$\mathbf{v_{\infty,\text{Lamb}}\;[\text{km s}^{-1}]}$ & $\mathbf{2\,740}$ & $\mathbf{2\,400}$ & $\mathbf{2\,890}$\\
			$\beta$ & 0.9 & 1.1 & 1.5\\
			\textbf{quasi-}$\mathbf{\beta}$ & $\mathbf{0.9}$ & $\mathbf{0.85}$ & $\mathbf{0.95}$\\
			$v_{\rm turb,ini}\;[\text{km s}^{-1}]$ & 10.0 & 10.0 & 10.0\\
			$\mathbf{v_{\rm turb,Lamb}\;[\text{km s}^{-1}]}$ & $\mathbf{10.0}$ & $\mathbf{7.0}$ & $\mathbf{10.0}$\\
			$v\sin i\;[\text{km s}^{-1}]$ & 210.0 & 94.0 & 100.0\\
			He/H & 0.16 & 0.15 & 0.15\\
			$[$C/C$_\odot]$ & 0.188 & 0.891 & 0.491\\
			$[$N/N$_\odot]$ & 9.510 & 4.710 & 2.000\\
			$[$O/O$_\odot]$ & 0.136 & 0.236 & 0.849\\
			$[$P/P$_\odot]$ & 0.343 & 1.000 & 1.010\\
			$[$Fe/Fe$_\odot]$ & 0.811 & 1.000 & 0.918\\
			$[$Ni/Ni$_\odot]$ & 0.962 & 1.000 & 1.010\\
			\hline
		\end{tabular}
	\end{table*}

\section{Results}\label{results}    

	\begin{figure*}[t!]
		\centering
		\includegraphics[width=0.45\linewidth]{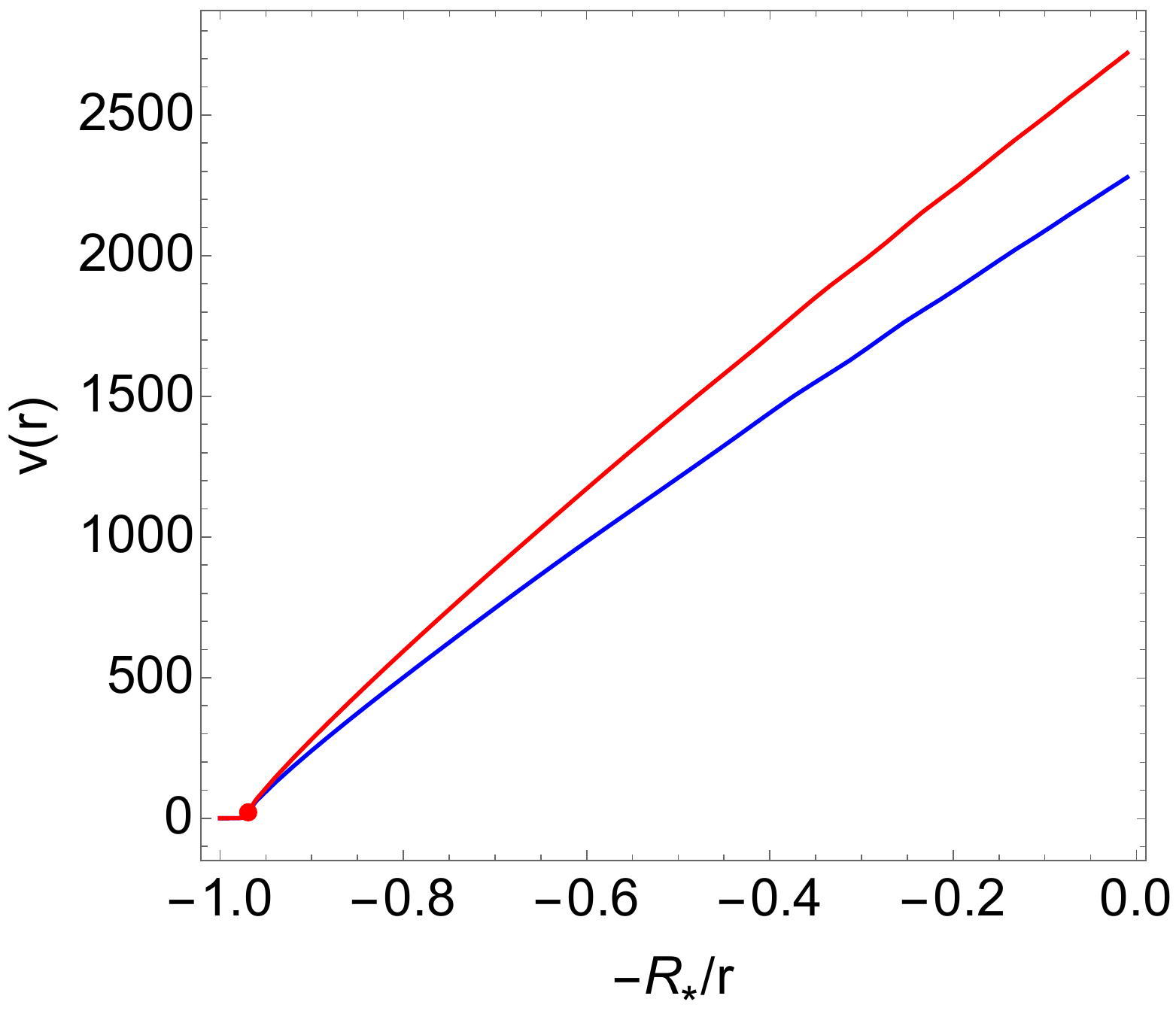}
		\includegraphics[width=0.45\linewidth]{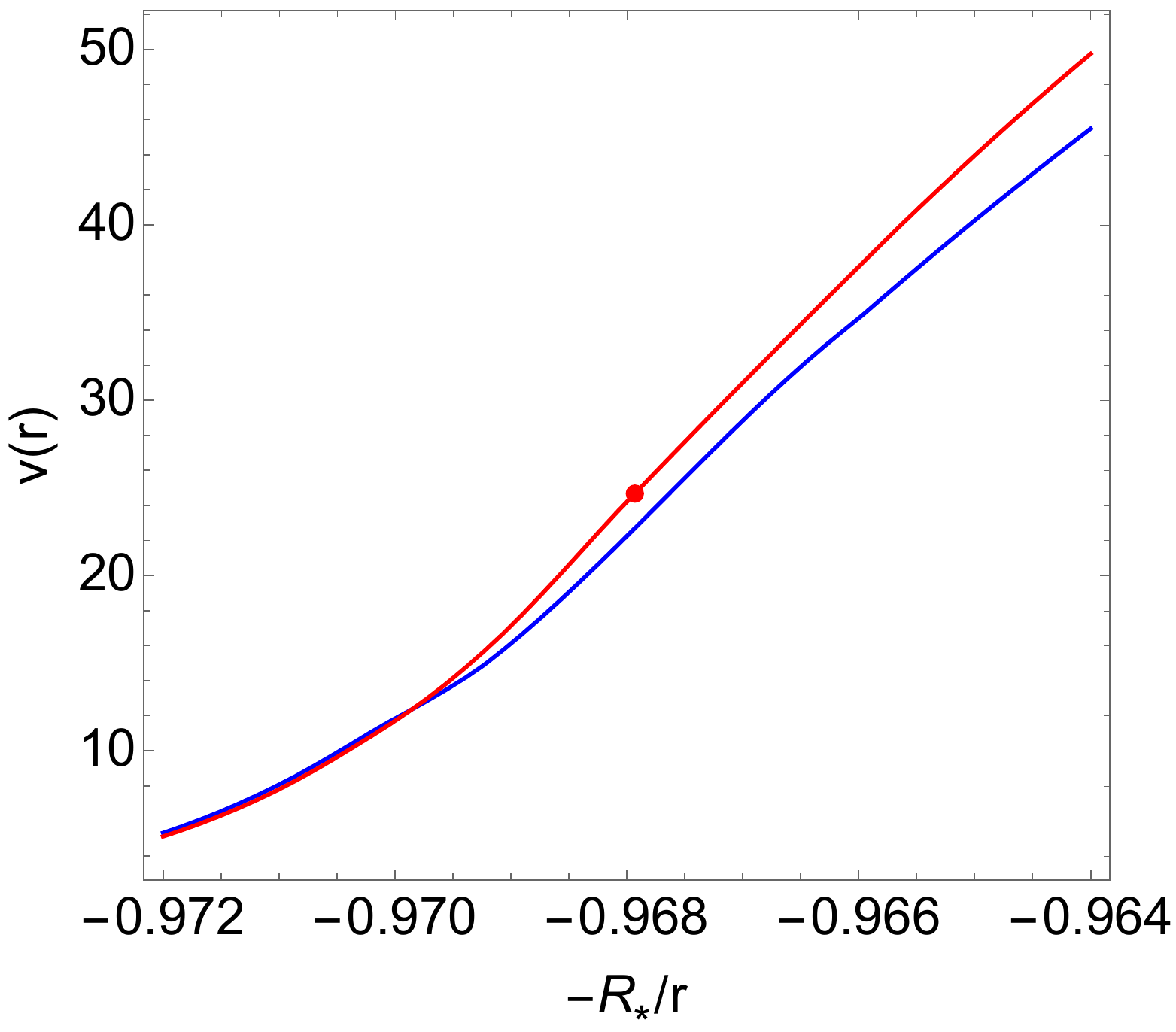}
		\caption{Left panel: differences between initial velocity field described with $\beta$-law (blue lines) and the converged velocity profile from Lambert-procedure (red lines) for $\zeta$-Puppis model. Right panel: zoom over the transition region is shown, displaying the sonic point with a red dot.}
		\label{oldvsnewvel}
	\end{figure*}
	
	Given that CMFGEN requires a large computational effort, and that this is an initial study, the present study is focused only on three O-type supergiants: $\zeta$-Puppis (HD 66811), HD 163758 and $\alpha$-Cam (HD 30614).
	The stellar parameters are listed in Table~\ref{initialmodel}.
	To evaluate the validity of the Lambert-procedure to perform hydrodynamic solutions, we take one CMFGEN model for each one of these stars, with the initial parameters shown in Table~\ref{initialmodel} (hereafter initial models).
	The initial stellar and wind parameters correspond to the values used in the literature that fit the star's spectra over a wide spectral range.
	Atomic species and their levels are specified in Appendix~\ref{species}.
	
	The star $\zeta$-Puppis was chosen because it has been long considered the prototypical O star, and was the main star studied in this work.
	It has been extensively analysed  \citep[e.g.,][(Paper I)]{kudritzki83,pauldrach94,bouret12,pauldrach12,sander17,alex19}.
	The initial stellar and wind parameters adopted for $\zeta$-Puppis correspond to those found by \citet[][Section 6.5]{bouret12} and \citet{marcolino17} from their spectral analysis at optical and infrared wavelengths respectively.
	HD 163758 and $\alpha$-Cam are late O-type stars with lower effective temperatures and surface gravities.
	Parameters for HD~163758 were also taken from \citet{bouret12}, whereas the parameters for $\alpha$-Cam are in the neighbourhood of those determined from the $L$-band spectroscopic study of \citet{najarro11}.
	
	Results shown in this section correspond to the final solution for each model once the convergence is achieved and once the unicity of the solution is verified.

\subsection{$\zeta$-Puppis}\label{zpuppis}
\begin{figure}[h!]
		\centering
		\includegraphics[width=0.95\linewidth]{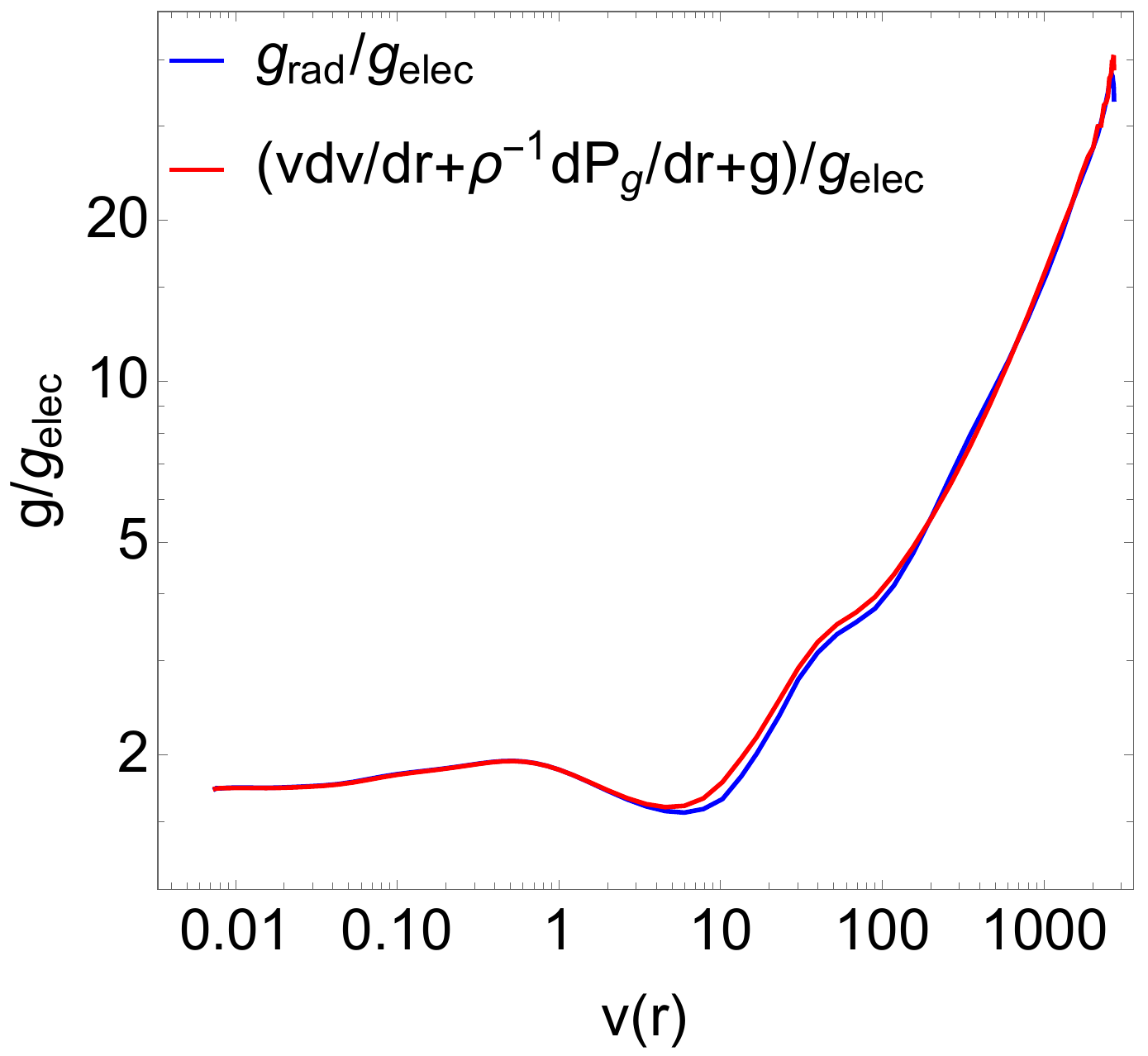}\\
		\caption{Radiative acceleration (divided by acceleration due to electron scattering) as a function of wind velocity for the final CMFGEN Lambert-model (blue) for $\zeta$-Puppis, compared with the expected value when equation of momentum is solved (red).}
		\label{finalglinev}
	\end{figure}

	Comparisons between the initial and converged self-consistent velocity fields are shown in Fig.~\ref{oldvsnewvel}.
	Both velocity profile and line-acceleration were evaluated to satisfy the threshold conditions from Eq.~\ref{thresholdv} and Eq.~\ref{thresholdg}.
	Since the new model is self-consistent, both the left hand and right hand sides of Eq.~\ref{momentumgrad} are in agreement as shown in Fig.~\ref{finalglinev} (see also Fig.~\ref{initialglinev}).
	
\begin{figure*}[htbp]
		\centering
		\includegraphics[width=0.45\linewidth]{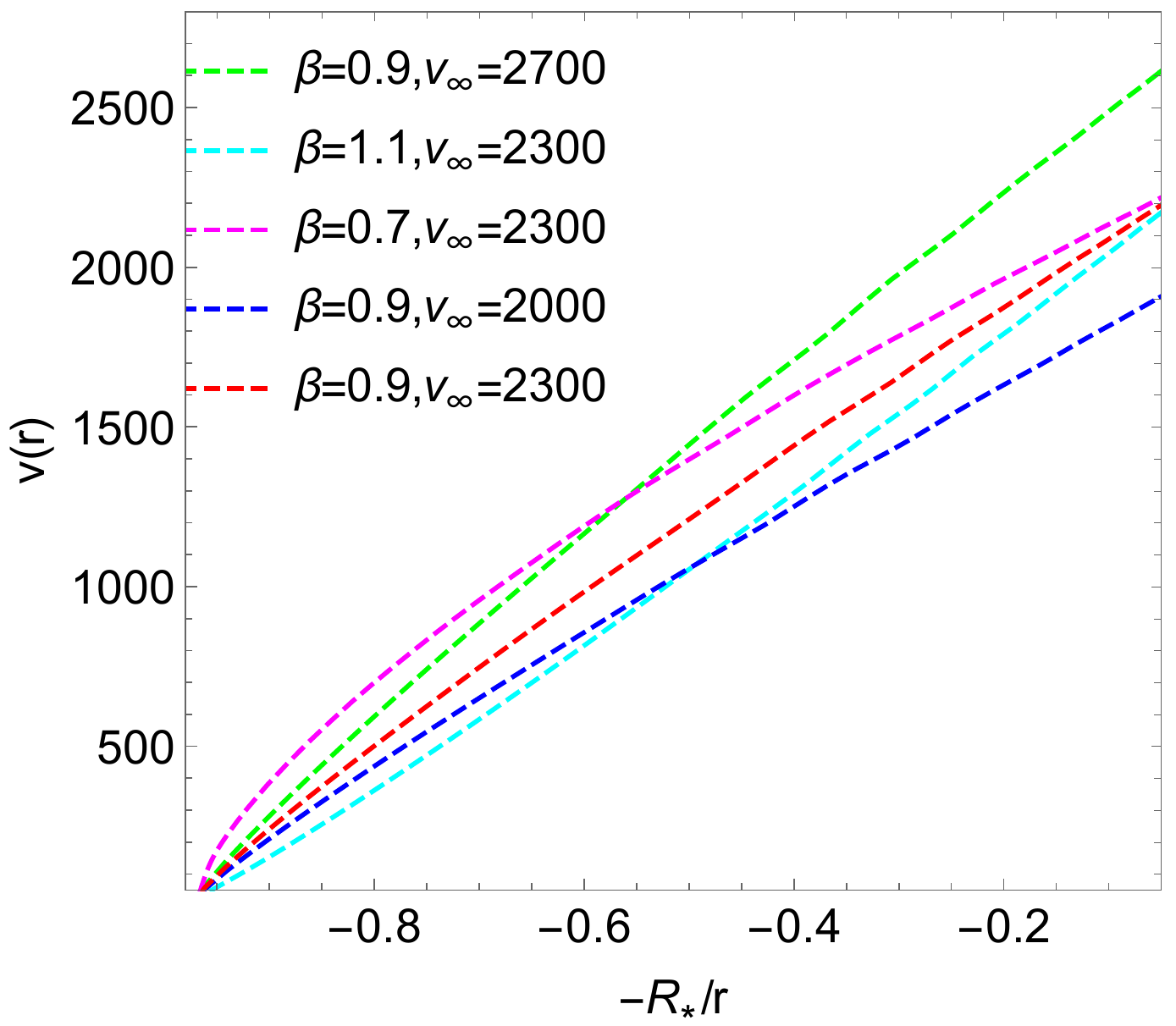}
		\includegraphics[width=0.45\linewidth]{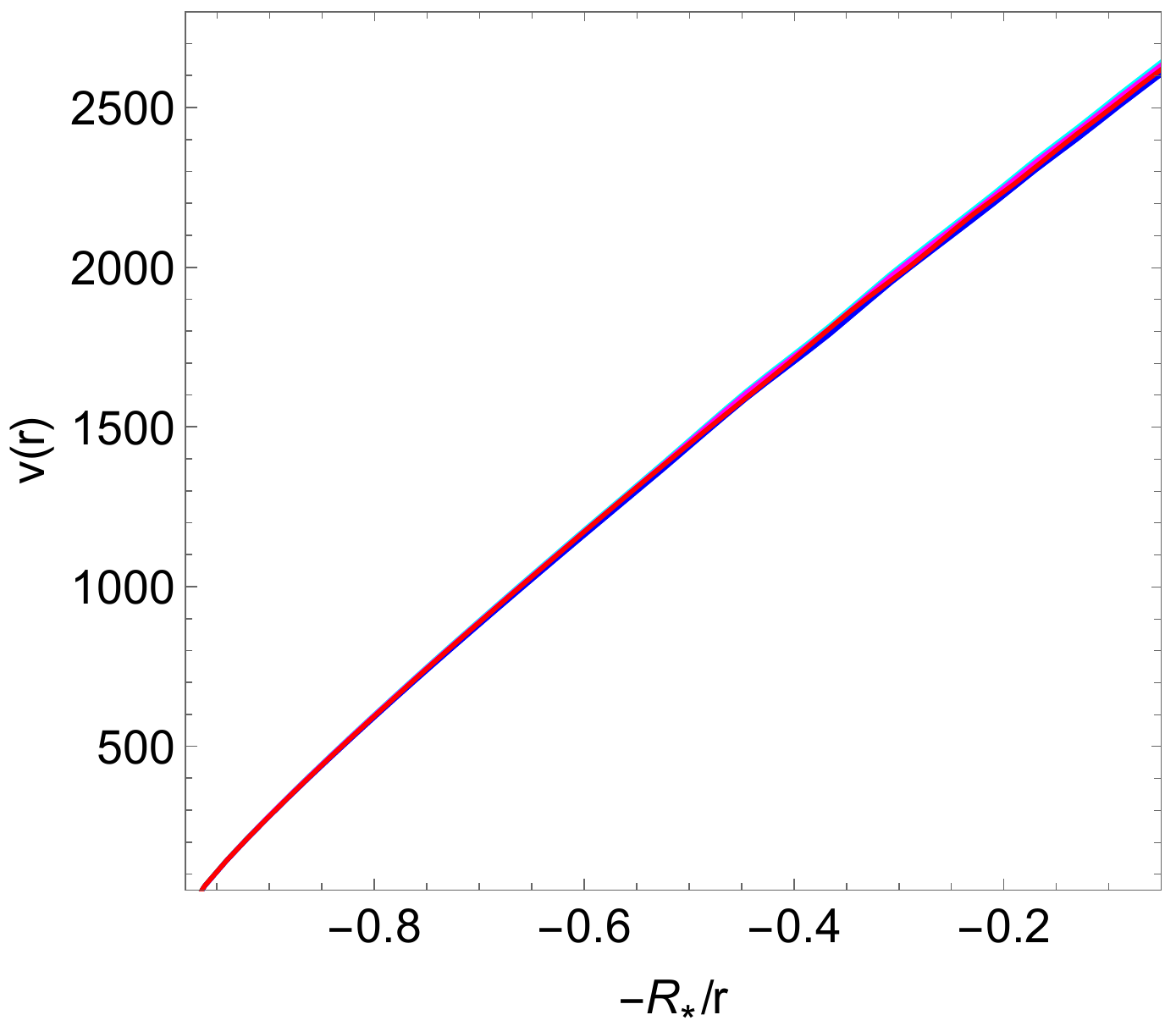}
		\caption{Left panel: velocity profiles (in km s$^{-1}$) for $\zeta$-Puppis using different initial values of $\beta$ and terminal velocity $v_\infty$. Right panel: converged Lambert-profiles for each one of these initial models.}
		\label{hydros_alt}
	\end{figure*}

	Unicity of the Lambert-solution is verified by running CMFGEN models with different initial values for $\beta$ and $v_\infty$.
	 Figure~\ref{hydros_alt} shows the unique convergence of all these parallel models: the left panel represents the initial velocity profiles with $\beta$-law, whereas the right panel shows the final converged Lambert velocity profiles (where the overlap demonstrates all the initial models converging into the same solution).
	 
\begin{figure}[htbp]
		\centering
		\includegraphics[width=0.95\linewidth]{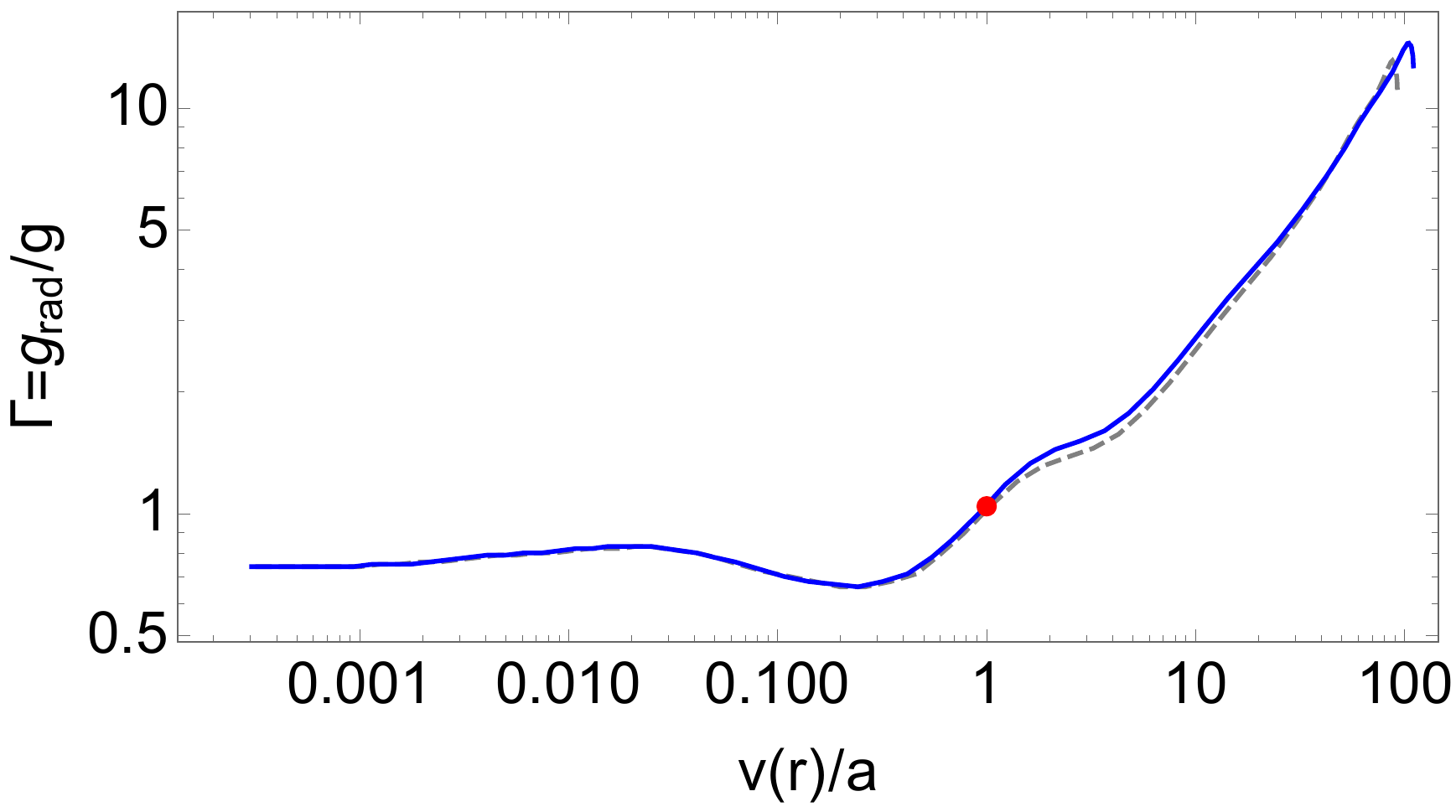}
		\caption{$\Gamma$ as a function of the normalised velocity profile for the Lambert-solution of $\zeta$-Puppis (blue) and its respective initial model (dashed grey). Red dot shows $\Gamma=1\pm0.1$ at $\hat v=1$.}
		\label{gamma_zpup}
	\end{figure}

	Because the condition $\Gamma=1$ was already satisfied in the initial model we did not modify the mass-loss rate nor the clumping factor.
	The run of $\Gamma$ as a function of normalised velocity is shown in Fig.~\ref{gamma_zpup}.
	
	With the converged self-consistent hydrodynamics the resulting terminal velocity has increased by a factor of $\sim1.2$ from 2\,300 to 2\,740 km/s.
	Besides, due to the rescaling in the subsonic region, there are not significant differences in the resulting velocity profile below the sonic point, just above $\sim12-13$ km s$^{-1}$.
	
	\begin{figure}[t!]
		\centering
		\includegraphics[width=0.95\linewidth]{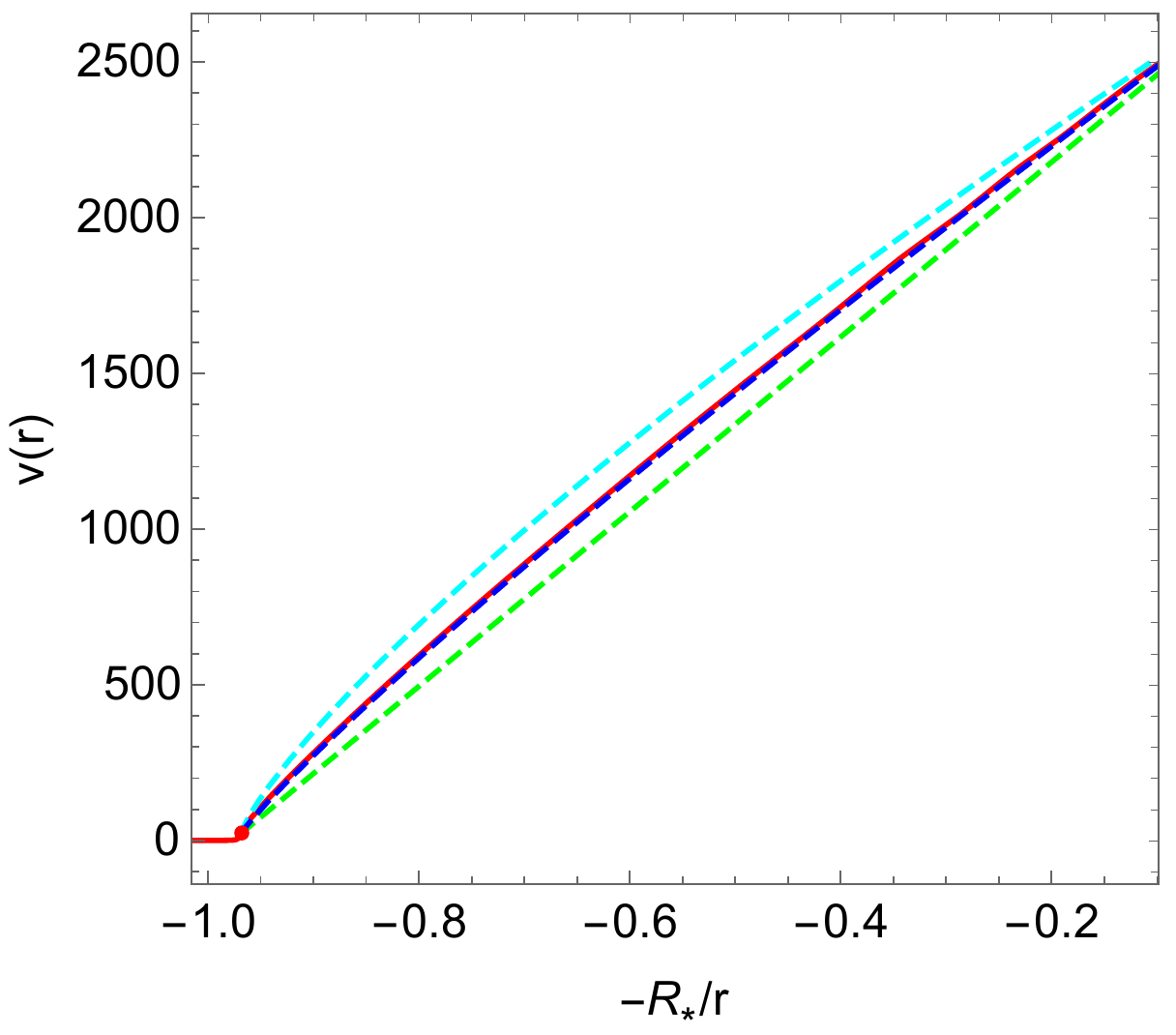}\\
		\includegraphics[width=0.95\linewidth]{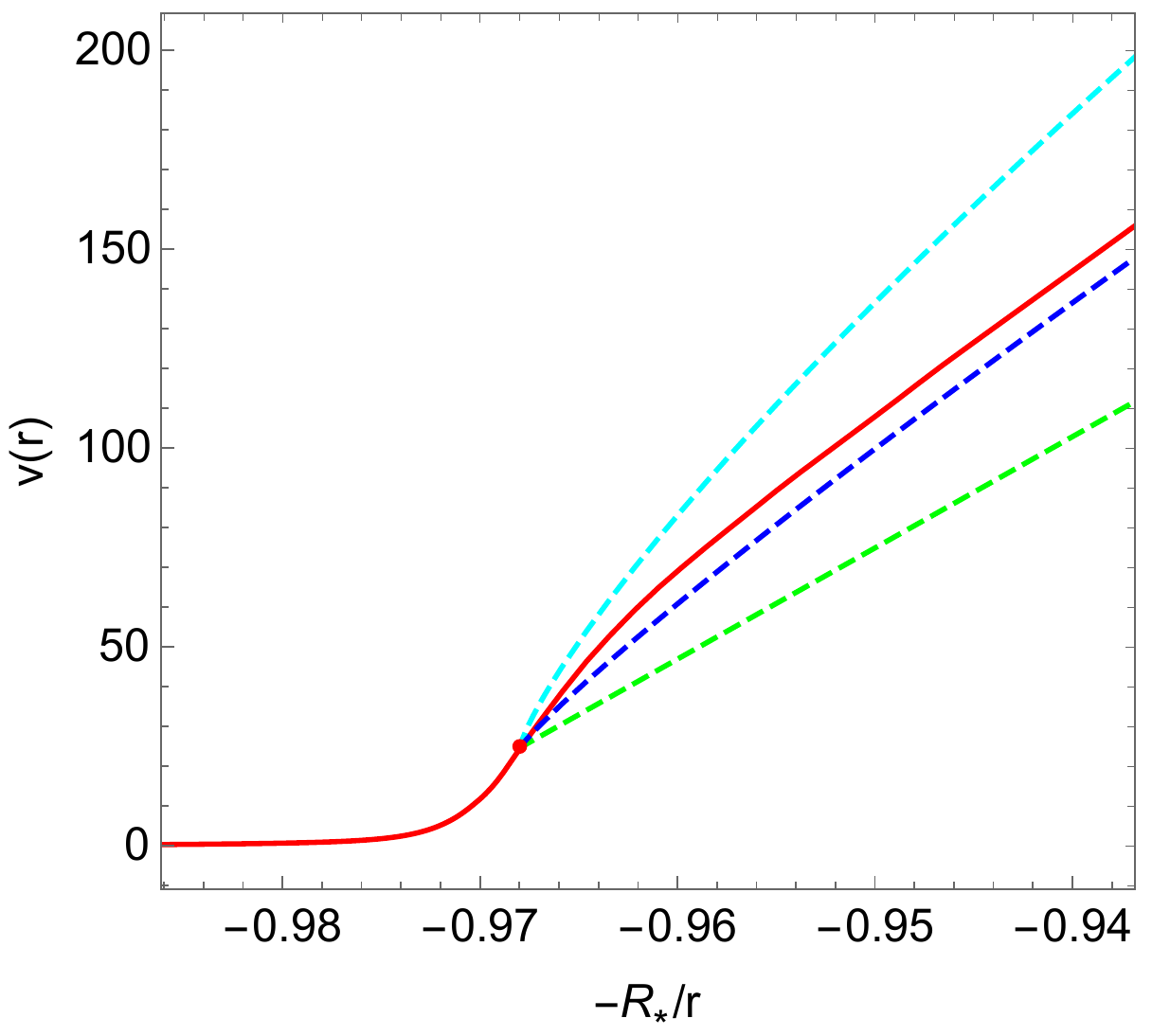}
		\caption{Upper panel: velocity profile of the converged Lambert-model (red), fitted with a $\beta$-law using $\beta=0.8$ (cyan), $\beta=1.0$ (green) and $\beta=0.9$ (blue). Lower panel: a zoom around the sonic point.}
		\label{oldvsnewvelbeta}
	\end{figure}

	Moreover, from Fig.~\ref{oldvsnewvelbeta} we observe that it is possible to find a $\beta$ value that can closely fit the new velocity field.
	This demonstrates that velocity profiles can be approximated by a $\beta$-law at large scale only, but around the critical point it would be necessary to use an alternative prescription to recover the shape of $v(r)$, reason why we defined this new exponent as ``quasi-$\beta$").
	Several such procedures are implemented in CMFGEN, one of which is very similar to that proposed by \citet{bjorklund21}.

\subsection{HD 163758}\label{hd163758}
\begin{figure}[h!]
		\centering
		\includegraphics[width=0.95\linewidth]{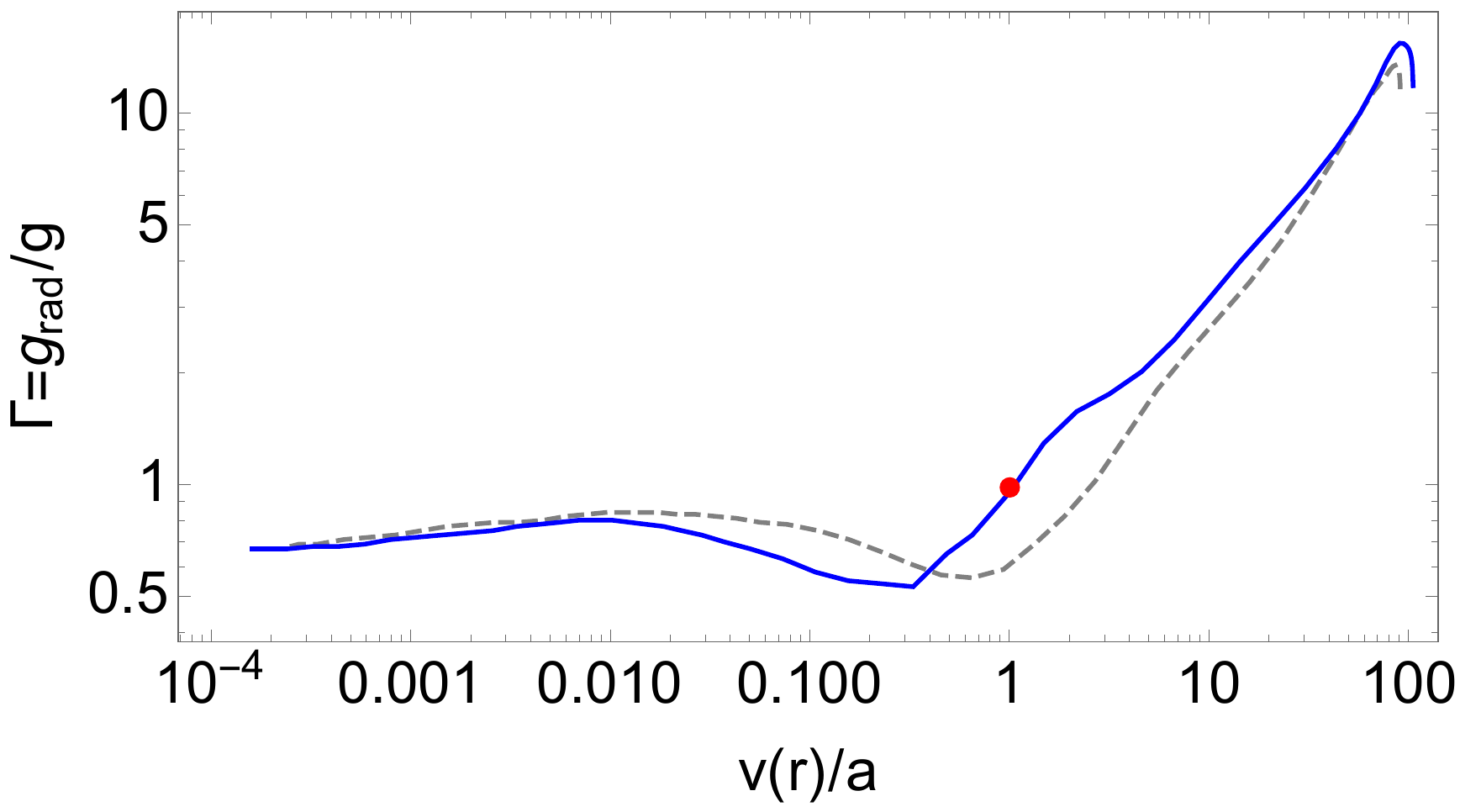}
		\caption{$\Gamma$ in function of the normalised velocity profile for the Lambert-solution of HD 163758 (blue) and its respective initial model (dashed grey). Red dot shows $\Gamma=1\pm0.1$ at $\hat v=1$.}
		\label{gamma_hd163758}
	\end{figure}

	A good solution for HD 163758 was found by decreasing the mass-loss rate compared with the initial value provided by \citet{bouret12} and tabulated in Table~\ref{initialmodel}.
	The run of $\Gamma$ as a function of normalised velocity is shown in Fig.~\ref{gamma_hd163758}.
	We observe, in this case, that our initial model for HD 163758 did not satisfy the criteria $\Gamma=1$ at the sonic point, even after adjustments of the velocity law in the neighbourhood of the sonic point.
	Thus we reduced the mass-loss rate from $1.5\times 10^{-6}$ to $1.2\times 10^{-6}$\,\Msunyr.

\begin{figure*}[t!]
		\centering
		\includegraphics[width=0.45\linewidth]{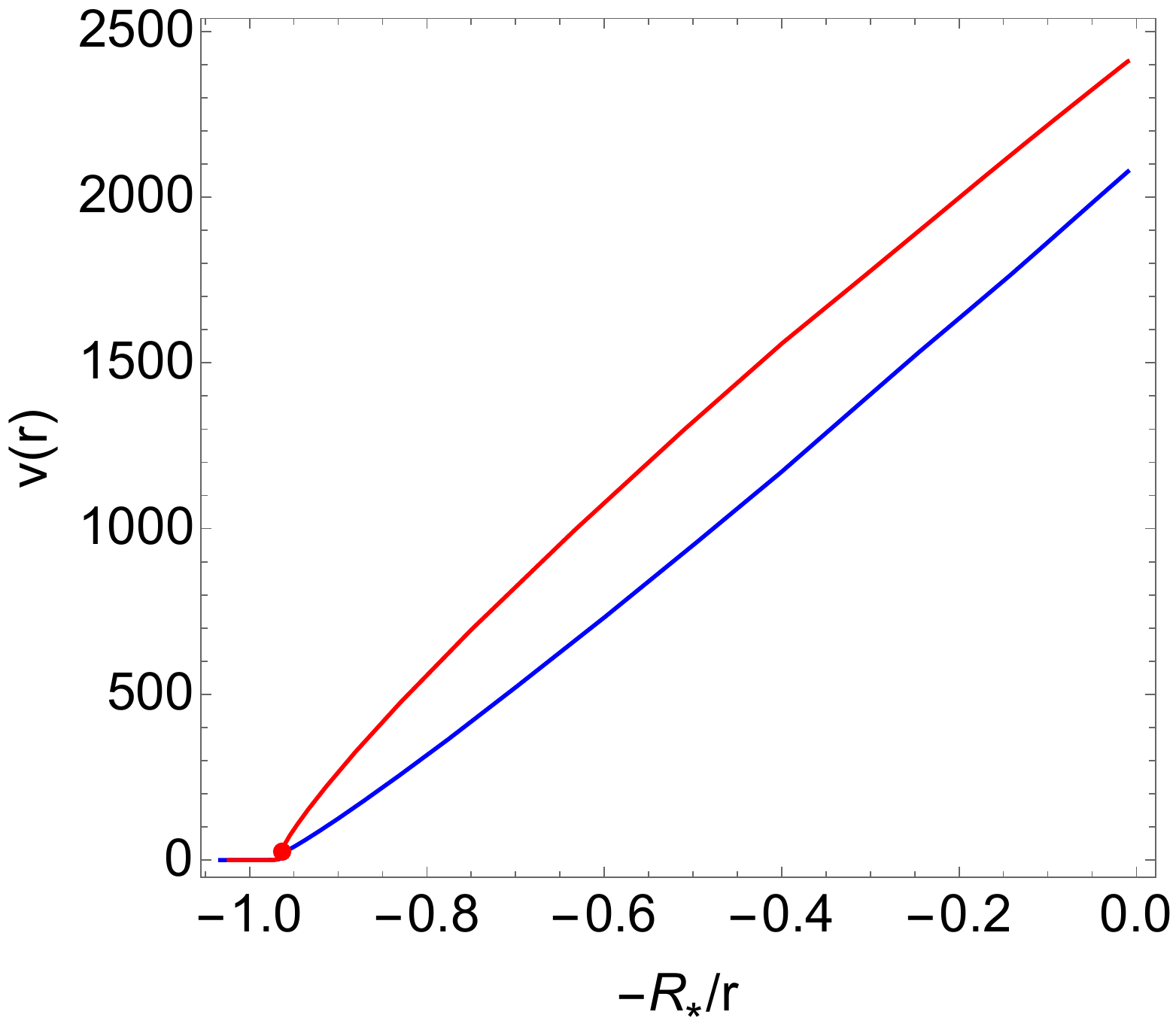}
		\includegraphics[width=0.45\linewidth]{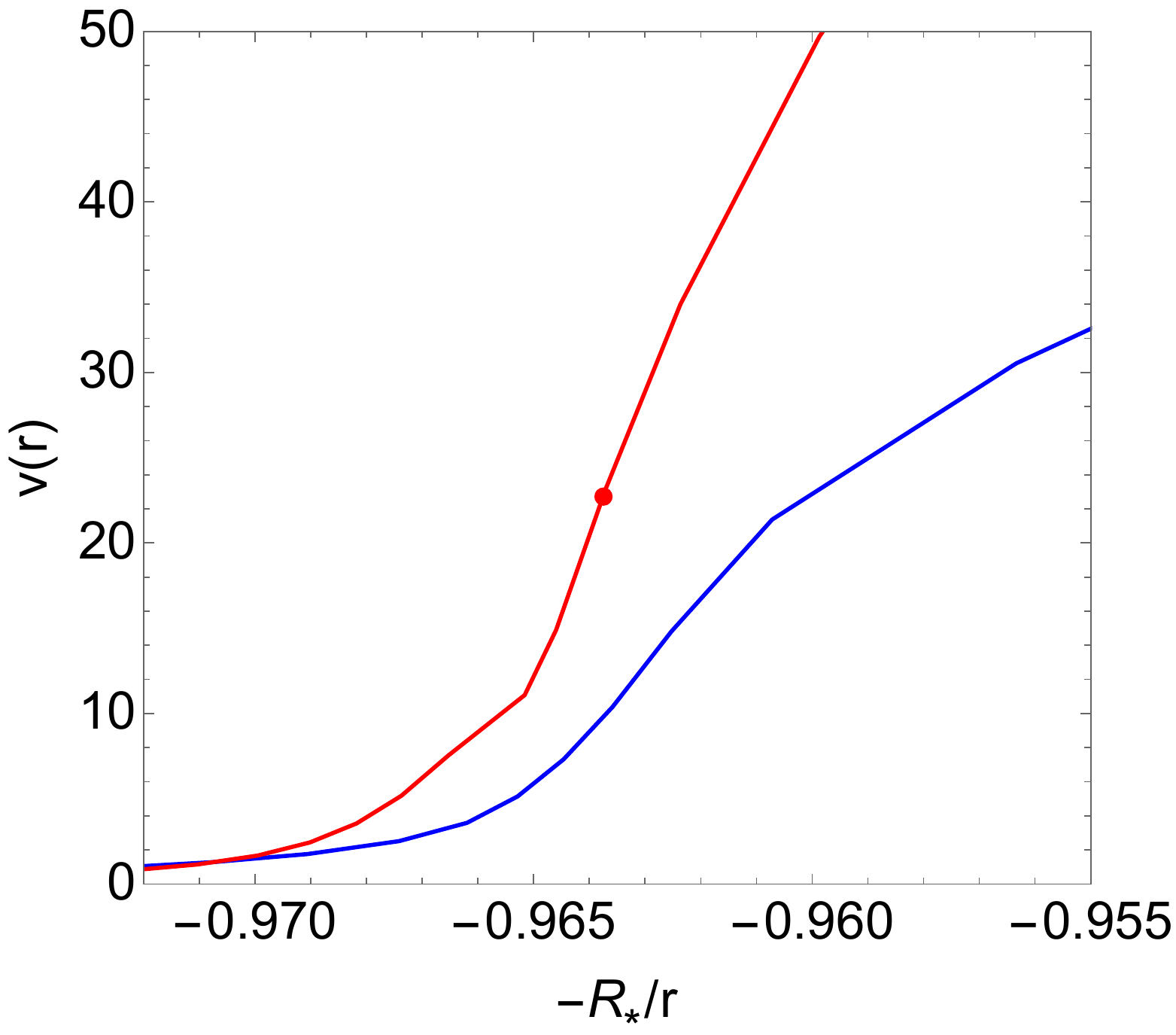}
		\caption{Left panel: differences between initial velocity field with $\beta$-law (blue) and the converged velocity profile from Lambert-procedure (red) for HD 163758 model. Right panel: zoom over the transition region is shown, displaying the sonic point with a red dot.}
		\label{hd163758oldvsnewvel}
	\end{figure*}

	Checks for the unicity of the solution were the same as for $\zeta$-Puppis.
	Comparisons between the old and the self-consistent new velocity profile are shown in Fig.~\ref{hd163758oldvsnewvel}.
	The increment in the terminal velocity for a Lambert-model is 1.14 times, from $2\,100$ to $2\,400$ km s$^{-1}$, lower than for $\zeta$-Puppis.
	However, we observe the ``departure" of the Lambert-velocity profile from the initial $v(r)$ at a much deeper zone, near $v\sim2$ km s$^{-1}$.
	This occurs because the new Lambert-velocity profile has a lower value for its quasi-$\beta$ (0.85) compared with the initial introduced $\beta$-parameter (1.1).
	
\begin{figure}[h!]
		\centering
		\includegraphics[width=0.95\linewidth]{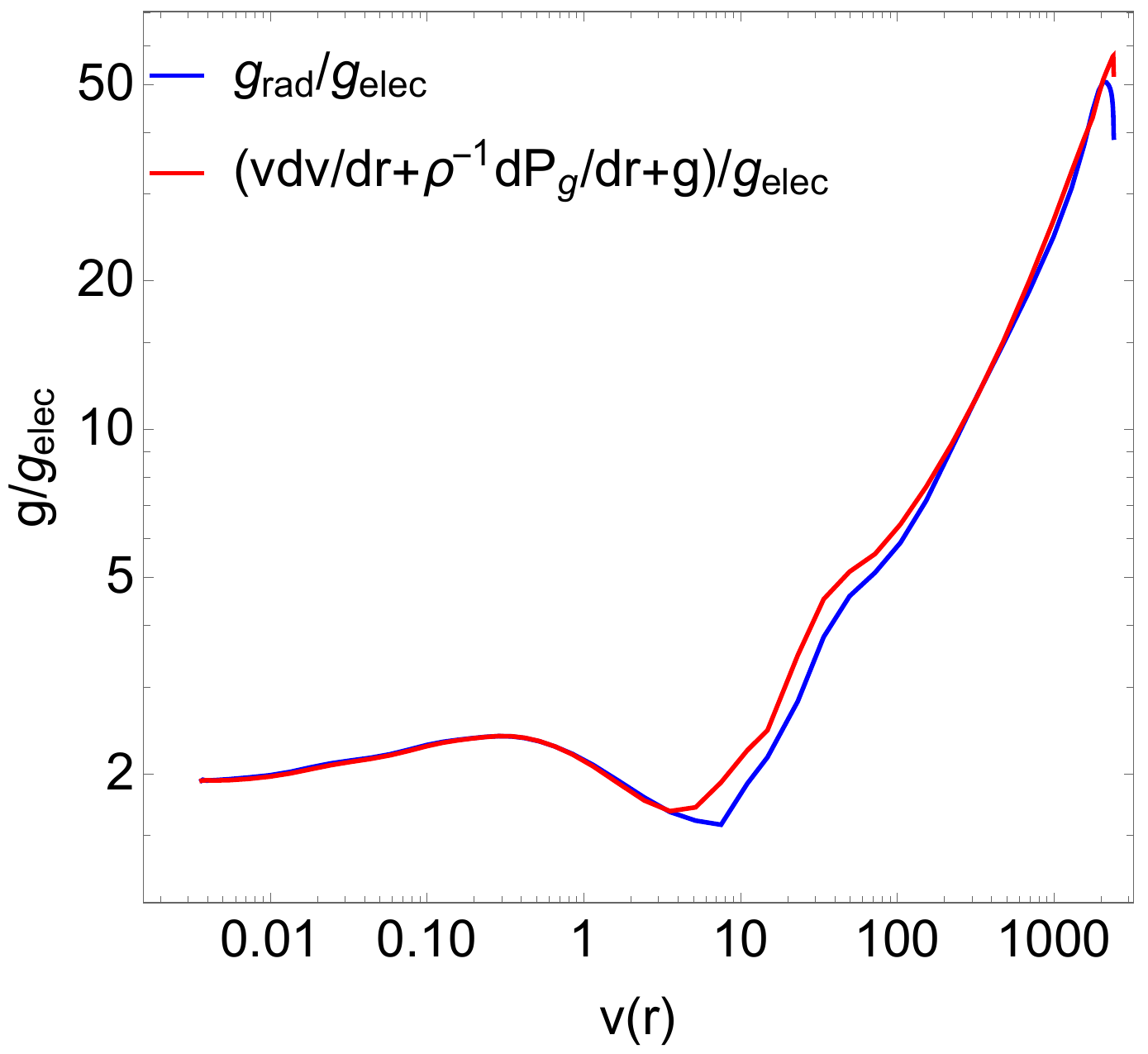}\\
		\caption{Radiative acceleration (divided by acceleration due to electron scattering) as a function of wind velocity for the final CMFGEN Lambert-model (blue) for HD 163758, compared with the expected value when equation of momentum is solved (red).}
		\label{hd163758finalglinev}
	\end{figure}

	Despite these encouraging result, the instability of the model could not be reduced to the same level as for $\zeta$-Puppis.
	As it can be seen from Fig.~\ref{hd163758finalglinev}, the agreement for the equation of motion this time is not satisfied as well as in the case for $\zeta$-Puppis.
	
	Particularly for this star, we tested the effects of the modification of the turbulent velocity $v_\text{turb}$.
	We found that its reduction from 10 to 7 km s$^{-1}$ reduced the error of the CMFGEN model, without any significant variation over the resulting velocity profile.
	
\begin{figure}[h!]
		\centering
		\includegraphics[width=0.95\linewidth]{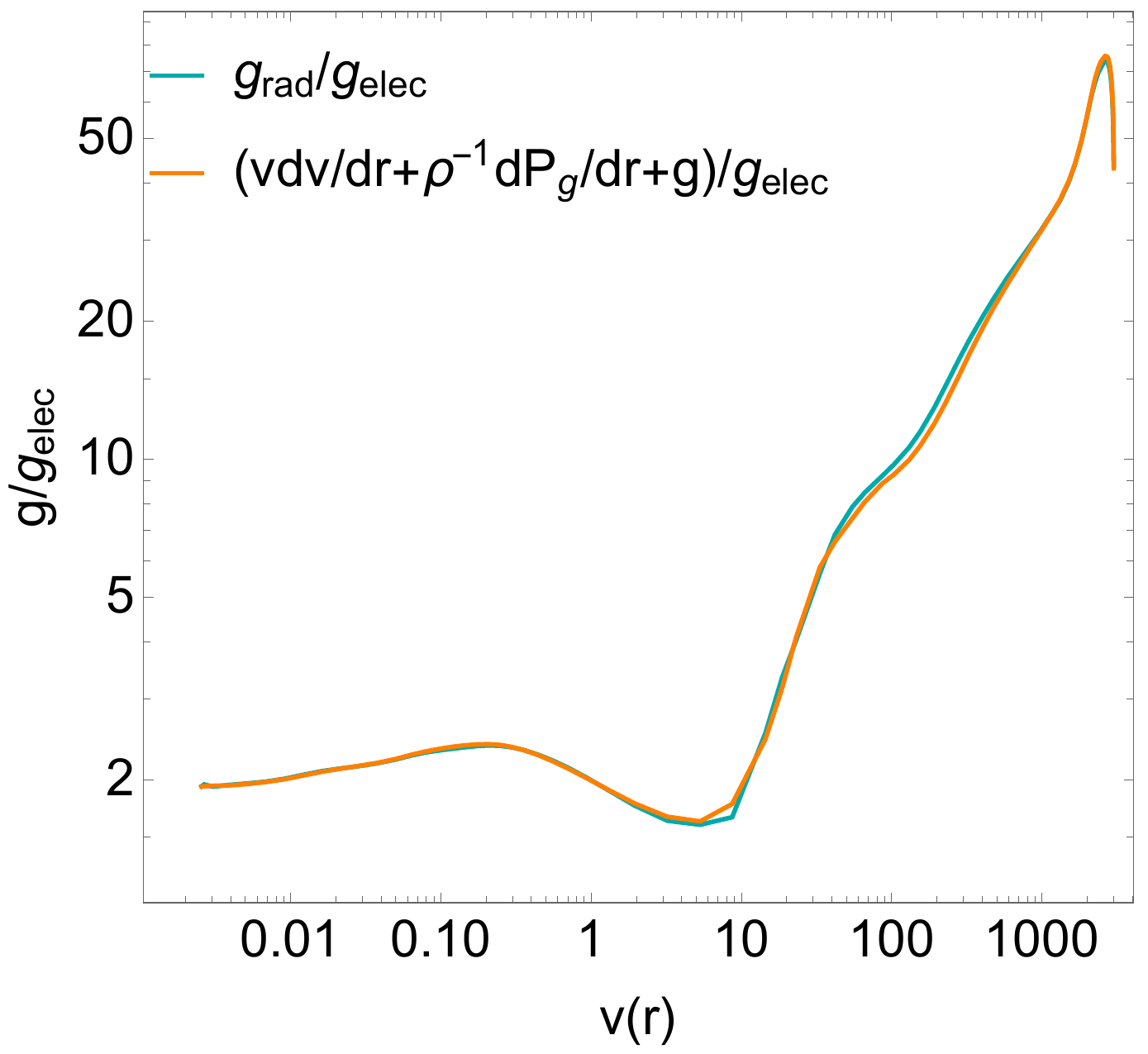}\\
		\caption{Radiative acceleration (divided by acceleration due to electron scattering) as a function of wind velocity for the alternative CMFGEN Lambert-model (cyan) of HD 163758, compared with the expected value when equation of momentum is solved (orange).}
		\label{hd163758finalglinevalt}
	\end{figure}

\begin{figure}[h!]
		\centering
		\includegraphics[width=0.95\linewidth]{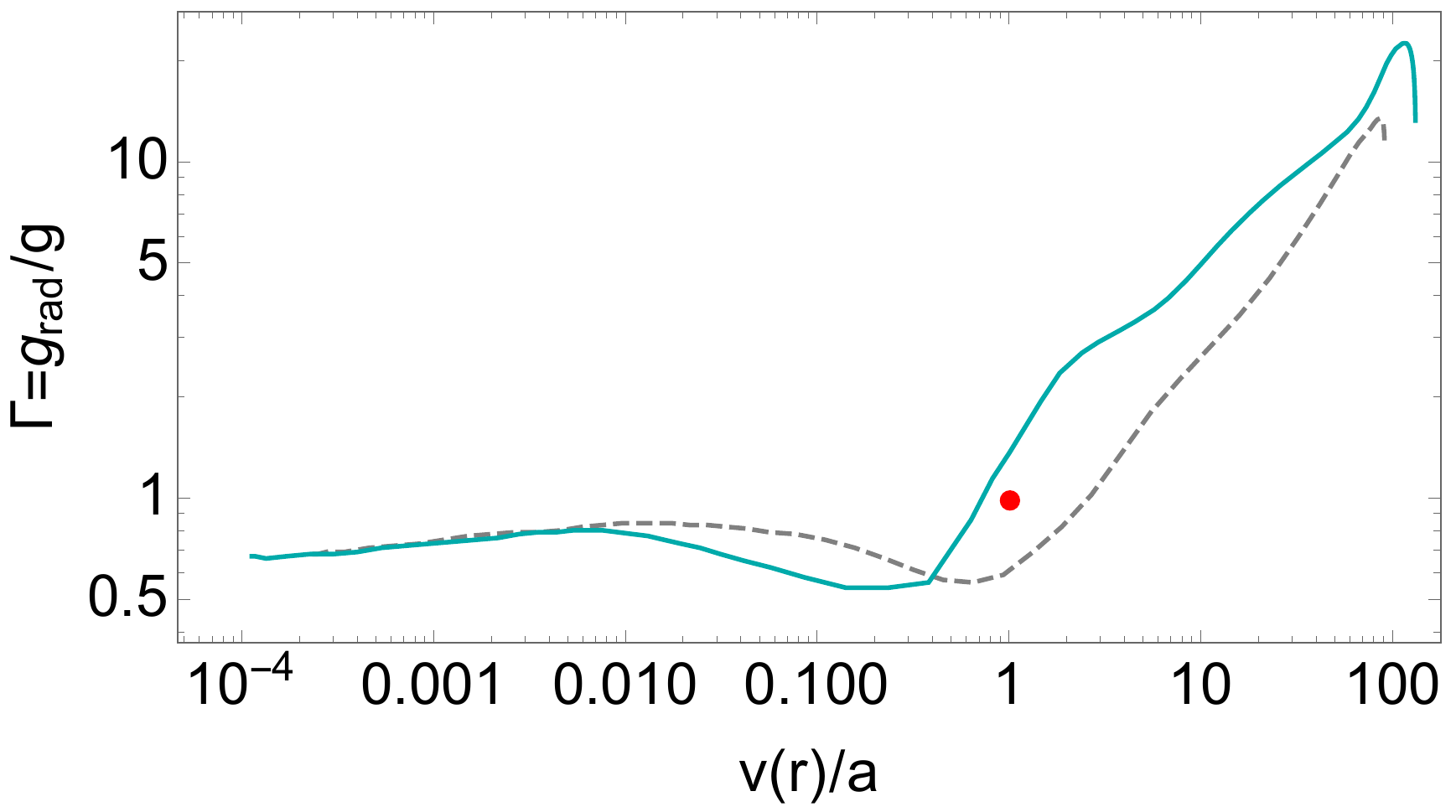}
		\caption{$\Gamma$ as a function of the normalised velocity profile for the Lambert-solution of the alternative model for HD 163758 (cyan) and its respective initial model (dashed grey). Red dot shows $\Gamma=1\pm0.1$ at $\hat v=1$.}
		\label{gamma_hd163758_alt}
	\end{figure}

	Alternatively, we have found that a significant decreasing of the mass-loss rate led to a more stable converged Lambert-model and therefore a better agreement of Eq.~\ref{momentumgrad}.
	This is observed in Fig.~\ref{hd163758finalglinevalt}, where both sides exhibit solid agreement.
	The derived $\dot M=8.3\times10^{-7}\Msunyr$ is almost a factor of 2 lower than the initial model, while $v_\infty=3\,100$ km s$^{-1}$ is factor 1.5 times higher.
	Such strong differences affect the determination of the Gamma.
	Fig.~\ref{gamma_hd163758_alt} shows that the condition $\Gamma=1$ is not satisfied for the sonic point now, which can be assumed as a reinforcement that alternative mass-loss rate is too low.
	This implies, minimisation of the CMFGEN error may not lead to a smooth solution through the sonic point.
	
\subsection{$\alpha$-Cam}\label{alphacam}
\begin{figure}[htbp]
		\centering
		\includegraphics[width=0.95\linewidth]{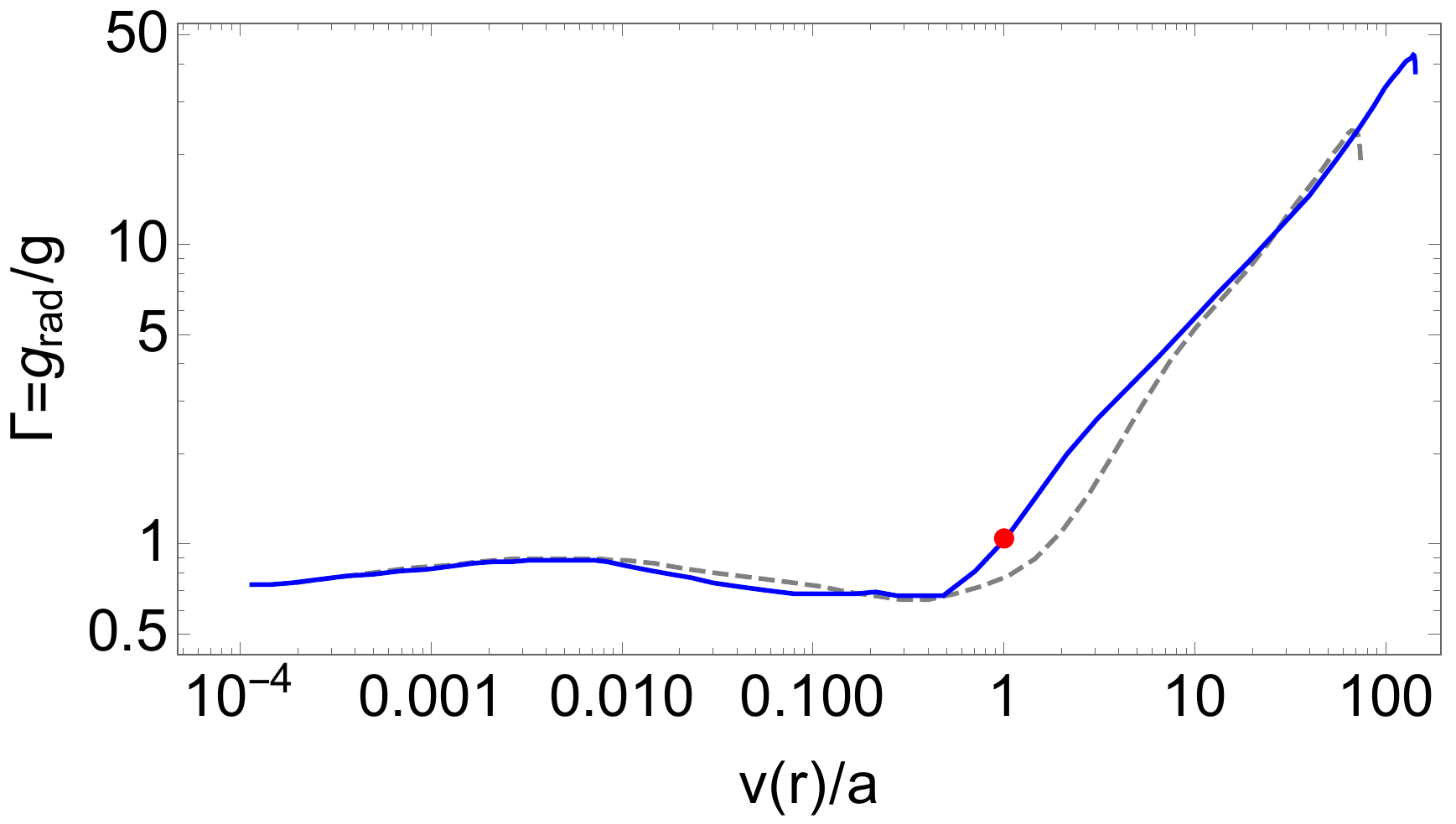}
		\caption{$\Gamma$ as a function of the normalised velocity profile for the Lambert-solution of $\alpha$-Cam (blue) and its respective initial model (dashed grey). Red dot shows $\Gamma=1\pm0.1$ at $\hat v=1$.}
		\label{gamma_alphacam}
	\end{figure}

	A solution for $\alpha$-Cam was found for a higher mass-loss rate and an increased volume filling factor (i.e., decreased clumping) compared to the initial values displayed in Table~\ref{initialmodel} that are similar to the ones provided by \citet{najarro11}.
	The run of $\Gamma$ as a function of normalised velocity is shown in Fig.~\ref{gamma_alphacam} where, as for HD 163758, $\Gamma=1$ at the sonic point was not satisfied in the initial model.
	
\begin{figure*}[t!]
		\centering
		\includegraphics[width=0.45\linewidth]{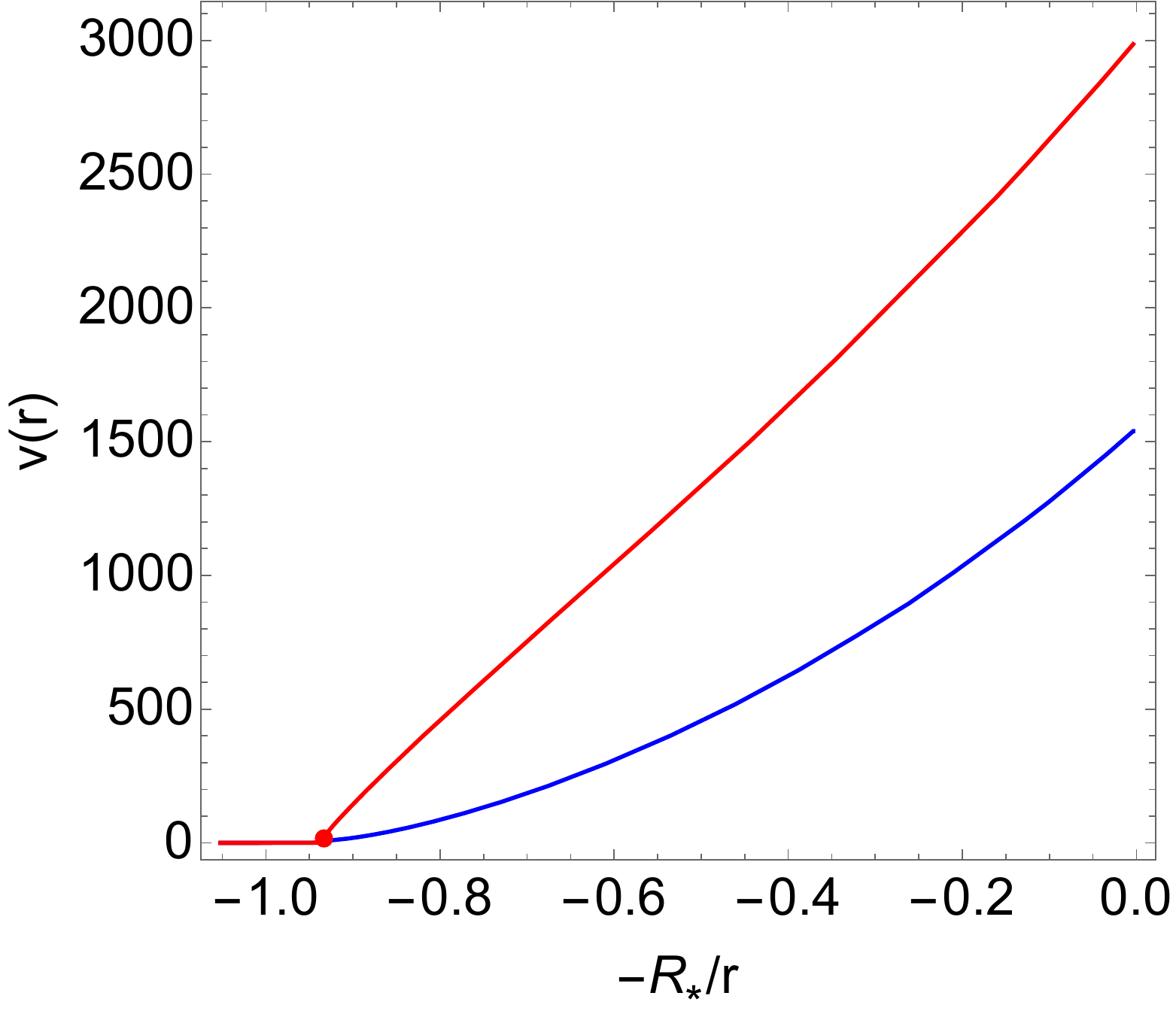}
		\includegraphics[width=0.45\linewidth]{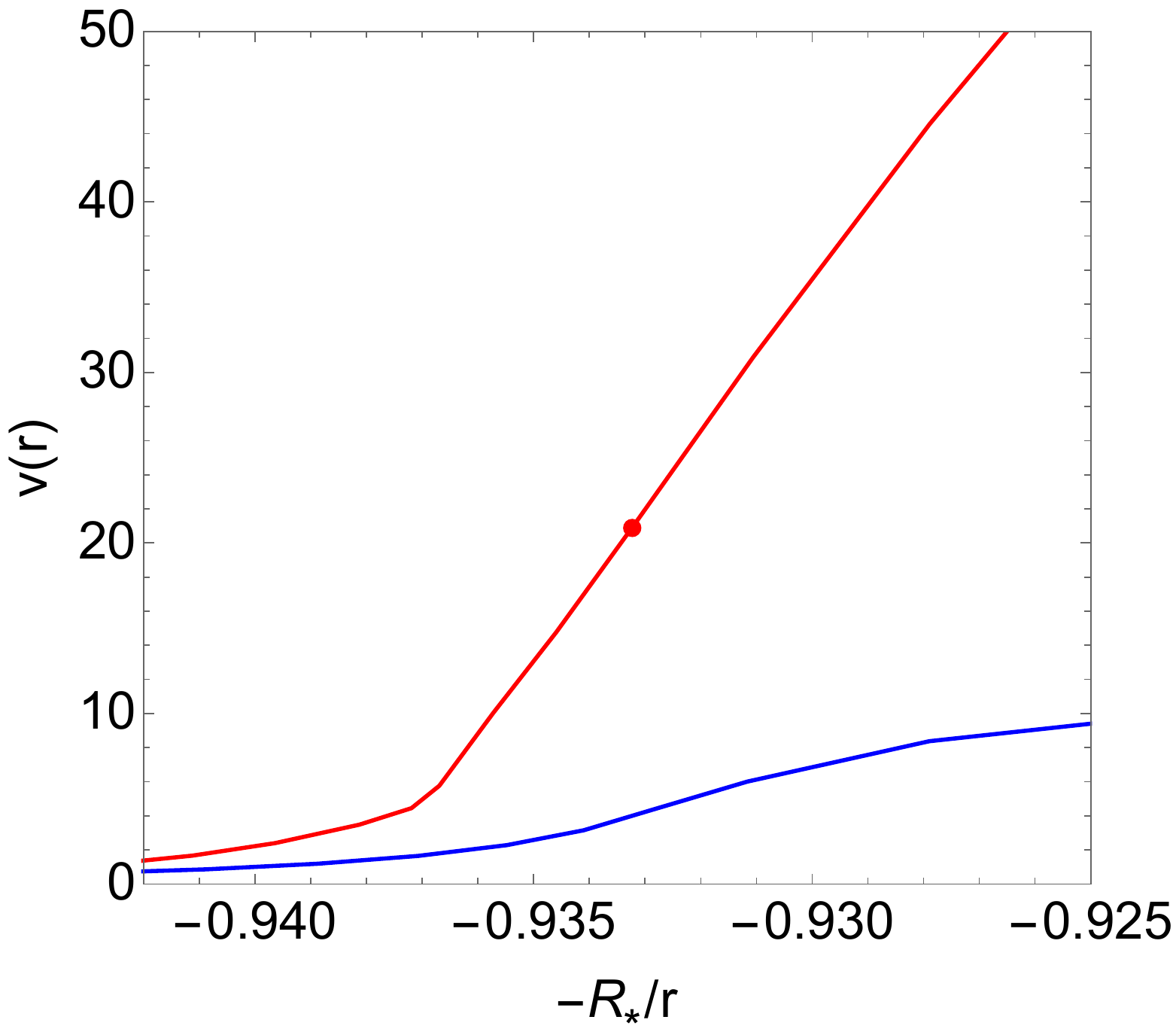}
		\caption{Left panel: differences between initial velocity field with $\beta$-law (blue) and the converged velocity profile from Lambert-procedure (red) for $\alpha$-Cam model. Right panel: zoom over the transition region is shown, displaying the sonic point with a red dot..}
		\label{acamoldvsnewvel}
	\end{figure*}

	Comparisons between the old and the new self-consistent velocity profile are shown in Fig.~\ref{acamoldvsnewvel}.
	This star presents the highest increment in the terminal velocity for a Lambert-model among the stars studied, from $1\,550$ to $3\,420$ km s$^{-1}$, a factor $\sim2.2$.
	The differences in the velocity profiles (for both the Lambert and initial model) go much deeper into the subsonic region ($v\lesssim1$ km s$^{-1}$), which again can be attributed to the large departure of the resulting quasi-$\beta$ (0.95) from the initial value of the $\beta$-parameter (1.5).
	
\begin{figure}[h!]
		\centering
		\includegraphics[width=0.95\linewidth]{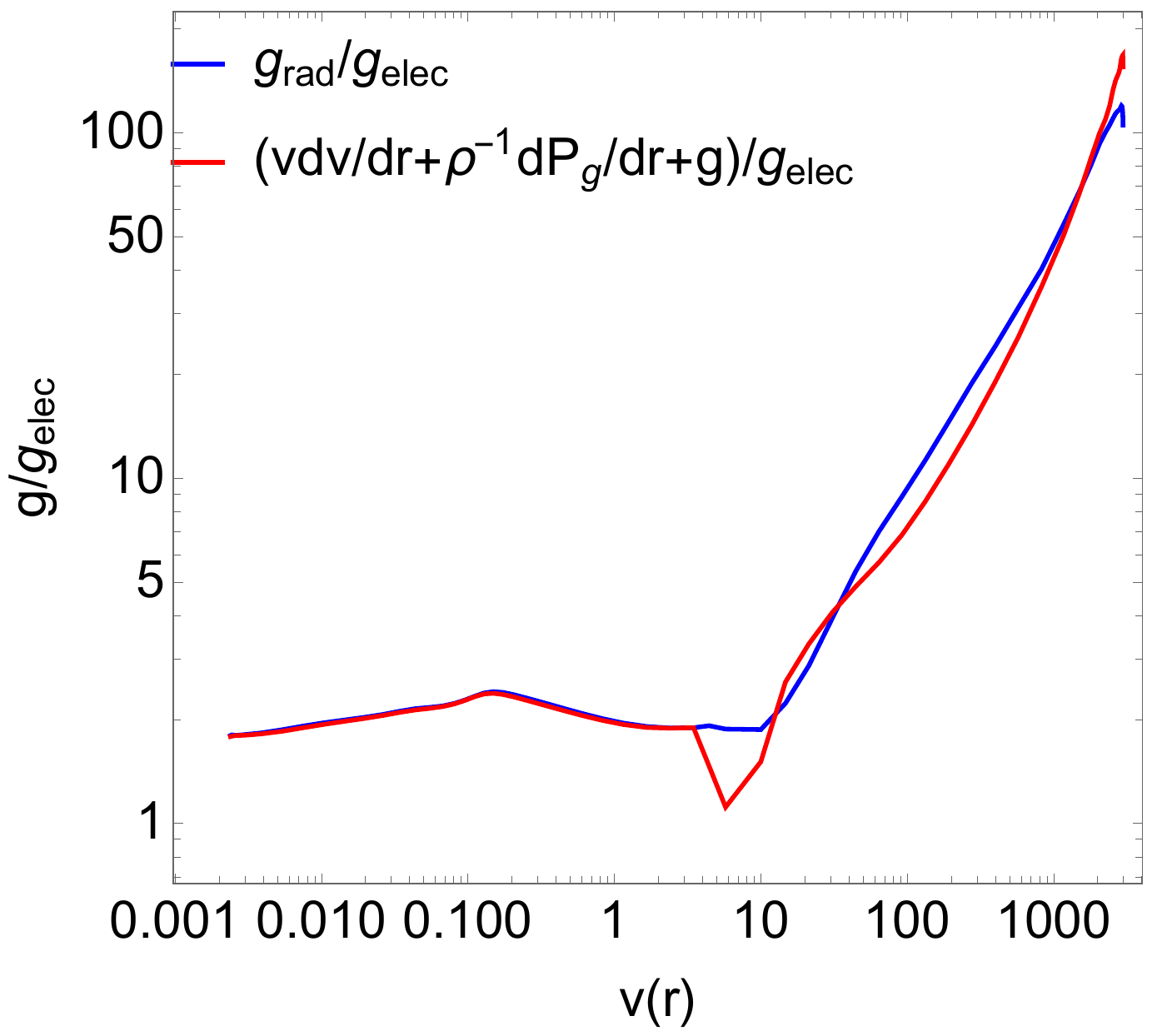}\\
		\caption{Radiative acceleration (divided by acceleration due to electron scattering) as a function of wind velocity for the CMFGEN Lambert-model (blue) for $\alpha$-Cam, compared with the expected value when equation of momentum is solved (red).}
		\label{acamfinalglinev}
	\end{figure}

	This great enhancement of the velocity profile also causes instabilities in the hydrodynamic structure of the model, making it impossible to obtain an accurate agreement for the equation of momentum, as illustrated in Fig.~\ref{acamfinalglinev}.
	
\begin{figure}[h!]
		\centering
		\includegraphics[width=0.95\linewidth]{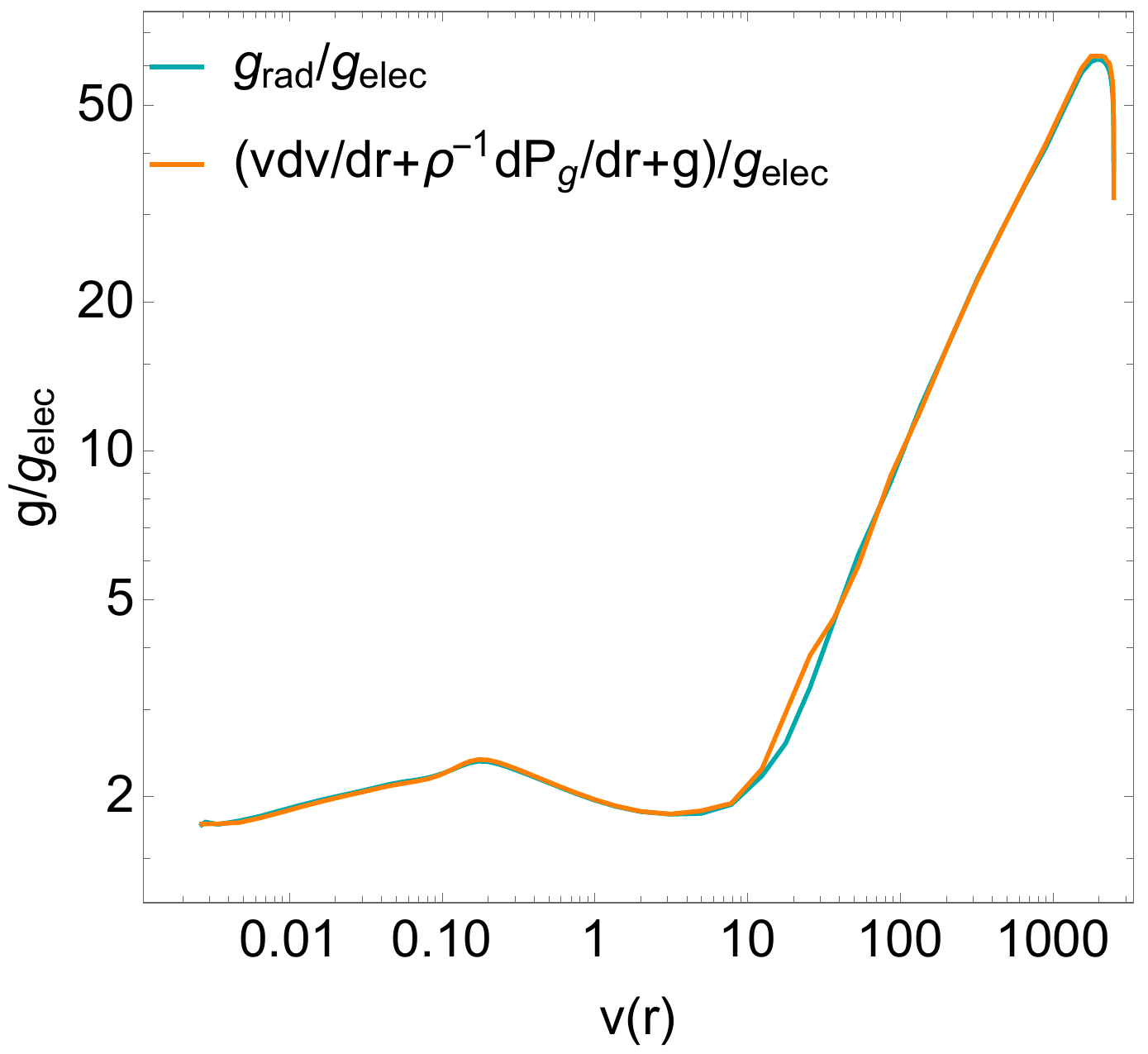}\\
		\caption{Radiative acceleration (divided by acceleration due to electron scattering) as a function of wind velocity for the alternative CMFGEN Lambert-model (cyan) for $\alpha$-Cam, compared with the expected value when equation of momentum is solved (orange).}
		\label{acamfinalglinevalt}
	\end{figure}

\begin{figure}[htbp]
		\centering
		\includegraphics[width=0.95\linewidth]{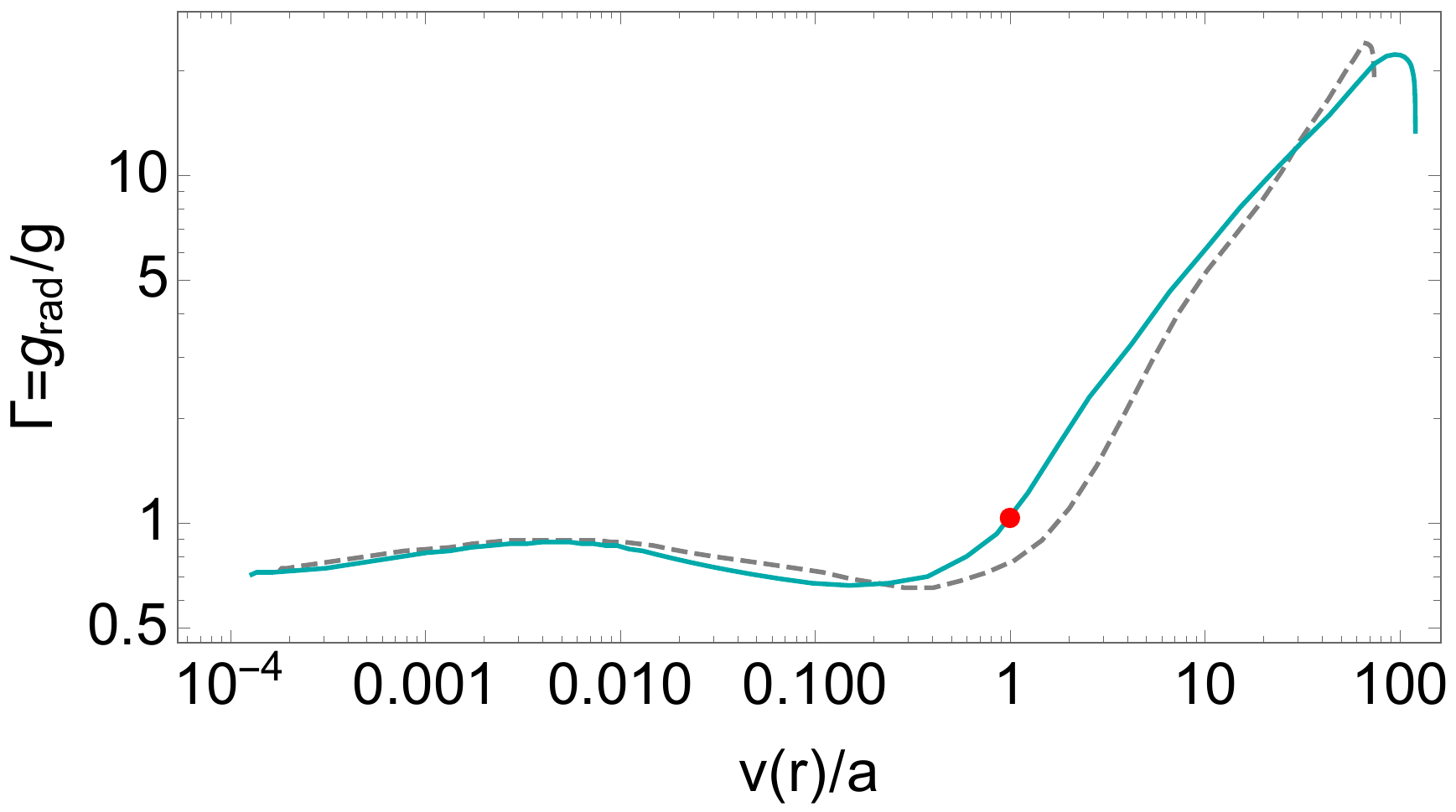}
		\caption{$\Gamma$ as a function of the normalised velocity profile of the Lambert-solution for the alternative $\alpha$-Cam (blue) and its respective initial model (dashed grey). Red dot shows $\Gamma=1\pm0.1$ at $\hat v=1$.}
		\label{gamma_alphacam_alt}
	\end{figure}

	For $\alpha$-Cam, we tested the effects of the modification of the $v_\text{cl}$ (see Eq.~\ref{fillingfactorv}).
	The modification from 37.5 to 100 km s$^{-1}$ reduced the error of the CMFGEN model but led to a Lambert-solution with a terminal velocity increased by $\sim$500 km s$^{-1}$.
	However, the increase in mass-loss rate to $8.5\times10^{-7}\Msunyr$ (a factor of $\sim1.2$ higher than the previous Lambert-model) led to a more stable model and a lower terminal velocity ($2\,650$ km s$^{-1}$).
	Fig.~\ref{acamfinalglinevalt} shows a more precise agreement for both sides of Eq.~\ref{momentumgrad}, and Fig.~\ref{gamma_alphacam_alt} shows that the condition $\Gamma=1$ for the sonic point is also satisfied.
	Thus, the alternative model satisfies better our two criteria, and therefore provides a more reliable value for the mass-loss rate.
	
\section{Observational data and synthetic spectra}\label{obsdata}
	Observed spectra, which are used for comparisons with the synthetic spectra have already been presented by the authors referenced previously (and were kindly provided to us).
	The optical spectrum for $\zeta$-Puppis was obtained using the FEROS instrument (resolution $\lambda/\Delta\lambda=48\,000$), while the optical spectrum for HD 163758 was retrieved from the UVES-POP database\footnote{\url{http://www.eso.org/sci/observing/tools/uvespop/field_stars_uptonow.html}} \citep[see table 2 of][for details]{bouret12}.
	For the case of $\alpha$-Cam, spectra comes from the Indo-US Library of Coudé Feed Stellar Spectra \citep{valdes04}. Ultraviolet spectra of the three stars were obtained from the high-resolution echelle (SWP) International Ultraviolet Explorer (IUE) spectra from MAST\footnote{\url{https://mast.stsci.edu/portal/Mashup/Clients/Mast/Portal.html}}.
	The SWP spectra cover the spectral range, $\lambda\lambda$1150-2000 $\AA$, at a resolution of $\lambda/\Delta\lambda=10\,000$.
	
	Comparisons between the observational data and the synthetic spectra calculated from the Lambert-models discussed in Section~\ref{results} are presented in Appendix~\ref{cmfgenspectra}.
	Because the stellar parameters (effective temperature, gravity surface, abundances) are not modified, we do not expect to improve the fits obtained from the CMFGEN models using the $\beta$-law.
	However the differences between them can provide insights into our understanding of stellar winds.
	In our discussion, we primarily focus on the H$_\alpha$ and the C IV $\lambda\lambda$ 1548, 1551 line profiles.
		
	For $\zeta$-Puppis, because the initial model already had an accurately constrained value for the mass-loss rate, and we see that there are no large differences between the initial model spectra, and the final Lambert-model spectra, for both the optical (Fig.~\ref{lamb022visiblefit}) and the UV (Fig.~\ref{lamb022uvfit}).
	The main exception arises from the higher terminal velocity of the Lambert model which yields excess absorption on the blue side of the C IV $\lambda\lambda$ 1548, 1551 doublet.
	
	For HD 163738, the decrease in mass-loss rate weakens the emission component of H$_\alpha$ (Fig.~\ref{lamb404visiblefit}), while the increase in terminal velocity modifies the blue side of C IV $\lambda\lambda$ 1548, 1551 (Fig.~\ref{lamb404uvfit}).
	The severely low value for the mass-loss rate of the alternative model is responsible for the absence of an emission component for H$_\alpha$ (Fig.~\ref{lamb833visiblefit}), whereas the high value of the terminal velocity increases the disagreement for the blue side of C IV $\lambda\lambda$ 1548, 1551 (Fig.~\ref{lamb833uvfit}).
	
	Finally, for $\alpha$-Cam, where the resulting parameters of the Lambert-model are very distant from the original parameters, the resulting spectrum, not surprisingly, looks very different.
	In this case, the H$_\alpha$ line does not present an emission component (Fig.~\ref{acam172visiblefit}) and the P-Cygni absorption component associated with C IV $\lambda\lambda$ 1548, 1551, has a much larger blueward extension (Fig.~\ref{acam172uvfit}).
	The increased mass-loss rate and clumping for the alternative model also does not produce any H$_\alpha$ emission component (Fig.~\ref{acam186visiblefit}), although the blue extension of C IV $\lambda\lambda$ 1548, 1551 is slightly decreased (Fig.~\ref{acam186uvfit}).
	
	Discrepancies between observational spectra and synthetic spectra computed from Lambert-solutions, which are particularly evident in the emission components of the H$_\alpha$ lines might suggest that stellar parameters of the models (effective temperature, stellar radius, surface gravity, abundances) should be revisited.
	However, another plausible reason is that the selected mass-loss rates for our Lambert-models are not accurately constrained.
	This suggests that the values of stellar and wind parameters derived in the literature for the stars studied in this paper, and are considered to give a good fit with the observations, may not be valid when consistent calculations of the velocity profile are included in the CMFGEN code.
	To achieve better agreement with the observed spectra using our Lambert-procedure would require the calculation of a small grid of models for different combinations of stellar parameters, including the mass-loss rate and clumping.
	Such a study is beyond the scope of the current paper and it will be done in future studies.

\section{Discussion}\label{discussion}

\subsection{Inherent Inconsistency of NLTE line formation calculations}\label{inconsistency}	

	\begin{figure}[t!]
		\centering
		\includegraphics[width=0.9\linewidth]{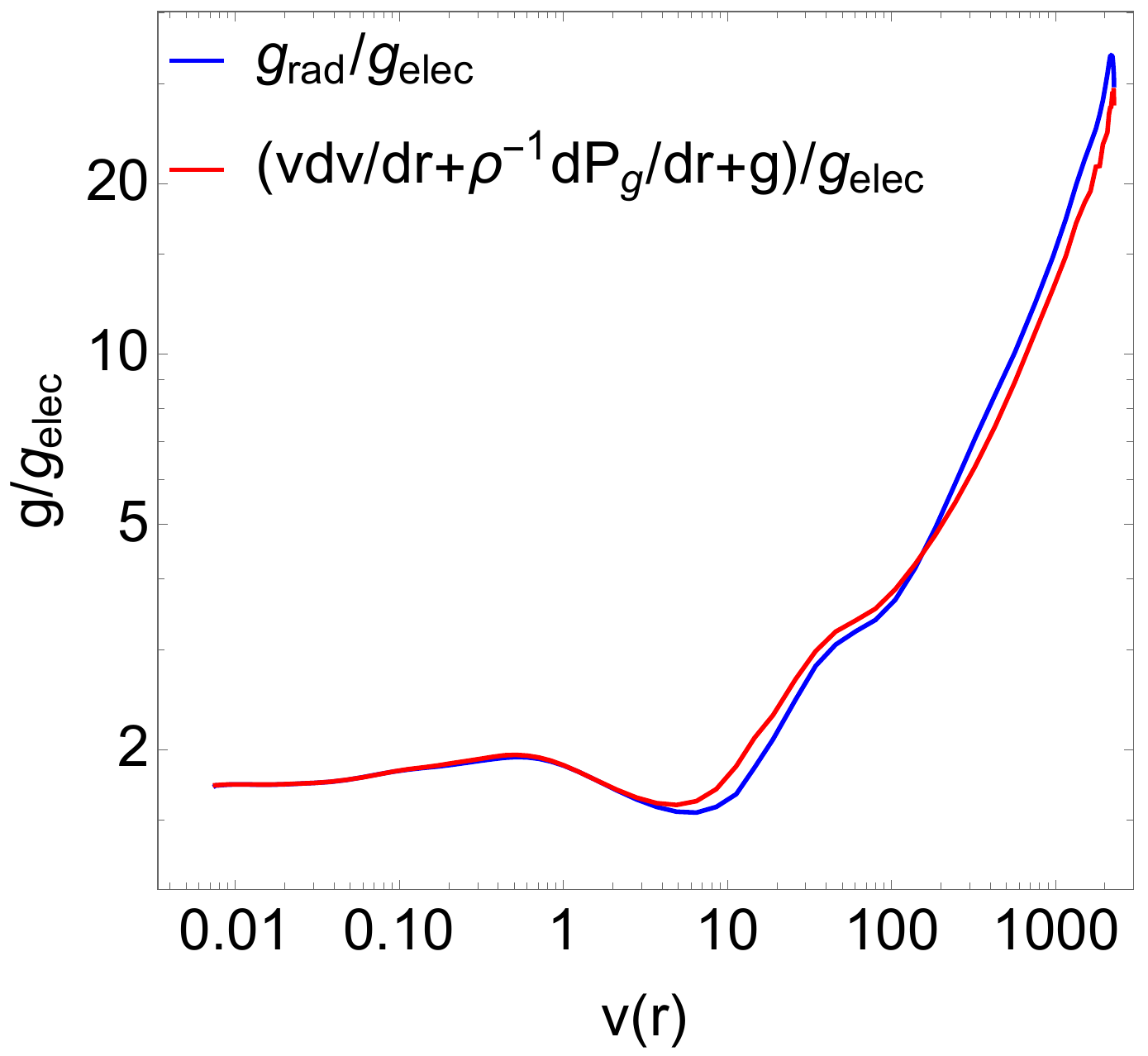}\\
		\caption{Comparison of left-hand side of Eq.~\ref{momentumgrad} (blue) and right-hand side (orange) for the hydrodynamics obtained from the CMFGEN model for $\zeta$-Puppis from \citet{bouret12}. Accelerations were rescaled by $g_\text{elec}$ (Eq.~\ref{gelec}) for illustrative reasons.}
		\label{initialglinev}
	\end{figure}
	
	An example of the discrepancy between observation and theory is presented in section 6.5 of the study of \citet{bouret12}, where they checked the consistency of the line-force for $\zeta$-Puppis.
	A truly self-consistent solution must satisfy the stationary equation of momentum (Eq.~\ref{momentumgrad}).
	Both the left and right hand side of this equation are plotted in Fig.~\ref{initialglinev}.
	This $g_\text{rad}(r)$ was calculated internally by CMFGEN starting from the initial parameters outlined in Table~\ref{initialmodel}.
	However, even when the choice of these parameters achieved a better agreement between the left-hand-side (LHS) and right-hand-side (RHS) for Eq.~\ref{momentumgrad} compared with those stellar and wind parameters determined by the spectral fit \citep[see the discussion in][]{bouret12}, both sides are still far from being in agreement.
	Hence, this particular CMFGEN model is not hydrodynamically self-consistent.
	
	The discrepancy shown in Fig.~\ref{initialglinev} arises because the velocity field is calculated from a $\beta$-law rather than from the line-acceleration.
	It could be argued that this discrepancy might be only in a specific star, but this was previously also noticed by \citet{puebla16}.
	Hence, the lack of consistency between hydrodynamics and radiative acceleration for a CMFGEN model with a velocity profile fixed by a $\beta$-law seems to be a general rule, and we can infer the same inconsistency for other radiative transfer codes (e.g., FASTWIND, PoWR) if they are also using the $\beta$-law.
	On the contrary, Lambert-solutions present a remarkably good agreement between both sides of Eq.~\ref{momentumgrad}, especially for those Lambert-models with the smallest error.
	In Fig.~\ref{errorgraph}, we compared the resulting Lambert-model for $\zeta$-Puppis with our other two stars HD 163758 and $\alpha$-Cam.
	Figure~\ref{errorgraph} is in concordance with that shown in Figures~\ref{finalglinev}, \ref{hd163758finalglinev}, \ref{hd163758finalglinevalt}, \ref{acamfinalglinev} and \ref{acamfinalglinevalt}.
	The lower the error, the better the agreement between the LHS and RHS of equation of momentum for the CMFGEN model.
	This provides evidence for the validity of the velocity profiles computed from the Lambert $W$-function (instead if using $\beta$-law), and also that equation of motion was properly satisfied for the CMFGEN model (Eq.~\ref{clumpedeom}), despite the condition $\Gamma=1$ not being fulfilled at $v=a$.
	In such a case the critical point for the equation is not the sonic point but in its close neighbourhood.

\subsection{Comparison with previous self-consistent studies}
	The search for a full self-consistent solution (coupling line-acceleration, hydrodynamics and radiative transfer) for the wind of massive stars has been approached previously by other studies \citep{puls00,kudritzki02,pauldrach12,kk17,sander17}.
	In particular, we emphasise the work done by \citet{sander17}, where another iterative procedure was implemented in order to obtain a self-consistent solution for the wind hydrodynamics of $\zeta$-Puppis under the non-LTE regime, using the radiative transfer code PoWR \citep{grafener02,hamann03}.
	In that study, radiative acceleration was also obtained from the output of the radiative transfer solution (PoWR for them, CMFGEN for us), and that acceleration was re-used to obtain a new hydrodynamic profile.
	
	Although their study shares a similar philosophy with ours, there are important differences.
	First, our new velocity profile is calculated by means of the Lambert $W$-function, whereas their $v(r)$ is recalculated by updating the stratification of the wind.
	The advantage of the Lambert-procedure is that it ensures the existence of a unique solution, i.e., the final converged Lambert-model does not depend on the initial velocity profile.
	Besides, our Lambert-procedure considers iterative changes in the velocity profile only, letting the stellar parameters (temperature, radius, mass, etc.) free and the other wind parameters (mass-loss rate and clumping factor) can be simultaneously modified to reduce the error in the models.
	Such relaxation imposed by us allows the search by eye inspection of the best spectral fit, reducing the number of free parameters (hydrodynamic will depend now from the initial stellar parameters and the constrained $\dot M$ and $f_\infty$).	
	Hence, we gain flexibility to constrain the mass-loss rate either by observations or by theoretical criteria, keeping in every case the self-consistency for the velocity profile.
	
	\begin{table*}[htbp]
		\centering
		\caption{Summary of self-consistent solutions performed by \citet{sander17}, \citet{alex19} and this present study (Table~\ref{initialmodel}). These three models assume the same abundances as \citet{bouret12}. Radiative transfer code used to perform the respective synthetic spectra is also marked, together with the code/methodology used to calculate the hydrodynamics. Equivalence between filling factor and clumping factor $D_\infty$ (used by \textsc{PoWR}) was made assuming the interclump medium is void, following \citet{sundqvist18}. Additionally we include from the grid of \citet{bjorklund21}, the closest parameters to $\zeta$-Puppis.}
		\label{zpuppissolutions}
		\begin{tabular}{ccccc}
			\hline\hline
			& \citet{sander17} & \citet{alex19} & \citet{bjorklund21} & This work\\
			\hline
			RT code & \textsc{PoWR} & FASTWIND & FASTWIND & CMFGEN\\
			Hydro method & – & \textsc{HydWind} & – & Lambert $W$-function\\
			$T_\text{eff}$ [kK] & 40.7 & 40 & 40.7 & 41\\
			$\log g$ & 3.63 & 3.64 & 3.65 & 3.6\\
			$R_*/R_\odot$ & 15.9 & 18.7 & 18.91 & 17.9\\
			$v_\infty$ [\,km s$^{-1}]$ & 2\,046 & 2\,700 & 2\,302 & 2\,740\\
			$\dot M$ [$10^{-6}M_\odot$ yr$^{-1}$] & $1.6$ & $5.2$ & $2.5$ & $2.7$\\
			$f_\infty$ [$1/D_\infty$] & 0.1 & 0.2 & – & 0.10\\
			\hline
		\end{tabular}
	\end{table*}
	
	Because of the different numerical approaches adopted in the literature, we select these two previous studies to compare with the results given by our Lambert-procedure.
	Comparison among these three self-consistent solutions for $\zeta$-Puppis are summarised in Table~\ref{zpuppissolutions}.
	In addition, we could have included the study made by \citet{pauldrach12}, who also analysed extensively $\zeta$-Puppis deriving self-consistent values for stellar and wind parameters of the star.
	However, they assumed a different value for the stellar radius of $\sim28~R_\odot$, far from the $\sim18~R_\odot$ adopted by \citet{sander17}, \citet{alex19} and the present work, and hence we decided not to include it in the comparisons.
	
	As an initial comment, we remark on the discrepancy between the wind parameters derived by \citet{sander17} and those derived by us, even when differences between stellar parameters are small.
	First, for the case of the terminal velocity, Sander's value of 2\,046 km s$^{-1}$ lies below typical values obtained by spectral fitting: 2\,250 km s$^{-1}$ by \citet{puls06} and 2\,300 km s$^{-1}$ by \citet{bouret12}.
	On the contrary, the final terminal velocity provided by the Lambert-procedure is $\sim20\%$ higher than the ``observed" $v_\infty$; this is seen, for example, in the fit for C IV $\lambda\lambda$ 1548, 1551 (Fig.~\ref{lamb022uvfit}).
	However, their terminal velocity is close to that determined by Paper I's m-CAK prescription \citep[see table~5 from][]{alex19}.
	
	The difference in methodology between the current paper, which uses the Lambert $W$-function, and Paper I, which uses m-CAK theory, can be summarised as it follows:
	Paper I:
	\begin{itemize}
		\item used a simplified "quasi-NLTE" scenario for the treatment of atomic populations, following formulations performed by \citet{mazzali93} and \citet{puls00} whereas in this work we solve the proper statistical equilibrium equations when CMFGEN is executed.
		\item used a flux field calculated by \textsc{Tlusty}, which uses the plane-parallel approximation whereas our flux field is calculated with CMFGEN.
		The radiation field computed in this case includes a full treatment of the line-blanketing and the multi-line scattering in spherical geometry \citep{puls87}.
		\item did not consider effects from clumping upon the resulting $g_\text{line}$ whereas, from Fig.~\ref{clumpinitial}, it is clear that line-acceleration changes when the mass-loss rate stays constant but the clumping factor is modified.
		\item creates hydrodynamic models starting from the photosphere of the star using \textsc{HydWind}, whereas with the Lambert-procedure velocity profiles are calculated from the sonic point outwards.
	\end{itemize}
	
	Therefore, these differences between the codes could easily explain the differences in wind parameters for $\zeta$-Puppis.

	Further, it is well known that rotation enhances the values for mass-loss rate on the equator of the star.
	Hydrodynamics calculated by \textsc{HydWind} in Paper I used a value for the normalised stellar angular velocity\footnote{Normalised stellar angular velocity is defined as
	$$\Omega=v_\text{rot}/v_\text{crit}\;,$$
	with $v_\text{crit}$ as defined in Eq.~\ref{changeofvariables2}.} of $\Omega=0.39$ in order to reproduce the known value of $v\sin i=210$ km s$^{-1}$ for $\zeta$-Puppis \citep{bouret12}, whereas the Lambert-procedure does not consider rotational effects because CMFGEN assumes spherical symmetry.
	According to \citet{venero16}, mass-loss rates increase their values by a factor of $\sim1.2-1.3$ for $\Omega=0.4$, which leads us to think that the self-consistent value for $\dot M$ obtained in Paper I would decrease down to $\sim4\times10^{-6}$ $M_\odot$ yr$^{-1}$ if $\Omega=0$, closer to the $\dot M$ determined by the Lambert-procedure.
	
	While that both m-CAK and Lambert prescriptions obtain self-consistent solutions, they work under different philosophies.
	In Paper I, the m-CAK prescription calculates its own self-consistent mass-loss rate, which is later tested by spectral analysis in order to \textit{check how near or far} it falls from the real solution.
	On the other hand, the Lambert-procedure presented in this work calculates a self-consistent solution for the wind hydrodynamics with the mass-loss rate as a free parameter, which is later (or can be) \textit{constrained by the spectral fitting}.
	
	We have also included in Table~\ref{zpuppissolutions} a comparison with the study performed by \citet{bjorklund21}, who provide a grid of self-consistent solutions following the iterative procedure using FASTWIND introduced in \citet{sundqvist19}.
	We selected the model with the stellar parameters closest to those assumed for $\zeta$-Puppis and we observe their predicted mass-loss rate is in agreement with the value set by us, although their value for terminal velocity is slightly lower than ours, and better matches the ``observational" value of $2\,300$ km s$^{-1}$ \citep{bouret12}.
	The difference from our result may be due to their neglect of clumping.
	Summary of the best hydrodynamical solution for HD 162758 and $\alpha$-Cam found in this study, compared with the closest parameters from the grid of \citet{bjorklund21}, are displayed in Tables~\ref{hd163758solutions} and \ref{alphacamsolutions} respectively.
	
	\begin{table*}[htbp]
		\centering
		\caption{Summary of the most reliable self-consistent solution for HD 163758 performed by this present study (Section~\ref{hd163758}), contrasted with the closest model from the grid of \citet{bjorklund21}.}
		\label{hd163758solutions}
		\begin{tabular}{ccc}
			\hline\hline
			& \citet{bjorklund21} & This work\\
			\hline
			RT code & FASTWIND & CMFGEN\\
			Hydro method & – & Lambert $W$-function\\
			$T_\text{eff}$ [kK] & 34.6 & 34.5\\
			$\log g$ & 3.43 & 3.4\\
			$R_*/R_\odot$ & 20.68 & 19.1\\
			$v_\infty$ [\,km s$^{-1}]$ & 2\,377 & 2\,400\\
			$\dot M$ [$10^{-6}M_\odot$ yr$^{-1}$] & $0.64$ & $1.2$\\
			$f_\infty$ [$1/D_\infty$] & – & 0.05\\
			\hline
		\end{tabular}
	\end{table*}
	\begin{table*}[htbp]
		\centering
		\caption{Summary of the most reliable self-consistent solutions for $\alpha$-Cam performed by this present study (Section~\ref{alphacam}), contrasted with the closest model from the grid of \citet{bjorklund21}.}
		\label{alphacamsolutions}
		\begin{tabular}{ccc}
			\hline\hline
			& \citet{bjorklund21} & This work\\
			\hline
			RT code & FASTWIND & CMFGEN\\
			Hydro method & – & Lambert $W$-function\\
			$T_\text{eff}$ [kK] & 28.4 & 28.2\\
			$\log g$ & 3.18 & 2.975\\
			$R_*/R_\odot$ & 23.11 & 30.3\\
			$v_\infty$ [\,km s$^{-1}]$ & 2\,972 & 2\,650\\
			$\dot M$ [$10^{-6}M_\odot$ yr$^{-1}$] & $0.18$ & $0.85$\\
			$f_\infty$ [$1/D_\infty$] & – & 0.05\\
			\hline
		\end{tabular}
	\end{table*}

\subsection{Equations of motion and Lambert $W$-function}

	The self-consistent velocity profiles introduced in this work have the advantage of having been analytically calculated from the equation of motion of the wind using the Lambert-procedure.
	However, the e.o.m. introduced in Eq.~\ref{motion3} assumes an isothermal and homogeneous wind.
	Such ``simplifications" were necessary to allow the e.o.m to be put in a form that can be solved using the Lambert $W$-function.

	The adopted iterative procedure (Appendix~\ref{iter_append}) allows a converged solution to be obtained for the wind velocity even though the full equation of motion is not satisfied everywhere (particularly in the neighbourhood of the sonic point).
	This was observed for HD 163758 and $\alpha$-Cam, where we could find two different new velocity profiles calculated by the Lambert $W$-function for each star, and where the initial choice of mass-loss rate led into different terminal velocities.
	
\subsection{Wind instabilities and mass-loss rates}\label{windinstabilities}

	\begin{figure}[t!]
		\centering
		\includegraphics[width=\linewidth]{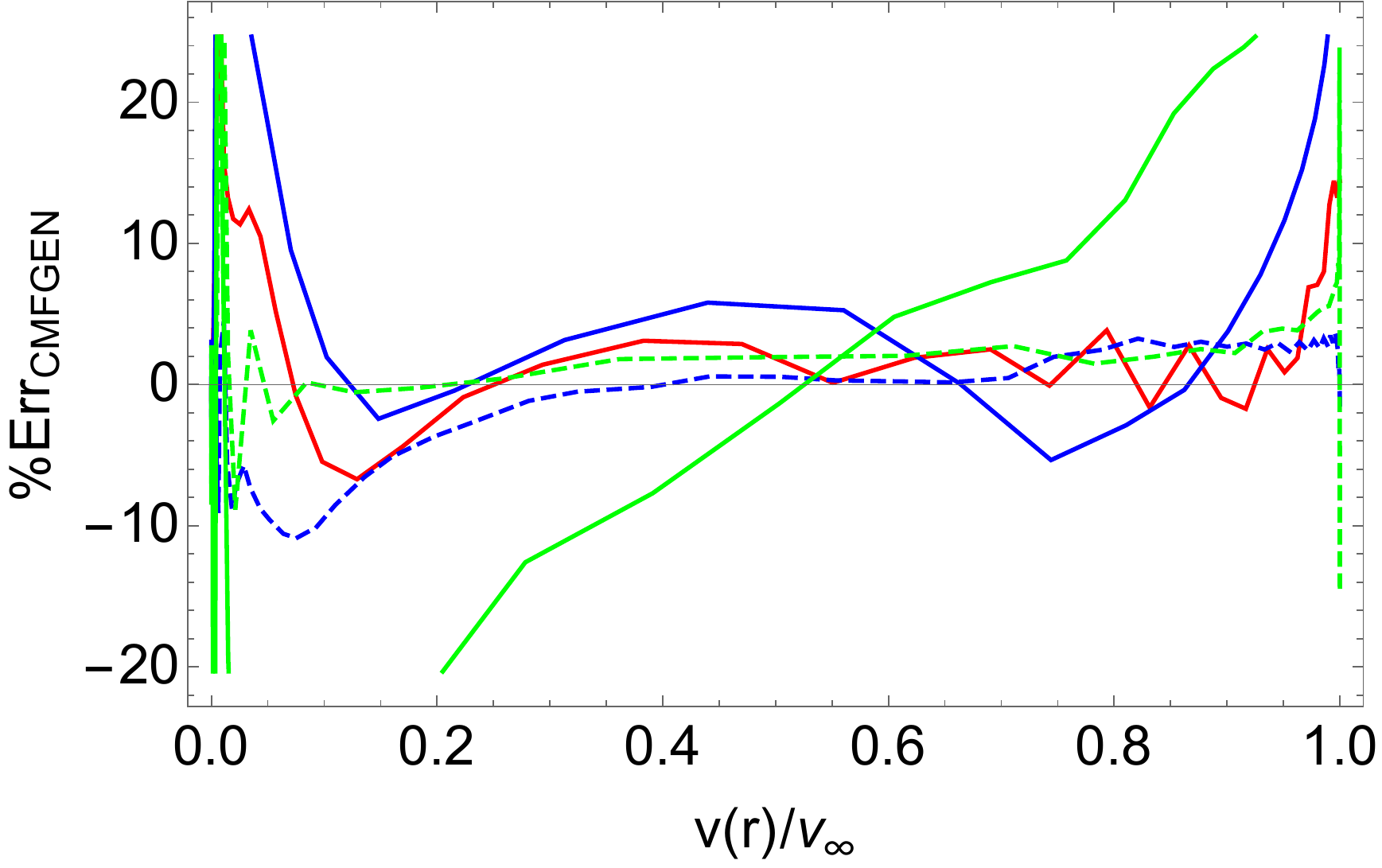}
		\caption{Numerical errors (Eq.~\ref{errorcmfgen}) of the final Lambert-models of $\zeta$-Puppis (red), HD 163758 (blue) and $\alpha$-Cam (green). Error for the alternative models of HD 163758 and $\alpha$-Cam are plotted in dashed lines.}
		\label{errorgraph}
	\end{figure}
	
	As we have stated, different values for the mass-loss rate of each star produce different grades of errors associated with the CMFGEN model along all the velocity structure.
	The Lambert-procedure can provide a converged solution for the velocity profile (i.e., we can recover the previous $v(r)$ after running the Lambert $W$-function) for any potential set of $\dot M$ and $f_\infty$ if the $g_\text{line}$ is given, but just one set of values for the mass-loss rate and clumping factor will provide the most stable line-acceleration from the CMFGEN model.
	However, one of the limitations of that strategy is that the error reduction can be compared only with the intrinsic error of the previous model but does not guarantee to achieve a standard threshold valid for any model (Fig.~\ref{errorgraph}).
	As a consequence, values for mass-loss rates presented in this work are those which lie close enough to the hypothetical \textit{true} value, and thus we talk about \textit{constraining} $\dot M$ instead calculate it.
	Also, CMFGEN models does not depend only in the mass-loss rate; this is particularly evident if we consider the extra features that affect the calculation of $g_\text{line}$.
		
	Hydrodynamic solutions from CMFGEN models depend not only on the selected wind parameters, but also on other physics which are not commonly considered in the standard model of line-driven winds.
	For instance, the inclusion of X-rays which modifies the ionisation balances of the atomic populations and therefore potentially affects the lines contributing to the line-acceleration.
	The turbulent velocity which affects the line-acceleration near the sonic point because of the enhancement of the Doppler width.
	This influence was extensively discussed by \citet{lucy07} and \citet{sundqvist19}, where different choices of $v_\text{turb}$ produced different wind parameters.
	Adopting a turbulent velocity also modifies the sound speed (Eq.~\ref{sound_plus_vturb}) and therefore the location of the critical point for the equation of motion.
	However, for HD 163758 a reduction of $v_\text{turb}$ did not produce any significant change in the final Lambert-solution.
	In part this is expected since the biggest change will occur around the sonic point.
	A more detailed study of the influence of the turbulent velocity on the Lambert $W$-function solution is warranted.
	
	Another important factor that influences both the resulting Lambert-models and the hydrodynamic solution obtained with CMFGEN is clumping.
	The models presented in this paper generally characterise the clumping via two parameters, the filling factor at infinity, $f_\infty$, and an onset velocity $v_\text{cl}$.
	Our functional form was adopted for simplicity – other forms may be more relevant, and in addition the clumping may not be a monotonic function \citep{puls06,najarro11}.
	The adopted value of $f_\infty$ is crucial for the wind dynamics, whereas the influence of $v_\text{cl}$ is more uncertain.
	Alternative exponential functions using Rosseland optical depth instead of the velocity field was proposed by \citet[][see their equation~32]{sander17}, but still there is an extra free parameter to be determined.
	CMFGEN currently neglects porosity, but the inclusion of porosity, which is probably more important in models with a ``low" mass-loss rate, could help to explain why $v_\infty$ in hydrodynamic models is sometimes larger than that observed.
	
	Given the numerous factors that influence the line-acceleration near the sonic point, it is extremely difficult to determine an accurate mass-loss rate from first principles.
	However using the Lambert-model to derive a velocity law does allow a range of mass-loss rates to be utilised in spectral fitting, and hence potentially provides tighter constraints on derived mass-loss rates and wind properties.
		
\section{Summary and conclusions}\label{conclusions}
	In this study we have presented a procedure to calculate self-consistent hydrodynamics beyond the m-CAK prescription presented in \citet[][Paper I]{alex19}.
	For this purpose, we have used the CMFGEN radiative transfer code and the Lambert $W$-function.
	This function allows us to analytically solve the equation of motion (Eq.~\ref{motion1}), using the line-acceleration $g_\text{line}$ provided by the solution of the radiative transfer equation in CMFGEN, to provide  a new velocity profile.
	Since the line acceleration computed in CMFGEN depends on $v(r)$ we use an iterative procedure (Lambert-procedure) until convergence is obtained.
	The hydrodynamic solutions given by Lambert-procedure is valid only in the supersonic region of the wind; the subsonic region needs to be rescaled in order to obtain a continuous solution that couples the hydrostatic structure of the wind.
	
	The Lambert-procedure has proved to converge to the same unique hydrodynamic solution, independent of the initial velocity profile (i.e., terminal velocity and $\beta$-value) chosen (Fig.~\ref{hydros_alt}).
	On the other hand, because Eq.~\ref{motion3} is explicitly independent of density, the mass-loss rate is not recovered by the self-consistent iterations – it needs to be set as an input free parameter.
	However, the dependence on density (and on mass-loss rate too) is \textit{implicitly included} in the modified equation of motion by means of the line-acceleration term (Fig.~\ref{mdotsinitial}).
	The line acceleration also depends on the clumping factor (Fig.~\ref{clumpinitial}), thus the final self-consistent hydrodynamic obtained by the Lambert-procedure depends on the initial values for $\dot M$ and $f_\infty$ introduced to the CMFGEN models.
	However, the choice of the mass-loss rate and clumping factor alters the internal computation of the e.o.m in CMFGEN (Eq.~\ref{clumpedeom}) and therefore it will affect the reliability of the calculated radiative acceleration.
	Improvements to the hydrodynamics in the neighbourhood of the sonic point, if desired, can be achieved iteratively by adjustments in the mass-loss rate and velocity near the sonic point.

	A full understanding of O star stellar winds will require 3D hydrodynamic numerical models with an accurate treatment of the photospheric layers.
	Although progress towards such calculations is being made \citep[for example, \textsc{HDust} from][]{carciofi06,carciofi08} they are prohibitively expensive.
	Thus insights and progress from 1D models are still needed.
	The Lambert procedure discussed in this paper is capable of providing accurate wind dynamics above the sonic point, and can thus be used for spectroscopic wind studies.
	If necessary, adjustments to the mass-loss rate can be made to improve the agreement in the neighbourhood of the sonic point, which appears to be the critical point for these stellar winds.
	However microturbulence and macroturbulence, and the possible existence of convection and clumping in the neighbourhood of the sonic point, will make it difficult to obtain accurate mass-loss rates from first principles.
	Thus the treatment of $\dot M$ as a free-parameter in spectral fitting will still be desirable.

	The main result of our work is the calculation of a self-consistent solution for the velocity profile beyond the $\beta$-law, that is unique given a specific set of stellar parameters and a specific mass-loss rate and clumping factor, enabling us to reduce the number of free parameters to be used in the spectral fitting.
	In a followup study we will explore in more detail a method to find mass-loss rates in conjunction with the Lambert-procedure; and therefore to provide a full NLTE prescription for self-consistent winds, complementary to the works from \citet{sander17} and \citet{sundqvist19}, and in the same line as the m-CAK work done by \citet{alex19}.
	
	Compared with the m-CAK theory, we find that both the mass-loss rate and the terminal velocity obtained by Lambert-procedure are lower, despite the stellar parameters and the atomic information being the same for both cases.
	Even though these differences can be partially explained by the differences in the self-consistent line-accelerations found (indicating that differences in final wind parameters depend on the difference of the methodologies to calculate $g_\text{line}$), a more detailed analysis is needed in order to support this hypothesis.
	It is also important to consider the already known differences between the codes FASTWIND and CMFGEN, as was pointed out by \citet{massey13}.
	
	Finally, despite the large computational effort required in an iterative loop involving CMFGEN, the Lambert-procedure has been demonstrated to be a useful methodology for finding hydrodynamically self-consistent solutions for a stellar wind. Follow-up research will be focused on the application of this prescription for a larger sample of stars.
	
\acknowledgments
	ACGM has been financially supported by the PhD Scholarship folio N$^{\rm o}$ 21161426 from National Commission for Scientific and Technological Research of Chile (CONICYT).
	ACGM and MC acknowledge support from Centro de Astrof\'isica de Valpara\'iso.
	M.C. thanks the support from FONDECYT project 1190485.
	This project has received funding from the European Unions Framework Programme for Research and Innovation Horizon 2020 (2014-2020) under the Marie Sk{\l}odowska-Curie grant Agreement N$^{\rm o}$ 823734.
	The Astronomical Institute Ond\v{r}ejov is supported by the project RVO:67985815.
	Partial support to DJH for this work was provided by STScI grants  HST-AR-14568.001, HST-GO-14683.002-A, and HST-AR-16131.001-A.
	F.N. acknowledges financial support through Spanish grants ESP2017-86582-C4-1-R and PID2019-105552RB-C41 (MINECO/MCIU/AEI/FEDER) and from the Spanish State Research Agency (AEI) through the Unidad de Excelencia “María de Maeztu”-Centro de Astrobiología (CSIC-INTA) project No. MDM-2017-0737.

\bibliography{lambertpaper} 
\bibliographystyle{aasjournal} 

\appendix
\section{Lambert $W$-function}\label{lambertdefinition}
	Equation of motion (Eq. \ref{motion3}) can be solved analytically with the help of a mathematical tool called the \textit{Lambert $W$-function} \citep[also named \textit{product logarithm},][]{corless93,corless96}, which is defined by the inverse of the function:
	
	\begin{equation}\label{invertlambert}
		z(w)=w e^w\;\;,
	\end{equation}
	with $z$ being any complex number.
	That means:
	
	\begin{equation}\label{lambertdef}
		W(z)e^{W(z)}=z\;\;,
	\end{equation}
	
\begin{figure}[htbp]
		\centering
		\includegraphics[width=0.45\linewidth]{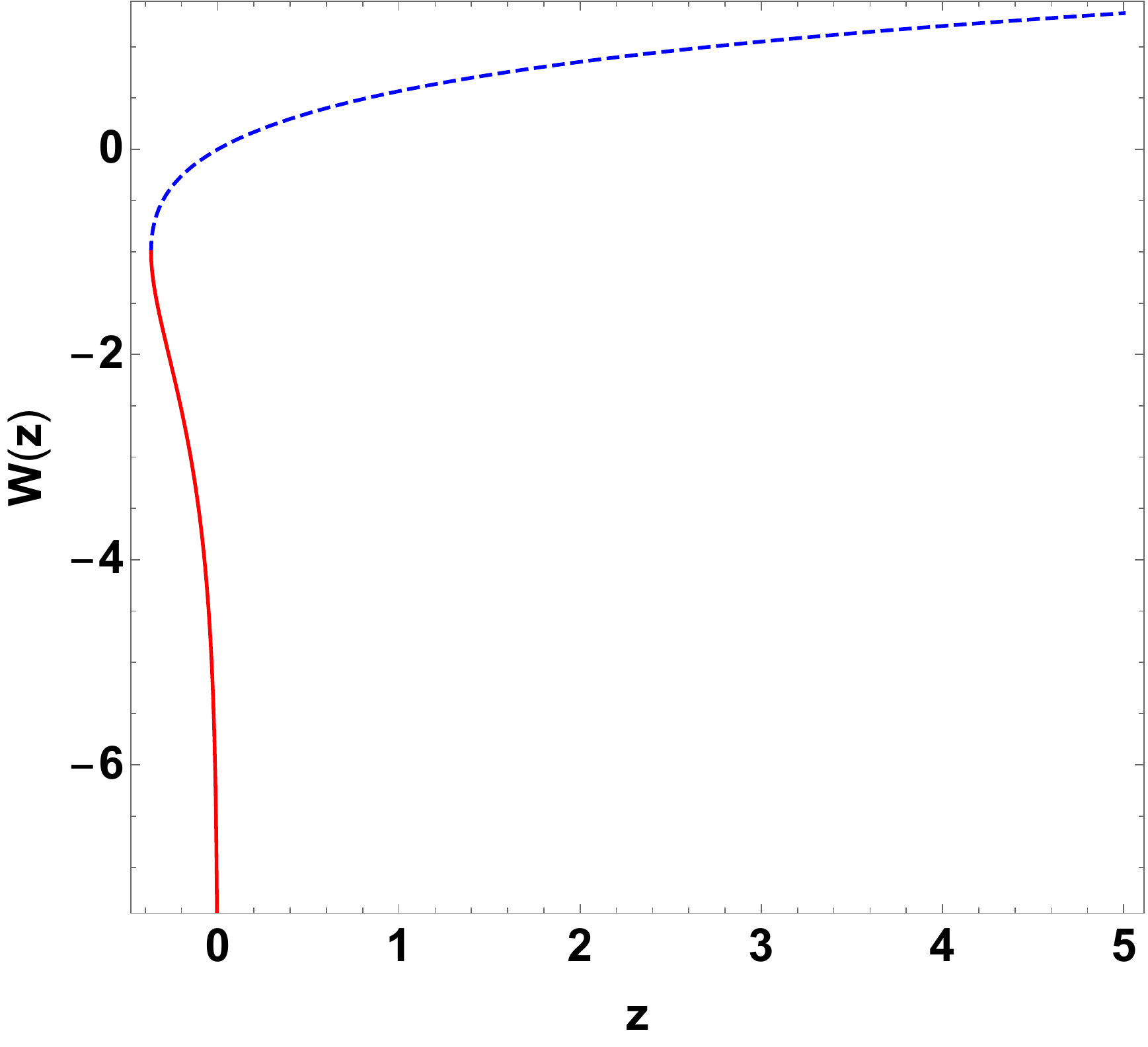}\\
		\caption{Real branches of the $W_k(z)$-Lambert function: $k=0$ (dashed blue line) and $k=-1$ (solid red line).}
		\label{lambertfunc}
	\end{figure}

	The Lambert $W$-function presents infinite solutions depending on all the possible non-zero values that $z$ may take.
	If we limit our search only for the physically relevant case, i.e., real values of $z$, we find that the $W(z)$ function is multivalued because $f(w)=we^w$ is not injective.
	Because of this reason, we split the Lambert $W$-function into two sections, corresponding them to the branches $W_0$ for $W(z)\ge-1$ and $W_{-1}$ for $W(z)\le-1$ (see Figure \ref{lambertfunc}), which are coupled in the lowest value for $z$:
	
	\begin{equation}
		W_0(-1/e)=W_{-1}(-1/e)=-1\;.
	\end{equation}
	
	From Eq.~\ref{invertlambert} and Fig.~\ref{lambertfunc}, it is clearly seen that the domains for $z$ and the codomains for the two branches are\footnote{Notation for open and closed intervals following definition given by the \textit{Encyclopaedia of Mathematics}:\\ \url{https://www.encyclopediaofmath.org/index.php/Interval_and_segment}}:
	
	\begin{equation}
		z\in\left[-1/e,0\right[\;\text{ for }W_{-1}\in[-1,-\infty[\;,
	\end{equation}
	
	\begin{equation}
		z\in\left[-1/e,+\infty\right[\;\text{ for }W_0\in[-1,+\infty[\;.
	\end{equation}
	
	This means, branch $W_{-1}$ diverges when $z$ approaches zero or, in other words, $z$ is an asymptote when $W_{-1}$ tends to $-\infty$.
	Because of this condition, $W_{-1}$ function can be used to solve any equation with the same asymptotic behaviour as $z$, such as the velocity field of a stellar wind.
	Then, the analytical solution for the equation of motion with $g_\text{line}(r)$ given is obtained once we reformulate Eq. \ref{motion1} in terms of the Lambert function, as it is demonstrated in Section~\ref{solutionmomentum}.

\section{Analytical solution of equation of motion}\label{analyticaleom}
	To solve analytically Eq.~\ref{motion3}, we integrate both sides by the normalised radius, from the critical point outwards for the supersonic region or inwards for the subsonic one.
	Because the final solution must be $\hat v(\hat r)$, we employ auxiliar variables $\hat r'$ and $\hat v'$ for the integration:
	
	\begin{equation}\label{integration}
		\int_{\hat r_c}^{\hat r}\left(\hat v'-\frac{1}{\hat v'}\right)\frac{d\hat v'}{d\hat r'}d\hat r'=\int_{\hat r_c}^{\hat r}\left(-\frac{\hat v_\text{crit}^2}{\hat r'^2}+\frac{2}{\hat r'}+\hat g_\text{line}(\hat r')\right)d\hat r'\;.
	\end{equation}
	
	Applying chain-rule on the left side:
	
	\begin{equation}\label{integration2}
		\int_1^{\hat v}\left(\hat v'-\frac{1}{\hat v'}\right)d\hat v'=\int_{\hat r_c}^{\hat r}\left(-\frac{\hat v_\text{crit}^2}{\hat r'^2}+\frac{2}{\hat r'}+\hat g_\text{line}(\hat r')\right)d\hat r'\;.
	\end{equation}
	
	Notice that we have also modified the limits of the left integral, specially considering that $\hat v(\hat r_c)=1$.
	
	This new RHS is:
	
	$$\text{RHS}=\frac{\hat v_\text{crit}^2}{\hat r'}\Bigr\rvert_{\hat r_c}^{\hat r}+2\ln\hat r'\Bigr\rvert_{\hat r_c}^{\hat r}+\int_{\hat r_c}^{\hat r}\hat g_\text{line}\,d\hat r'$$
	\begin{equation}\label{rhs}
		\Rightarrow\text{RHS}=\hat v_\text{crit}^2\left(\frac{1}{\hat r}-\frac{1}{\hat r_c}\right)+2\ln\left(\frac{\hat r}{\hat r_c}\right)+\int_{\hat r_c}^{\hat r}\hat g_\text{line}\,d\hat r
	\end{equation}
	
	Whereas LHS:
	
	$$\text{LHS}=\int_{1}^{\hat v}\hat v'\,d\hat v'-\int_1^{\hat v}\frac{d\hat v'}{\hat v'}$$
	$$\Rightarrow\text{LHS}=\frac{\hat v'^2}{2}\Bigr\rvert_1^{\hat v}-\ln\hat v'\Bigr\rvert_1^{\hat v}$$
	\begin{equation}\label{lhs}
		\Rightarrow\text{LHS}=\left(\frac{\hat v^2}{2}-\frac{1}{2}\right)-\ln\hat v
	\end{equation}
	
	Then, joining Equations \ref{lhs} $\&$ \ref{rhs}:
	
	$$\frac{\hat v^2}{2}-\ln\hat v=\frac{1}{2}+\hat v_\text{crit}^2\left(\frac{1}{\hat r}-\frac{1}{\hat r_c}\right)+2\ln\left(\frac{\hat r}{\hat r_c}\right)+\int_{\hat r_c}^{\hat r}\hat g_\text{line}\,d\hat r$$
	
	Multiplying by 2:
	
	$$\Rightarrow\hat v^2-\ln\hat v^2=1+2\,\hat v_\text{crit}^2\left(\frac{1}{\hat r}-\frac{1}{\hat r_c}\right)+4\ln\left(\frac{\hat r}{\hat r_c}\right)+2\int_{\hat r_c}^{\hat r}\hat g_\text{line}\,d\hat r$$
	
	Inverting signs:
	
	$$\Rightarrow-\hat v^2+\ln\hat v^2=-1-2\,\hat v_\text{crit}^2\left(\frac{1}{\hat r}-\frac{1}{\hat r_c}\right)-4\ln\left(\frac{\hat r}{\hat r_c}\right)-2\int_{\hat r_c}^{\hat r}\hat g_\text{line}\,d\hat r$$
	
	Applying exponential function:
	
	$$\hat v^2e^{-\hat v^2}=\left(\frac{\hat r_c}{\hat r}\right)^4\exp\left[-1-2\,\hat v_\text{crit}^2\left(\frac{1}{\hat r}-\frac{1}{\hat r_c}\right)-2\int_{\hat r_c}^{\hat r}\hat g_\text{line}\,d\hat r\right]$$
	\begin{equation}\label{v2ev2}
		\Rightarrow-\hat v^2e^{-\hat v^2}=-\left(\frac{\hat r_c}{\hat r}\right)^4\exp\left[-1-2\,\hat v_\text{crit}^2\left(\frac{1}{\hat r}-\frac{1}{\hat r_c}\right)-2\int_{\hat r_c}^{\hat r}\hat g_\text{line}\,d\hat r\right]
	\end{equation}
	
	Therefore, we have obtained an expression to be evaluated as Lambert $W$-function, with the RHS being $z$ and $W=-\hat v^2$.
	If we define the final RHS as $x(\hat r)$, the Lambert velocity profile is given by
	
	\begin{equation}\label{lambertvprofile}
		\hat v(\hat r)=\sqrt{-W(x(\hat r))}\;\;,
	\end{equation}
	which is the same expression for the velocity profile outlined by \citet{muller08} and \citet{araya14}.
	Notice that, because of how the variables $\hat r$ and $\hat v$ are defined in the integration of Eq.~\ref{integration} they can take values both smaller and larger than $\hat r_c$ for the case of the dimensionless radius (and smaller and larger than 1 for the case of the dimensionless velocity).

\section{Iteration and convergence of Lambert $W$-function}\label{iter_append}
	Here we outline the steps that assure convergence of the Lambert-procedure.
	\renewcommand{\theenumi}{\roman{enumi}}
	\renewcommand{\labelenumi}{\theenumi}
	\begin{enumerate}
		\item We obtain $x(\hat r)$ by solving the RHS of Eq.~\ref{v2ev2}, from our critical point determined by Eq.~\ref{sonicpoint} outwards.
		The output of CMFGEN gives us $\hat g_\text{line}(\hat r)$, whereas $\hat v_\text{crit}$ is calculated as shown in Eq.~\ref{changeofvariables2}.
		\item A new velocity profile for the supersonic region, $\hat v_\text{super}$, is then calculated using Eq.~\ref{lambertvprofile}.
		\item We couple this new $\hat v_\text{super}$ with the old velocity profile for the subsonic region, $\hat v_\text{sub}$.
		As shown in Fig.~\ref{coupling}, the critical point calculated from Eq.~\ref{sonicpoint} (the sonic point) may not coincide with the sonic point coming from the old velocity profile.
		In this step, it is necessary to perform a quantitative analysis on the difference between both sonic points.
		We define the \textit{difference factor}:
		\begin{equation}\label{diffsonicpoint}
			d_\text{sonic}=\log|\hat r_\text{c,new}-\hat r_\text{c,old}|\;,
		\end{equation}
		with $\hat r_\text{c,new}$ being the critical point determined from Eq.~\ref{sonicpoint}, and $\hat r_\text{c,old}$ the point where $v=a$ in the old velocity profile.
		Low values of $d_\text{sonic}$ ($\lesssim-2$) imply the coupling between both regions can be made only with a slight rescaling of the radius (radius is never modified more than a $0.3\%$).
		The difference factor also provides a hint about the validity, or not, of the adopted values for $\dot M$ and $f_\infty$ used for the CMFGEN models, although it is not possible to establish a direct relation between $d_\text{sonic}$ and mass-loss rate.
		\item The new velocity profile, rescaled in the subsonic region and calculated by the Lambert $W$-function for the supersonic regions, is set as input for a new CMFGEN model.
		\item When CMFGEN finishes, we take the $g_\text{line}$ from the output and step i is repeated.
	\end{enumerate}
	
\begin{figure}[t!]
		\centering
		\includegraphics[width=0.6\linewidth]{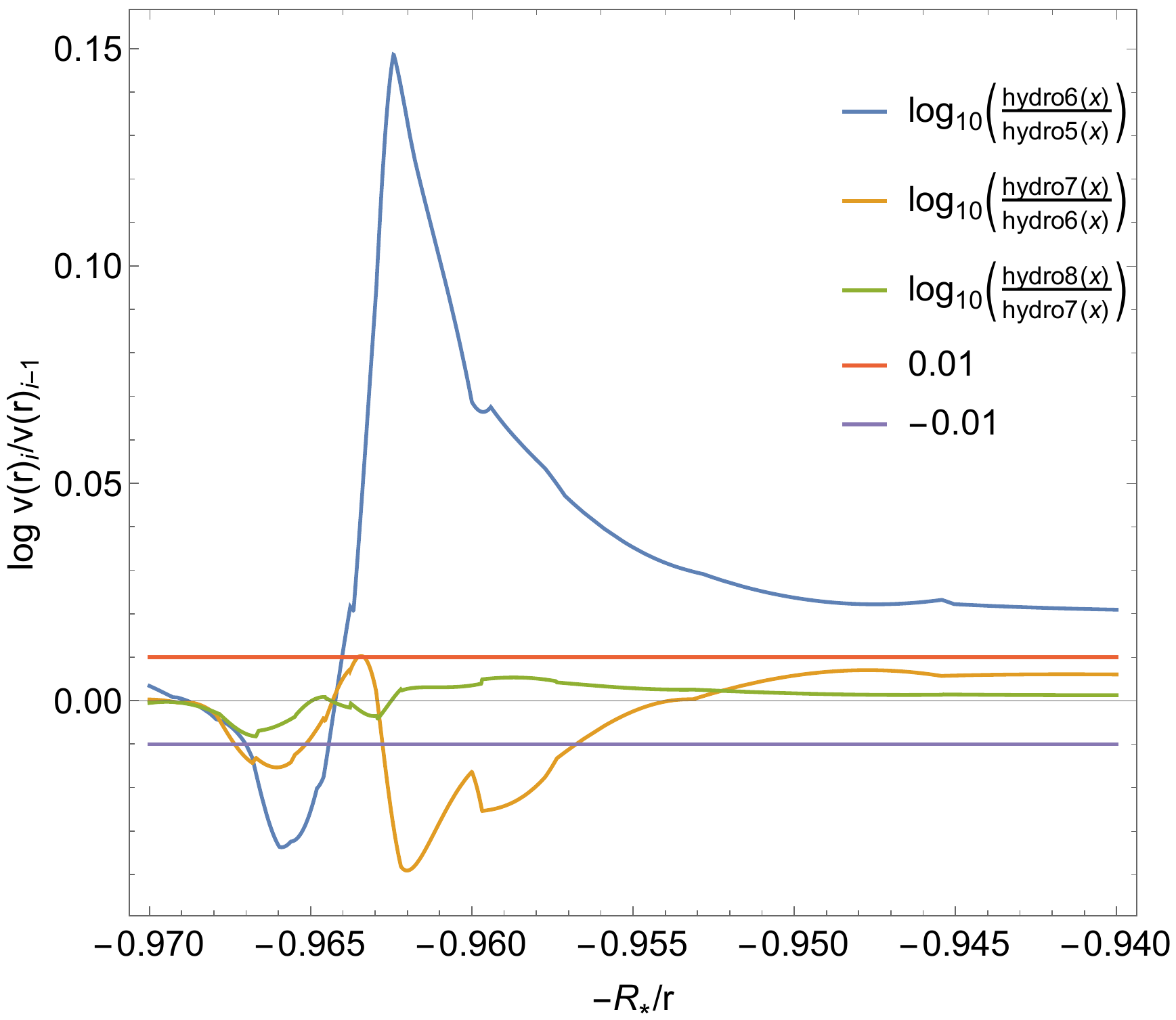}
		\caption{Plot of the last iterations of Eq.~\ref{thresholdv} for the CMFGEN models of HD 163758 with $\dot M=1.2\times10^{-6}$ and $f_\infty=0.05$. Threshold of $\pm0.01$ is represented by the straight lines.}
		\label{iterationsHD163758}
	\end{figure}

	This procedure continues the until convergence criteria are satisfied.
	As an example, we illustrate the convergence for a model of HD 163758 (Fig.~\ref{iterationsHD163758}).
	Because the ratio between the last two iterations (7 and 8) lies inside the thresholds, we consider the Lambert-procedure for HD 163758 with $1.2\times10^{-6}$ for mass-loss rate and $f_\infty=0.05$ for filling factor to be Lambert-converged.

\section{Synthetic spectra for Lambert-models}\label{cmfgenspectra}

	\begin{figure*}[t!]
		\centering
		\includegraphics[width=0.9\linewidth]{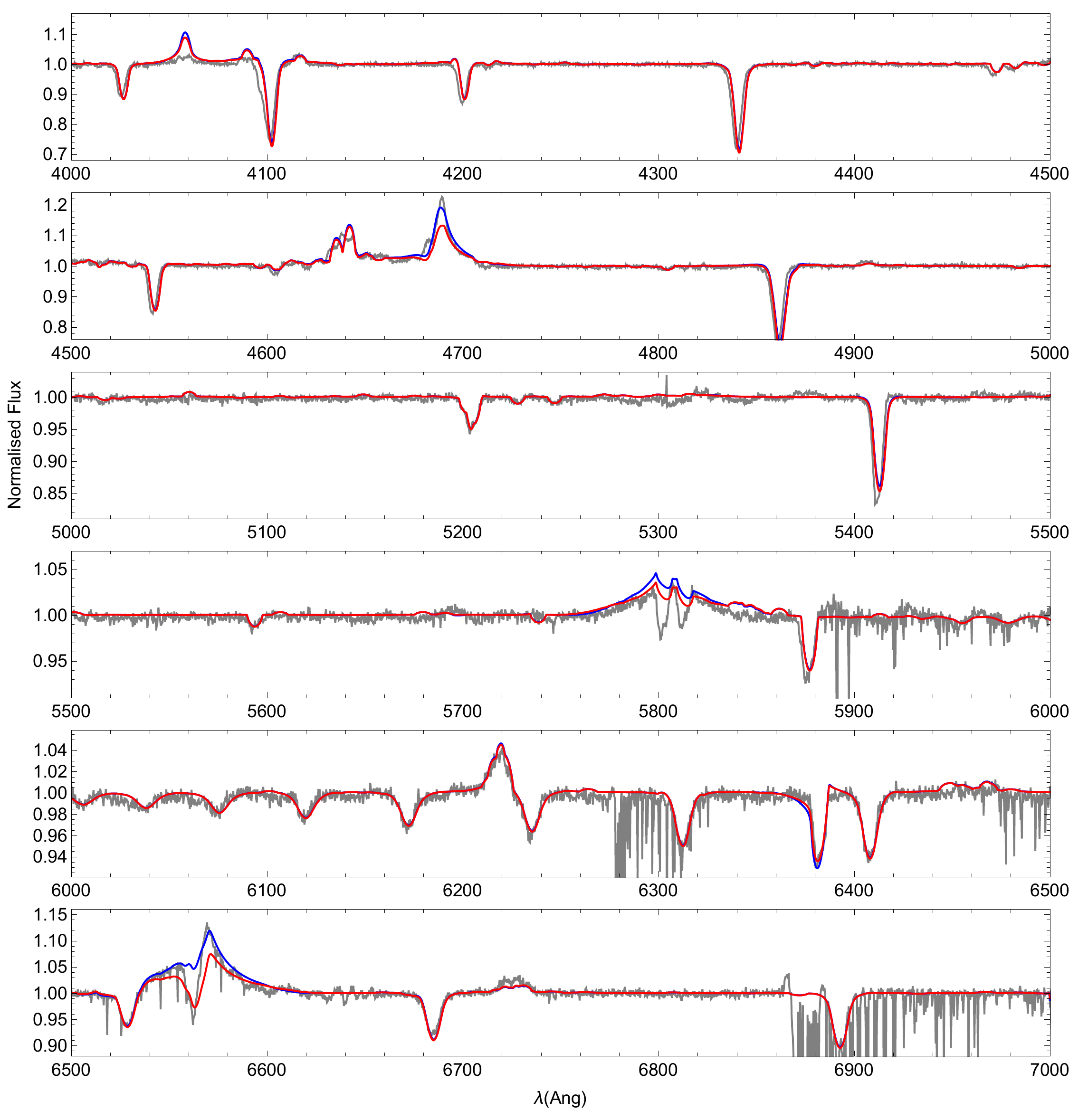}
		\caption{Comparison of the Lambert-solution for $\zeta$-Puppis (red), compared with the initial CMFGEN spectrum (blue) with the observed FEROS spectrum for the visible range from 4\,000 to 7\,000 $\AA$. These spectra were computed using CMF\_FLUX \citep{busche05}, and rotational broadening was taken into account using a standard convolution procedures. However the H$_\alpha$ profile (and particularly the absorption component), for example, is modified when rotational is explicitly calculated using a formal solution of the transfer equation \citep{hillier12}.}
		\label{lamb022visiblefit}
	\end{figure*}
	
	\begin{figure*}[t!]
		\centering
		\includegraphics[width=0.9\linewidth]{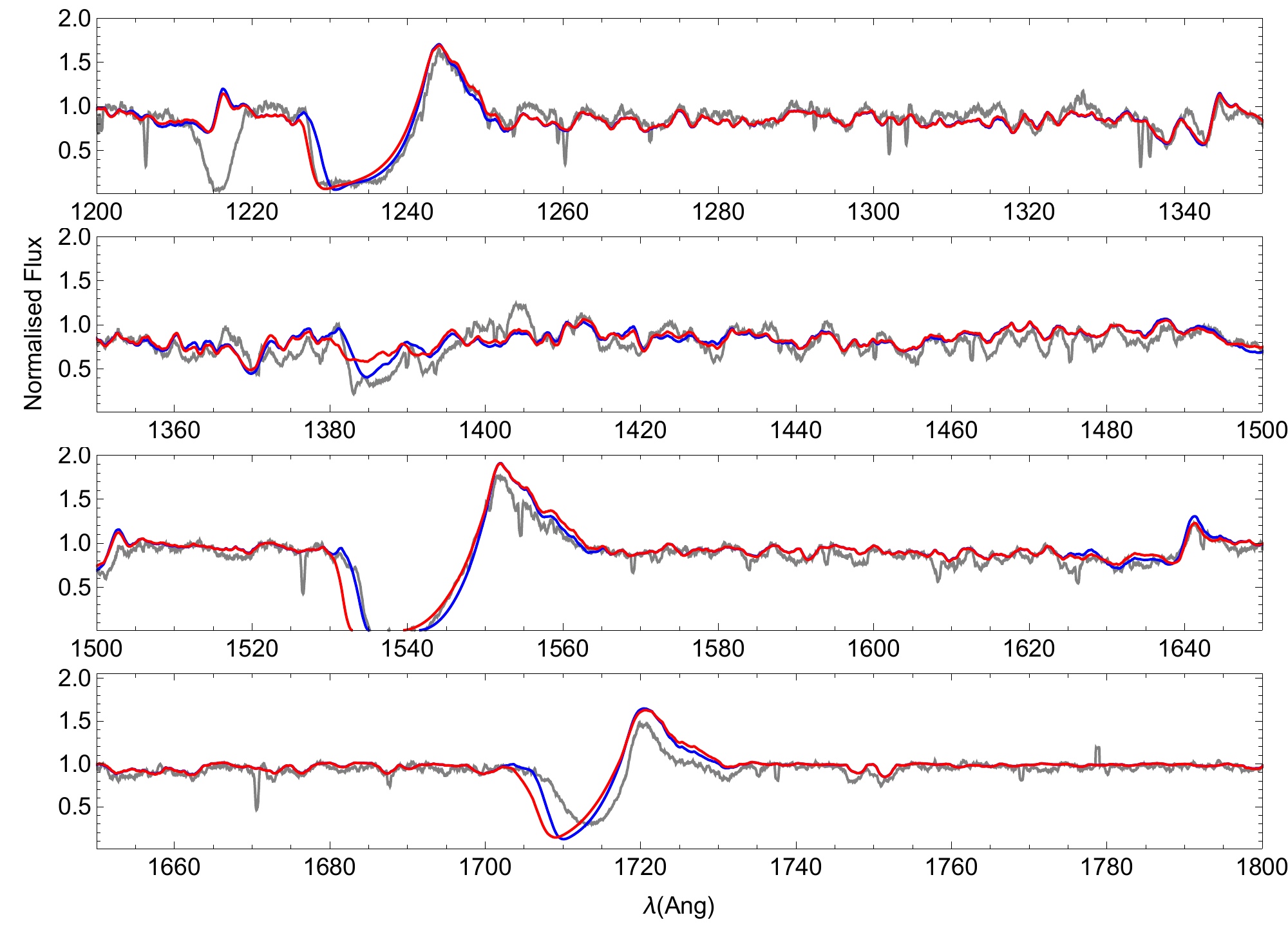}
		\caption{Comparison of the Lambert-solution for $\zeta$-Puppis (red), compared with the initial CMFGEN spectrum (blue) with the observed IUE spectrum for the ultraviolet range from 1\,200 to 1\,800 $\AA$.}
		\label{lamb022uvfit}
	\end{figure*}
	
	\begin{figure*}[t!]
		\centering
		\includegraphics[width=0.9\linewidth]{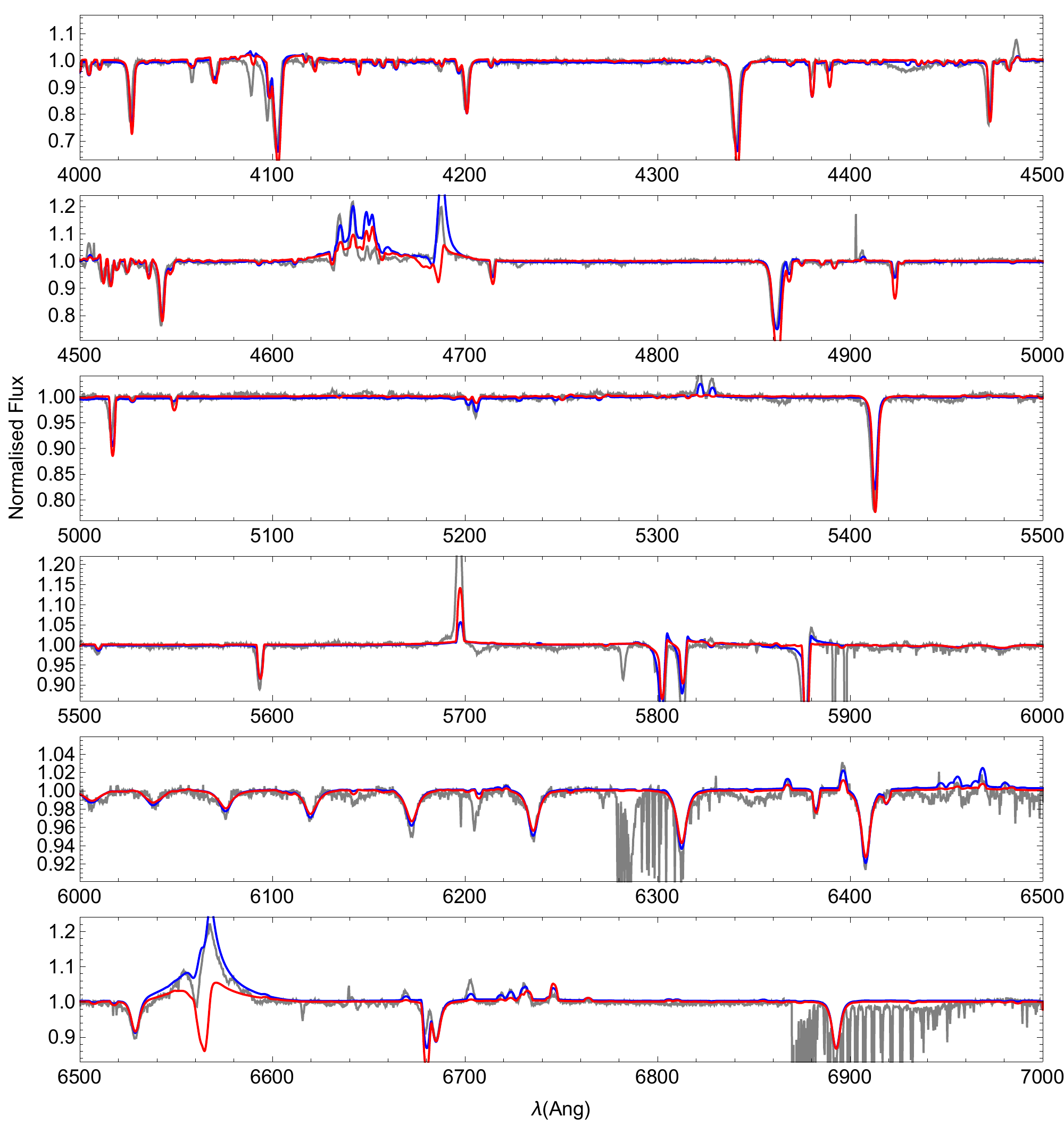}
		\caption{Comparison of the Lambert-solution for HD 163758 (red), compared with the initial CMFGEN spectrum (blue) with the observed FEROS spectrum for the visible range from 4\,000 to 7\,000 $\AA$.}
		\label{lamb404visiblefit}
	\end{figure*}
	
	\begin{figure*}[t!]
		\centering
		\includegraphics[width=0.9\linewidth]{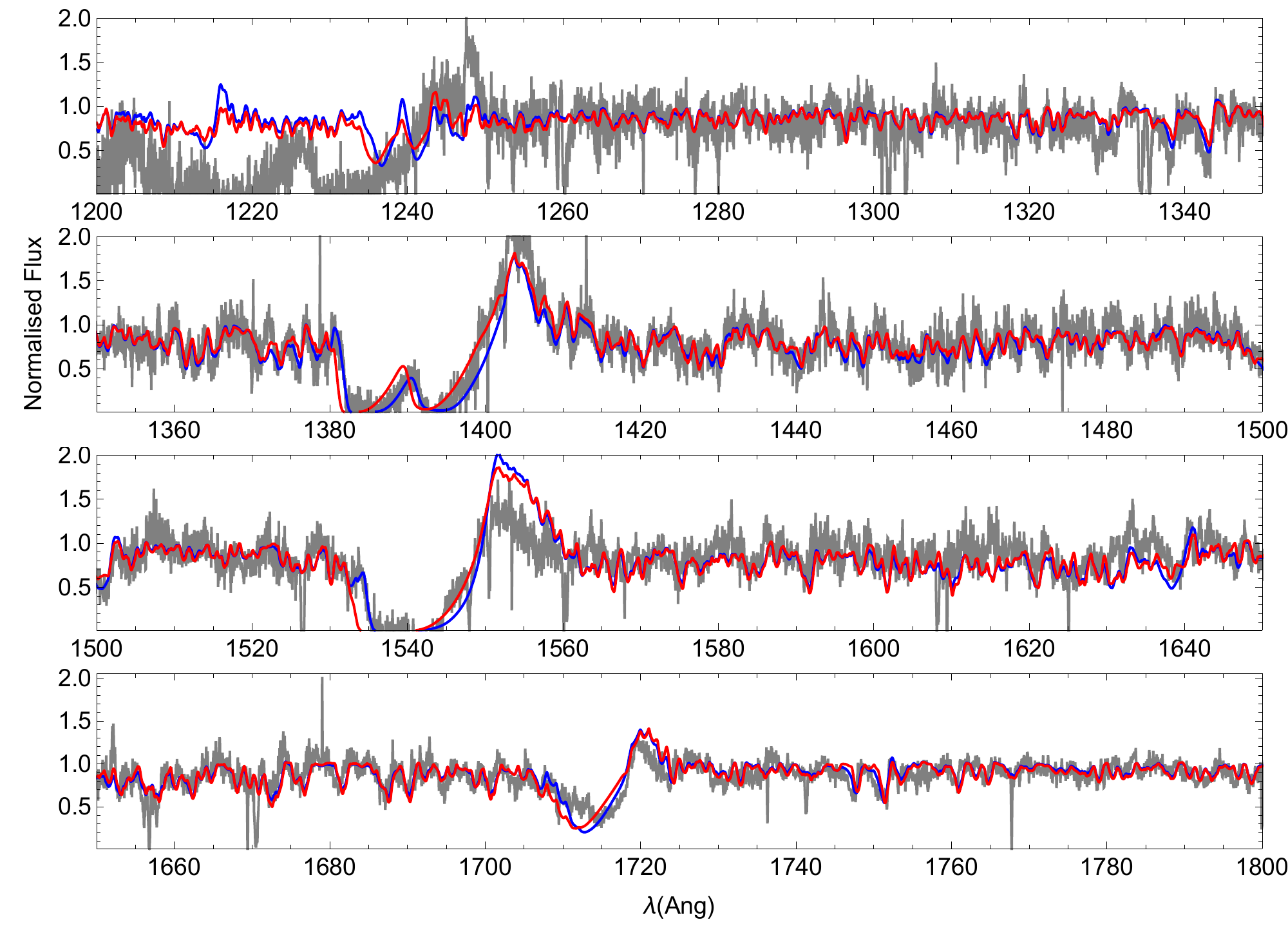}
		\caption{Comparison of the Lambert-solution for HD 163758 (red), compared with the initial CMFGEN spectrum (blue) with the observed IUE spectrum for the ultraviolet range from 1\,200 to 1\,800 $\AA$.}
		\label{lamb404uvfit}
	\end{figure*}
	
	\begin{figure*}[t!]
		\centering
		\includegraphics[width=0.9\linewidth]{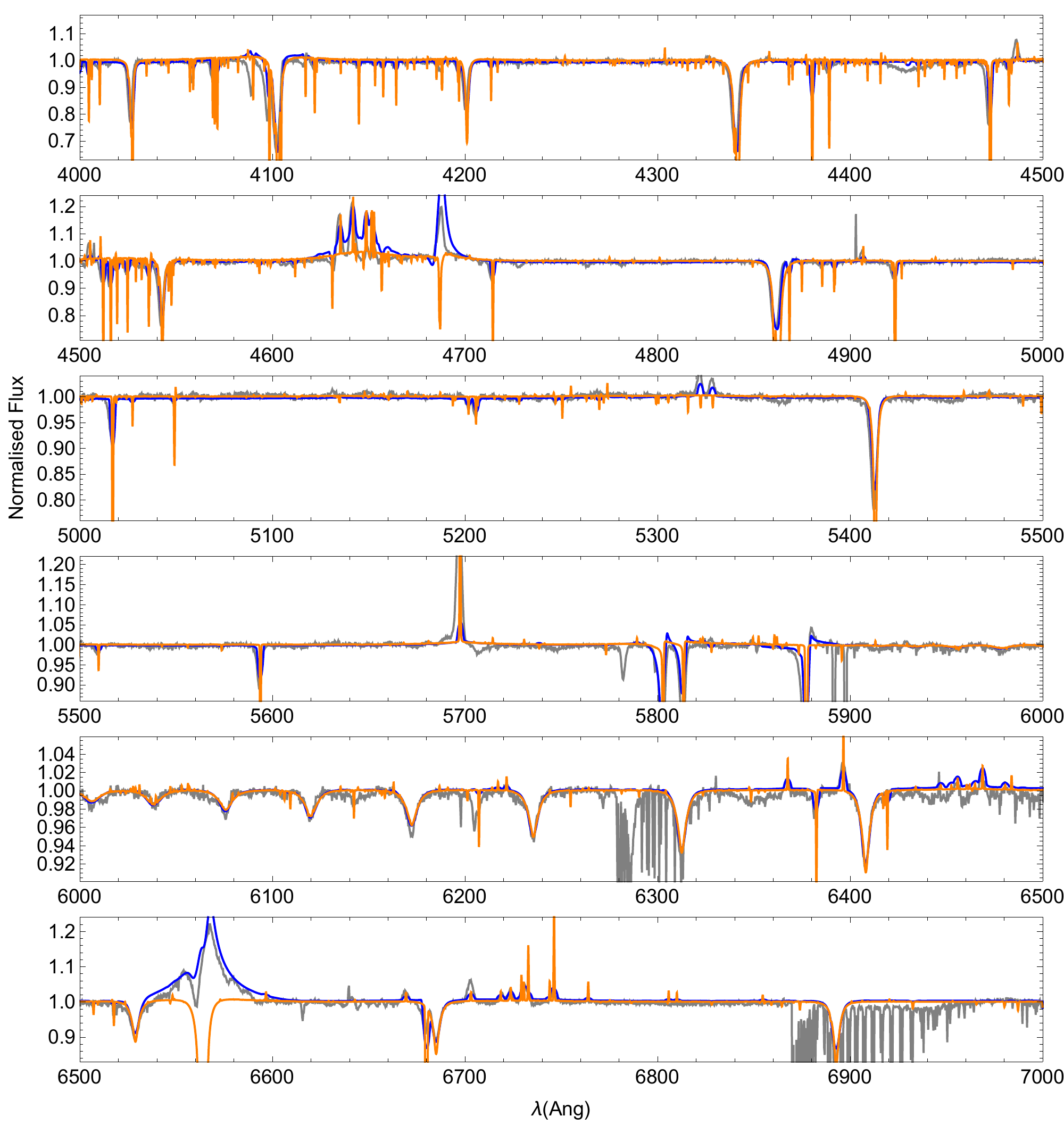}
		\caption{Comparison of the alternative Lambert-solution for HD 163758 (orange), compared with the initial CMFGEN spectrum (blue) with the observed FEROS spectrum for the visible range from 4\,000 to 7\,000 $\AA$.}
		\label{lamb833visiblefit}
	\end{figure*}
	
	\begin{figure*}[t!]
		\centering
		\includegraphics[width=0.9\linewidth]{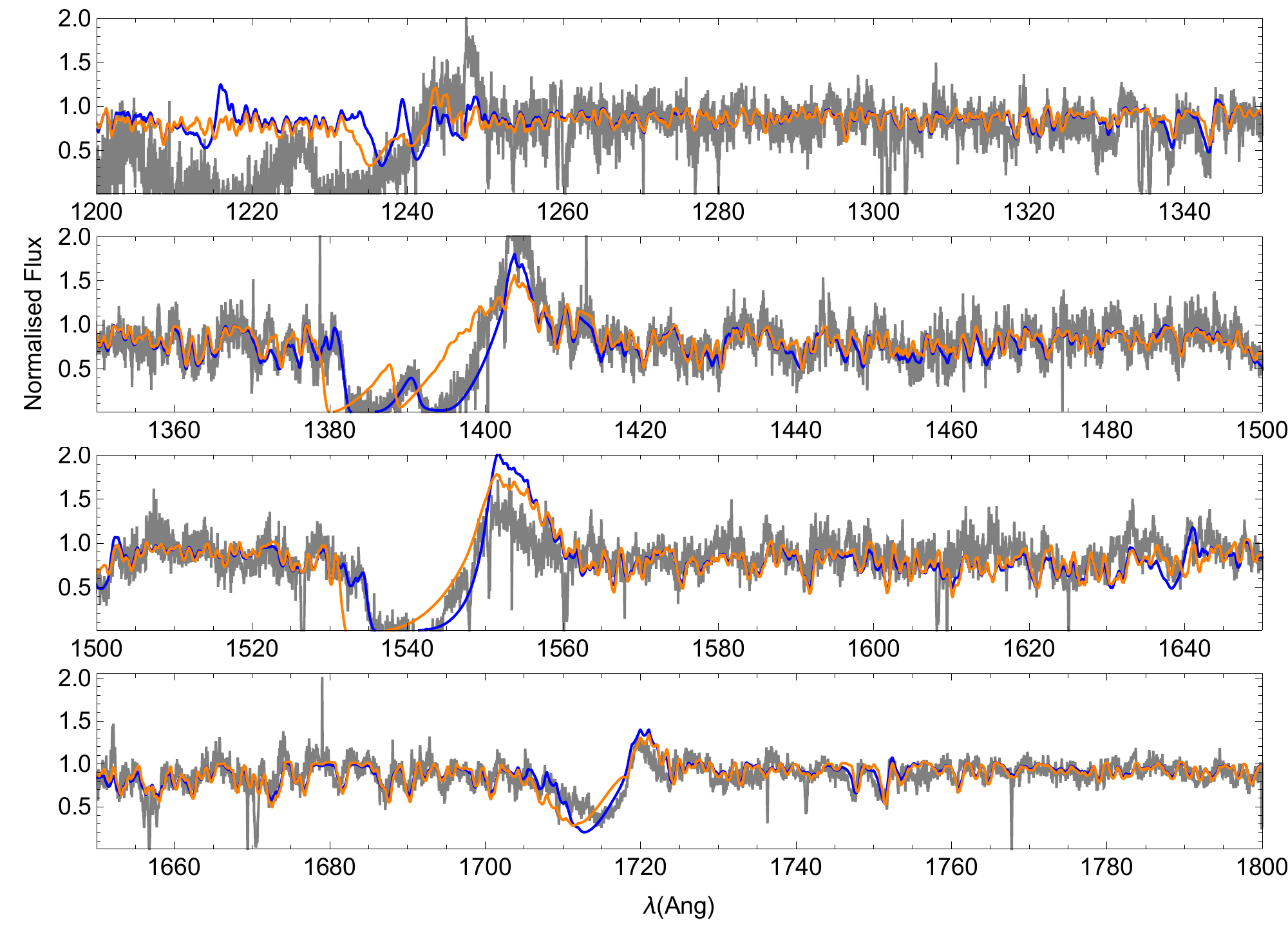}
		\caption{Comparison of the alternative Lambert-solution for HD 163758 (orange), compared with the initial CMFGEN spectrum (blue) with the observed IUE spectrum for the ultraviolet range from 1\,200 to 1\,800 $\AA$.}
		\label{lamb833uvfit}
	\end{figure*}
	
	\begin{figure*}[t!]
		\centering
		\includegraphics[width=0.9\linewidth]{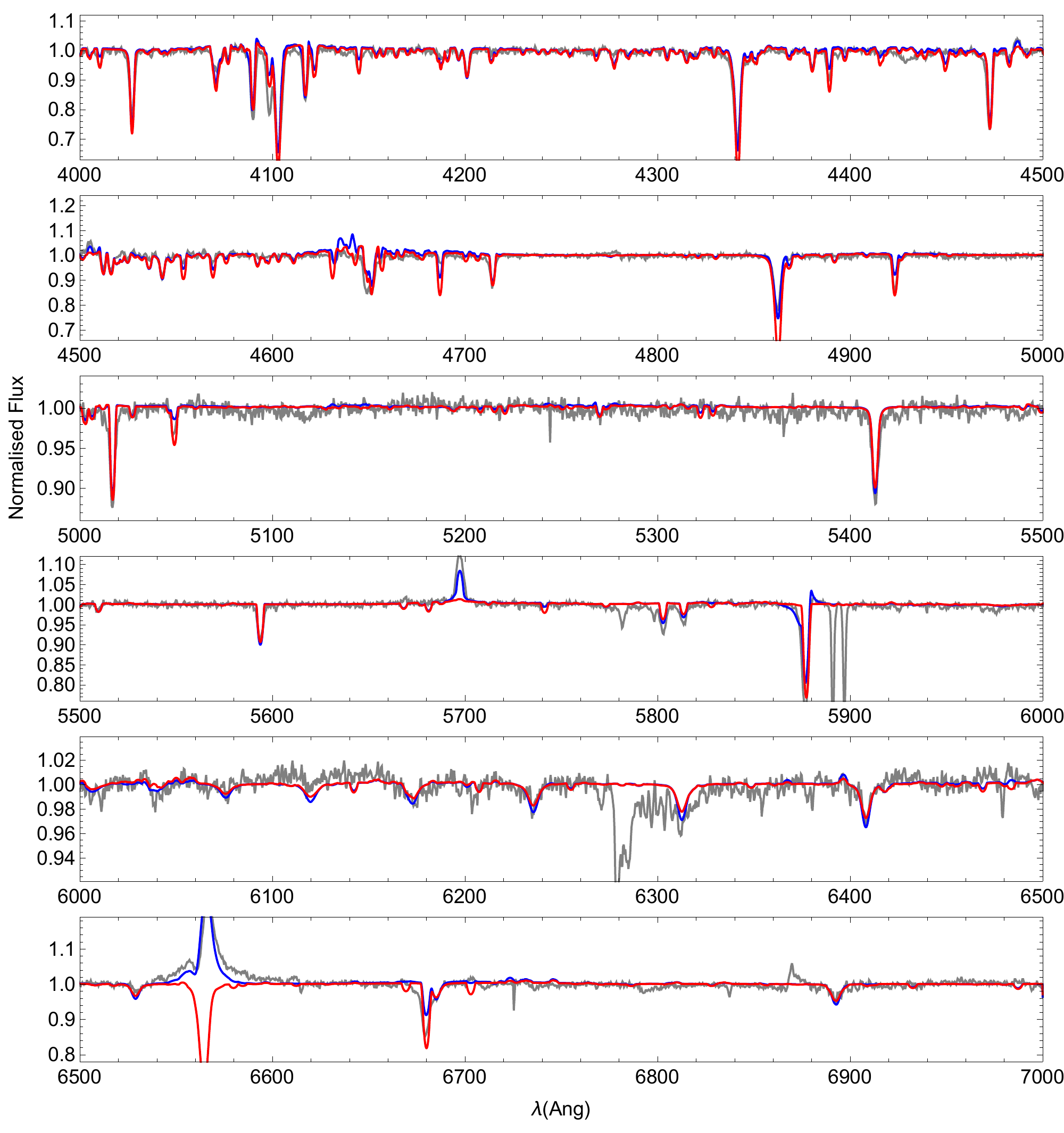}
		\caption{Comparison of the Lambert-solution for $\alpha$-Cam (red), compared with the initial CMFGEN spectrum (blue) with the observed INDO-US library spectrum for the visible range from 4\,000 to 7\,000 $\AA$.}
		\label{acam172visiblefit}
	\end{figure*}
	
	\begin{figure*}[t!]
		\centering
		\includegraphics[width=0.9\linewidth]{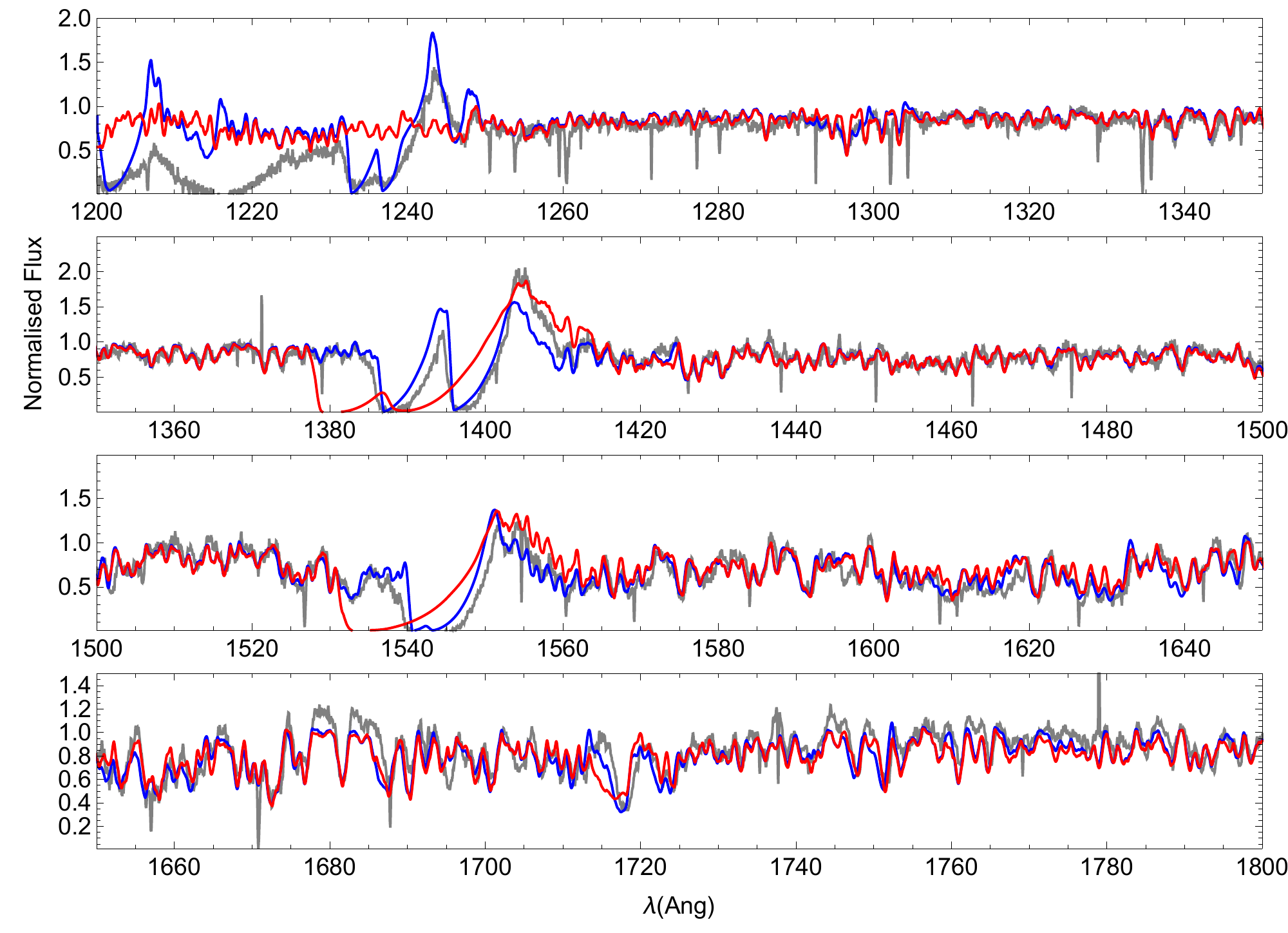}
		\caption{Comparison of the Lambert-solution for $\alpha$-Cam (red), compared with the initial CMFGEN spectrum (blue) with the observed IUE spectrum for the ultraviolet range from 1\,200 to 1\,800 $\AA$.}
		\label{acam172uvfit}
	\end{figure*}
	
	\begin{figure*}[t!]
		\centering
		\includegraphics[width=0.9\linewidth]{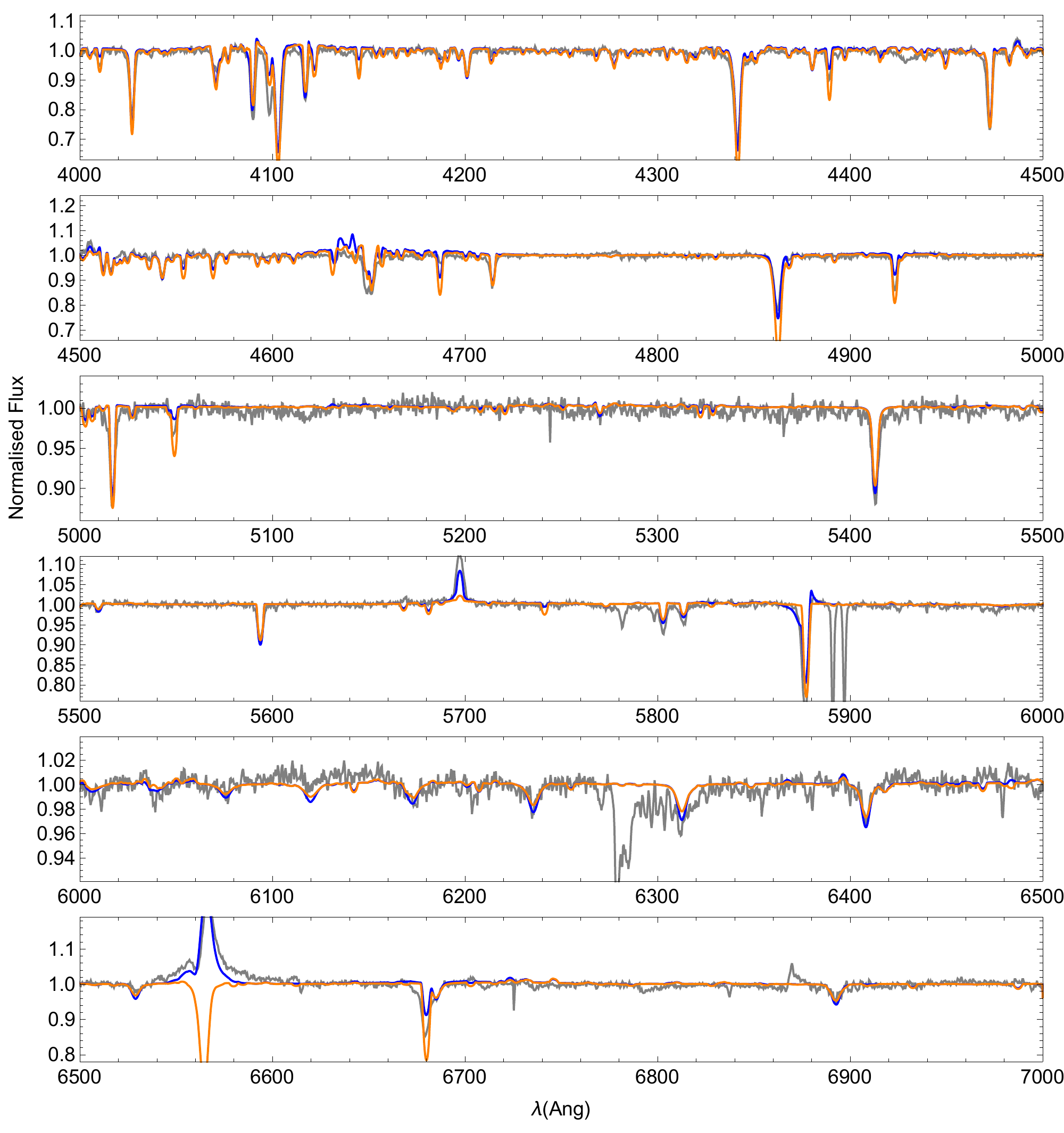}
		\caption{Comparison of the the alternative Lambert-solution for $\alpha$-Cam (orange), compared with the initial CMFGEN spectrum (blue) with the observed INDO-US library spectrum for the visible range from 4\,000 to 7\,000 $\AA$.}
		\label{acam186visiblefit}
	\end{figure*}
	
	\begin{figure*}[t!]
		\centering
		\includegraphics[width=0.9\linewidth]{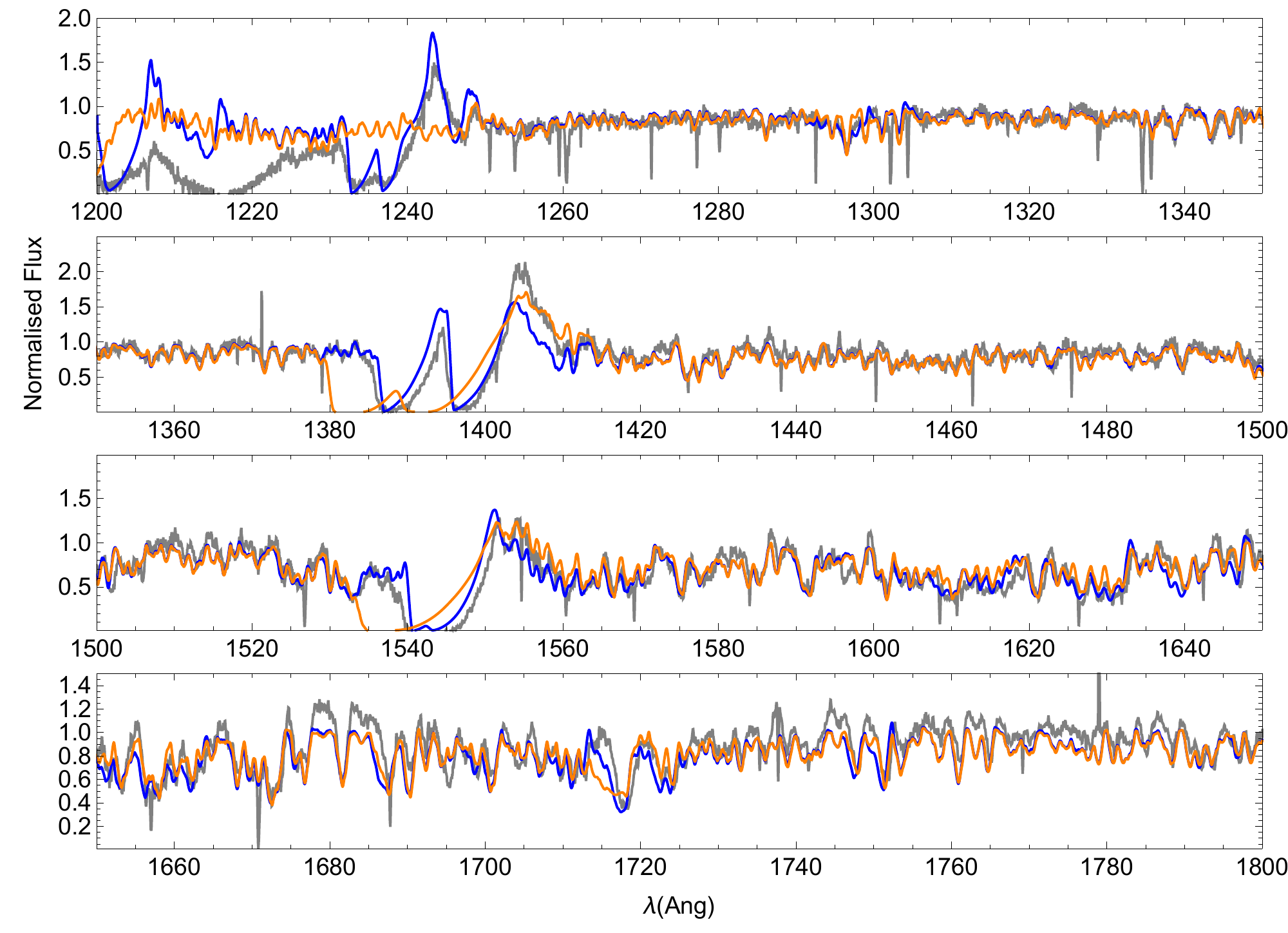}
		\caption{Comparison of the the alternative Lambert-solution for $\alpha$-Cam (orange), compared with the initial CMFGEN spectrum (blue) with the observed IUE spectrum for the ultraviolet range from 1\,200 to 1\,800 $\AA$.}
		\label{acam186uvfit}
	\end{figure*}

\newpage
\section{List of species for CMFGEN models}\label{species}
	Species used in the CMFGEN models are presented below in Tables~\ref{atomicspecies} and ~\ref{atomicspeciescont}.
	
	\begin{table*}[h!]
		\caption{Number of energy levels of the atomic species used for the CMFGEN models.}
		\label{atomicspecies} 
		\centering
		\begin{tabular}{lrrr}
			\hline\hline
			Ion & $\zeta$-Puppis & HD 163758 & $\alpha$-Cam\\
			\hline
			H I  & 30 & 30 & 30\\
			He I & 69 & 69 & 69\\
			He II & 30 & 30 & 30\\
			C II & – & 48 & 45\\
			C III & 33 & 99 & 140\\
			C IV & 43 & 64 & 59\\
			N II & – & 45 & 38\\
			N III & 117 & 41 & 62\\
			N IV & 69 & 44 & 140\\
			N V & 33 & 41 & 9\\
			O II & – & 54 & 81\\
			O III & 86 & 88 & 86\\
			O IV & 48 & 38 & 96\\
			O V & 41 & 41 & 19\\
			O VI  & 25 & 25 & –\\
			Ne III & 23 & 32 & 55\\
			Ne IV & 17 & 17 & 30\\
			Ne V & 37 & 37 & 31\\
			Mg II & 18 & 27 & 37\\
			Mg III & 29 & 29 & 29\\
			Mg IV & 27 & 27 & –\\
			Al III & – & – & 27\\
			Si III & 44 & 33 & 47\\
			Si IV & 22 & 55 & 22\\
			P IV & 30 & 36 & 13\\
			P V & 16 & 16 & 17\\
			S III & 13 & 24 & 31\\
			S IV & 51 & 51 & 65\\
			S V & 40 & 40 & 31\\
			Cl IV & 40 & – & 35\\
			Cl V & 41 & 26 & 26\\
			Cl VI & 32 & 18 & 18\\
			Cl VII & 37 & 37 & –\\
			Ar III & 10 & 10 & 31\\
			Ar IV & 31 & 31 & 31\\
			Ar V & 38 & 38 & 46\\
			Ar VI & 21 & 21 & 21\\
			Ar VII & 30 & 30 & –\\
			Ar VIII & 28 & 28 & –\\
			Ca III & 33 & 33 & 44\\
			Ca IV & 43 & 43 & 34\\
			Ca V & 73 & 73 & 46\\
			Ca VI & 47 & 47 & –\\
			Ca VII & 48 & 48 & –\\
			\hline
		\end{tabular}
	\end{table*}
	
	\begin{table*}[h!]
		\caption{Number of energy levels of the atomic species used for the CMFGEN models (cont.).}
		\label{atomicspeciescont} 
		\centering
		\begin{tabular}{lrrr}
			\hline\hline
			Ion & $\zeta$-Puppis & HD 163758 & $\alpha$-Cam\\
			\hline
			Cr IV & 29 & 29 & 30\\
			Cr V & 30 & 30 & 32\\
			Cr VI & 30 & 30 & –\\
			Mn III & – & 31 & –\\
			Mn IV & 39 & 39 & 39\\
			Mn V & 16 & 16 & 16\\
			Mn VI & 23 & 23 & 23\\
			Mn VII & 20 & 20 & –\\
			Fe III & – & 104 & 104\\
			Fe IV & 110 & 110 & 110\\
			Fe V & 139 & 139 & 101\\
			Fe VI & 59 & 158 & 44\\
			Fe VII & 29 & 29 & –\\
			Ni IV & 116 & 37 & 111\\
			Ni V & 152 & 46 & 140\\
			Ni VI & 62 & 37 & 37\\
			Ni VII & 37 & 37 & –\\
			\hline
		\end{tabular}
	\end{table*}

\end{document}